\documentclass[english,showpacs,twocolumn,superscriptaddress,showpacs,showkeys,longbibliography]{revtex4-1}
\usepackage{graphicx}
\usepackage{chngcntr}
\usepackage[export]{adjustbox}
\usepackage{rotating}
\usepackage{dcolumn}
\usepackage{bm}
\usepackage[usenames,dvipsnames]{color}
\usepackage{amsmath}
\usepackage{amssymb}
\usepackage{subcaption}
\usepackage[utf8]{inputenc}
\usepackage{fancybox}
\usepackage{orcidlink}
\usepackage[colorlinks=true,linkcolor=blue,citecolor=green,urlcolor=magenta]{hyperref} 
\usepackage[nameinlink]{cleveref} 
\setcitestyle{super}

\graphicspath{{./Files/}{./NewPlots/}}

\begin{document}


\title{The Role of Helical and Non-Helical Drives on the evolution of Self-Consistent Dynamos}
\author{Shishir Biswas$^{\orcidlink{0000-0002-4879-8889}}$} 
\email{shishirbeafriend@gmail.com}
\email{shishir.biswas@ipr.res.in}
\affiliation{Institute for Plasma Research, Bhat, Gandhinagar, Gujarat  382428, India}
\affiliation{Homi Bhabha National Institute, Training School Complex, Anushaktinagar, Mumbai 400094, India}

\author {Rajaraman Ganesh$^{\orcidlink{0000-0002-5095-3067}}$}
\affiliation{Institute for Plasma Research, Bhat, Gandhinagar, Gujarat 382428, India}
\affiliation{Homi Bhabha National Institute, Training School Complex, Anushaktinagar, Mumbai 400094, India}
\date{\today}
%


\begin{abstract}
In the self-consistent dynamo limit, the magnetic feedback on the velocity field is sufficiently strong to induce a change in the topology of the magnetic field. Consequently, the magnetic energy reaches a state of non-linear saturation. Here, we investigate the role played by helical and non-helical drives in the triggering and the eventual saturation of a self-consistent dynamo. Evidence of small-scale dynamo (SSD) activity is found for both helical and non-helical forcing, driven at the largest possible scale. Based on the spectrum analysis, we find that the evolution of kinetic energy follows Kolmogorov’s $k^{-\frac{5}{3}}$ law while that of magnetic energy follows Kazantsev’s $k^{\frac{3}{2}}$ scaling. Also, we have verified that the aforementioned  scalings remain valid for various magnetic Prandtl numbers ($P_m$). Statistical analysis is found to support our numerical finds. 
\end{abstract} 

\maketitle

\section{Introduction}
A substantial portion of the universe is magnetized. Multi-scale magnetic fields are detected in a variety of astrophysical entities such as the solar system, planets, galaxy clusters, interstellar medium, star cores, accretion disks, and spiral galaxies. One pertinent issue that arises is: what are the sources of this multi-scale magnetic field? The turbulent dynamo mechanism, which is the primary means of converting the kinetic energy of the turbulent motion of conducting fluids into magnetic energy, is thought to amplify and sustain magnetic fields in the majority of these systems. The dynamo process naturally necessitates a weak seed field, which eventually becomes stronger and comparable to the turbulent kinetic energy \cite{Larmor:1919,Parker_APJ:1955,Parker:1979,MOFFATT:1978}. One possible explanation for the formation of these seed magnetic fields is the Biermann battery mechanism \cite{Biermann:1950}, wherein a fragile magnetic field is produced by the separation of ions and electrons. This mechanism can occur during several astrophysical events such as cosmic reionization in photogalaxies, and during the development of large-scale structures. Alternatively, Weibel instability \cite{Shukla_Weibel:2003} is known to produce magnetic fields from counter-streaming plasmas, which are also regarded the source of seed magnetic fields.


Dynamos are categorized into two distinct flavours based on the feedback strength of the magnetic field onto the flow field. For example, a linear dynamo is one in which the magnetic field dynamics does not ``back react'' with the velocity field and the velocity field is either given or it obeys the Navier–Stokes equation. A nonlinear dynamo or self-consistent dynamo is when the nonlinear effects start to change the flow (once the magnetic field is large enough) to stop further magnetic field growth, that is, the flow and B-field ``back react'' on each other, typically leading to nonlinear saturation.
Several reviews discuss the theoretical, numerical, and observational aspects of dynamo activity \cite{Widrow_RMP:2002, Brandenburg_Physics_Report:2005, Kulsrud_Review:2008, Federrath_Review:2016, Rincon:2019, Tobias_Review:2021, Brandenburg_Evangelia:2023}.

In addition to the classification previously described, dynamos are further categorized into two subgroups based on the length scales involved. The first type is the large scale dynamo (LSD) or mean field dynamo, which refers to the increase of magnetic field on a scale that exceeds the forcing length scale of fluid motion. Large scale dynamo activity necessitates symmetry breaking, such as fluid helicity (mirror symmetry breaking), differential rotation, density lamination, etc. Numerical investigations have shown that significant scale separation between the energy-containing scale and the largest scale of the flow can lead to the generation of large-scale dynamo activity through the helical drive, which is not mirror symmetric. The Large Scale Dynamo (LSD) activity is sometimes known as the $\alpha$-dynamo. It has been
determined that the $\alpha$-dynamo effect is proportional to the inverse of the magnetic Reynolds
number ($R_m$), expressed as $\alpha \propto \frac{1}{R_m}$. The scale separation enhances the inverse cascade of magnetic helicity, leading to the generation of large-scale magnetic fields \cite{Meneguzzi_PRL:1981,Balsara_PoP:1999, Brandenburg_APJ:2001, Gomez:2004, Haugen_Brandenburg_PRE:2004, Boldyrev_PRL:2005, Alexakis_APJ:2006, Mininni_PRE:2007, Brandenburg_APJ:2009, Ponty_PRL:2011, Bhat_MNRAS:2016, Bhat_MNRAS:2021,Bermudez_Alexakis_PRL:2022, Zhou_Blackman_PRE:2024}. Large-scale dynamo action is crucial for understanding the magnetic field in the solar photosphere. The helical motion in the solar convection zone creates the $\alpha$ effect, which can generate a poloidal field that transforms into a toroidal field \cite{Cattaneo_APJ:1999}. It is a crucial element in the 11-year solar cycle, leading to the emergence of the solar butterfly diagram \cite{Stix_Sun_Butterfly:2004}.


Here we focus on the second kind of dynamo, namely the small-scale or fluctuation dynamo (SSD), which is defined as the growth of magnetic-fluctuation energy at or below the forcing length scale of the fluid motion \cite{Kazantsev:1968, Meneguzzi_PRL:1981, Kulsrud_APJ:1992, Cattaneo_APJ:1999,Schekochihin_APJ:2002, Schekochihin_PRL_FEB1:2004, Schekochihin_PRL_FEB2:2004, Haugen_Brandenburg_Dobler_PRE:2004, Schekochihin_Cowley_APJ:2004, Schekochihin_APJ:2005, Mininni_Ponty_APJ_JUN:2005, Ponty_Mininni_PRL_APR:2005, Mininni_Montgomery_PRE:2005,Mininni_POP:2006, Ponty_NJP:2007, Schekochihin_NJP:2007, Iskakov_PRL:2007, Brandenburg_APJ:2009, Schober_PRE:2012, Brandenburg_MNRAS:2018, Seta_MNRAS:2020, Seta_PRF:2020, Seta_PRF:2021, Skoutnev_APJ:2021, Seta_MNRAS:2022, Biswas_Scripta:2023, Warnecke_Nature:2023}. Fluid helicity is believed to not to impose any constraints for this type of dynamo action. Small scale dynamos (SSD) are essential in numerous astrophysical scenarios such as spiral galaxies \cite{Kulsrud_Review:2008}, elliptical galaxies \cite{Elliptic_Galaxy_APJ:1996}, galaxy clusters \cite{Galaxy_cluster_MNRAS:2017}, and stars \cite{Cattaneo_APJ:1999}. In addition, small-scale dynamos influence the dynamic processes of the Sun, such as angular momentum transport, the formation of differential rotation, and contribute to coronal heating. Accretion disk physics involves the presence of small-scale magnetic fields, which introduce a new paradigm for angular momentum transport. Therefore, it is evident that small-scale magnetic fields are ubiquitous in nature, similar to turbulence.

The previous discussion indicates that homogeneous isotropic turbulence lacking mirror symmetry leads to $\alpha$-dynamo. Additionally, flow shear and differential rotation have a pervasive impact in a wide range of astrophysical situations. The paradigm of rapid increase in magnetic field resulting from the interplay between small-scale velocity fluctuations and a large-scale velocity shear is known as the ``shear dynamo problem'' \cite{Tobias_Nature:2013, Biswas_POP:2023}. The interaction of dynamo action with flow shear can create the so-called $\omega$-dynamo, while helical turbulence combined with shear can result in an oscillatory dynamo or a $\alpha-\omega$ dynamo. Additionally, dynamo instability is affected by the Coriolis force as a consequence of differential rotation. In a rapidly rotating system, the flow becomes effectively 2-dimensional because the Coriolis force suppresses fluctuations in the direction of rotation \cite{Seshasayanan_Alexakis_JFM1:2016, Seshasayanan_Alexakis_JFM2:2016}. The transformation of the flow into two dimensions modifies statistical features of the background turbulence, such as altering the direction of the energy cascade from forward to inverse. 

\begin{figure*}
	\centering
	\includegraphics[scale=0.28]{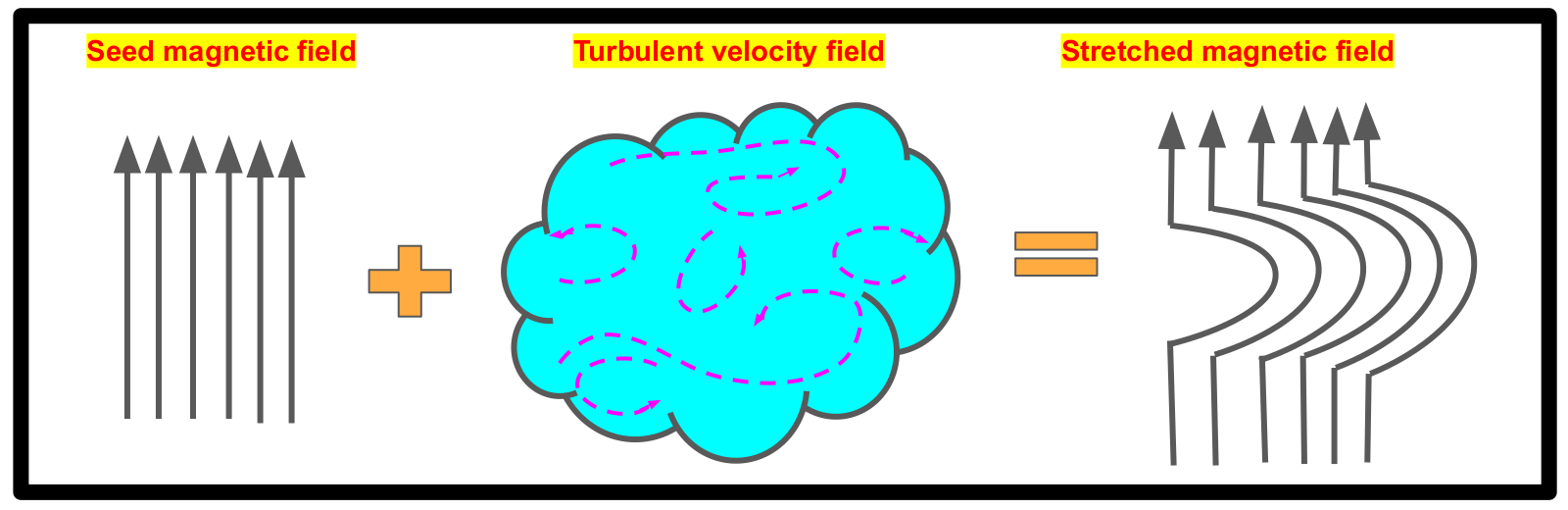}
	\caption{Schematic illustrating the interaction between the seed magnetic field and turbulent velocity field leading to the stretching of the magnetic field.} 
		\label{basic dynamo}
\end{figure*}
\begin{figure*}
		\centering
		\includegraphics[scale=0.28]{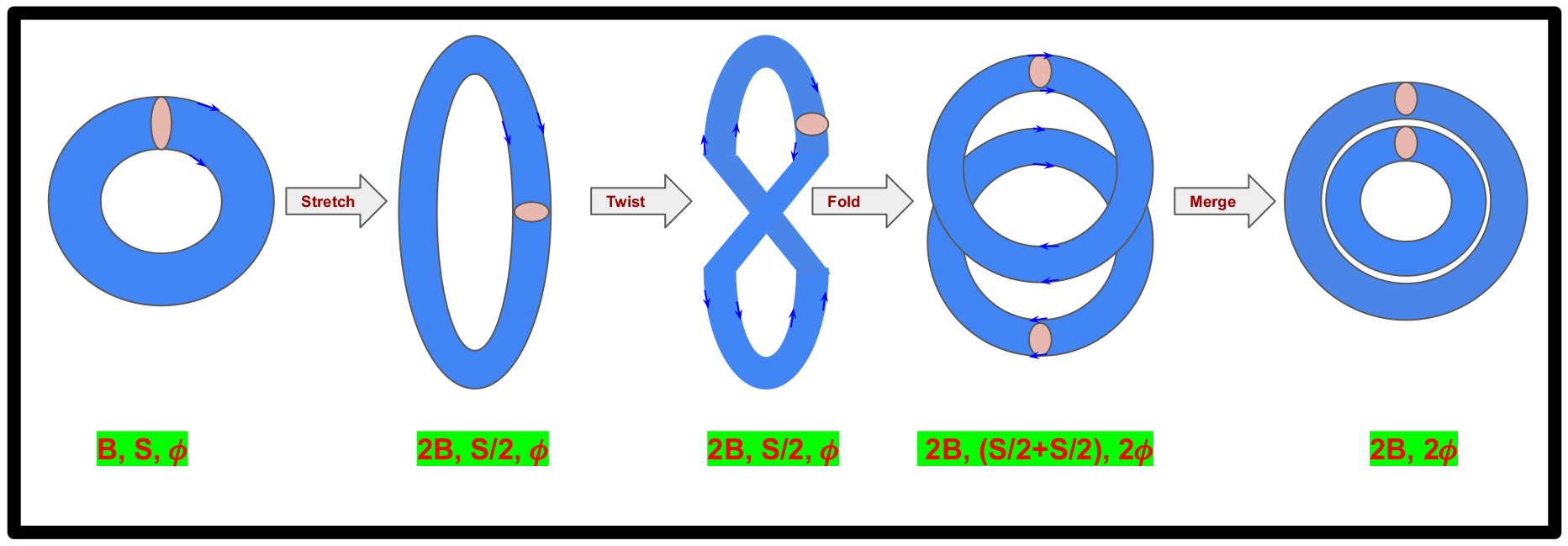}
	\caption{Schematic diagram illustrating the small-scale turbulent dynamo, showing the stretching, twisting, and folding (STF) of field lines caused by turbulence. Stretching the magnetic field increases its strength while maintaining a steady flux. The stretched field lines become twisted, necessitating movement into the third dimension. During the final stage, the twisted field lines are folded, leading to an increase in the flux. The graphic depicts symbols wherein $B$ denotes magnetic field, $S$ represents the cross-sectional area of a flux tube, and $\phi$ symbolizes magnetic flux.} 
		\label{STF Process}
\end{figure*}

Now let us return to small-scale or fluctuations dynamos. The fluctuation dynamo produces small-scale magnetic fields through the random stretching of field lines by turbulent velocity fields, as seen in Figure \ref{basic dynamo}. The primary process believed to be  responsible for this turbulent amplification of the magnetic field is the stretch-twist-fold (STF) mechanism \cite{Zeldovich:1972}. Small-scale turbulence initially elongates magnetic field lines, enhancing the field strength while maintaining a steady magnetic flux. The elongated field lines are subsequently twisted. This stage is important to highlight as it involves three spatial dimensions. Ultimately, the twisted field lines are folded, enhancing the magnetic flux. It is evident that rapid growth in the strength of the magnetic field would result from the repetition of this procedure. This process is illustrated using a cartoon diagram (Refer to Fig. \ref{STF Process}). Exponential growth of the field is possible as long as the dynamo remains in the linear (kinematic) regime, where the magnetic field is sufficiently weak that the tension force of the field lines is significantly smaller than the force contained in the small-scale turbulent eddies. As the field strength increases, the tension force strengthens, causing the dynamo to shift to the non-linear regime where the tension force is similar in magnitude to the forces produced by turbulence. Therefore, the energy contained in turbulence becomes insufficient to continue the process of folding field lines, causing the dynamo to attain saturation.


Both helical and non-helical dynamos are significant in different astrophysical scenarios. The evolution of both types of dynamos has been studied numerically in the presence of a variety of flow drives, including Arnold-Beltrami-Childress (ABC) flow \cite{Alexakis_PRE_AUG:2011}, Galloway-Proctor (GP) flow \cite{Galloway_nature:1992}, Roberts flow \cite{Radler_Brandenburg-PRE:2003, Mininni_Montgomery_PRE:2005}, Cats Eye flow \cite{Cats_Eye:2005}, Taylor-Green (TG) flow \cite{Ponty_Politano_PRL:2004}, Archotis flow \cite{Archontis_AA:2007,Sur_Archontis_MNRAS:2009}, Ponomarenko flow \cite{Ponomarenko_dynamo_POF:2003}, interacting vortex tubes \cite{SMITH_TOBIAS_JFM:2004}, Precession driven flow \cite{Precession_Driven_Flow:2023}, and Yoshida-Morrison (YM) flow \cite{EPI2D:2017, Biswas_Scripta:2023}. These flow drives are appreciated for their symmetry and significance in astrophysical scenarios and liquid metal research in laboratory environments \cite{Monchaux_Dynamo_exp_PRL:2007, DRESDYN_project:2019, MRI_PPPL_PRL:2022}.


The magnetic Prandtl number ($P_m$) is a crucial parameter in dynamo instability, defined as the ratio of viscosity to resistivity or the ratio of magnetic Reynolds number to kinetic Reynolds number ($P_m = \frac{\nu}{\eta} = \frac{R_m}{R_e}$). $P_m$ is defined as $10^{-5} \frac{T^4}{n}$ for a fully ionized medium, where $T$ is measured in Kelvin and $n$ is the particle concentration in $cm^{-3}$. Most astrophysical systems exhibit either high or low values of $P_m$. The former limit is suitable for low-density, warm and diffuse plasmas, including the intracluster medium of galaxy clusters, protogalaxies, high-temperature phases of the interstellar medium, the early universe, and so forth. The latter limit takes place in dense environments such as liquid-metal cores of planets, star convective zones, protostellar disks, and in liquid-metal (mercury, sodium, gallium) investigations. The range of $P_m$ has a significant influence on dynamo instability and is a crucial parameter for investigation. Therefore, several numerical studies \cite{Schekochihin_APJ:2002, Schekochihin_PRL_FEB1:2004, Schekochihin_PRL_FEB2:2004, Mininni_Montgomery_PRE:2005, Mininni_POP:2006, Mininni_PRE:2007, Brandenburg_APJ:2009, Alexakis_PRE:2011, Sadek_PRL:2016, Seta_MNRAS:2020, Seta_PRF:2020, Seta_MNRAS:2022, Warnecke_Nature:2023} have been conducted at various magnetic Prandtl values to simulate diverse astrophysical and laboratory situations related to dynamo instability.


The present study examines the effects of helical and non-helical drives on dynamos operating at small scales. By employing a self-consistent dynamo model, we have examined the effect of helical and non-helical dynamos. We simulate the velocity field in our model using the Navier-Stokes equation along with a drive. We have incorporated a recently reported three-dimensional Yoshida–Morrison flow (abbreviated YM flow) \cite{EPI2D:2017} into our simulation as the flow driver. The mirror symmetry (kinetic helicity) of YM flow can be manipulated by adjusting certain specific flow parameters \cite{Biswas_Scripta:2023}. YM flow exhibits similarities to the Arnold-Beltrami-Childress (ABC) flow in the maximal helicity limit. In the non-helical limit, it is often referred to as EPI2D flow \cite{EPI2D:2017}. While the dynamo action using ABC flow drive has been extensively studied, the usefulness of EPI2D flow drive for dynamo action is only starting to be investigated.




A case study is carried out to evaluate the impact of controlled helicity injection on small-scale dynamos in the nonlinear limit. Numerical investigation is conducted on the impact of magnetic Prandtl numbers ($P_m$) on small-scale helical and non-helical dynamos. In addition, we have developed various diagnostic tools such as those utilized to monitor the variations in kinetic and magnetic energy over time [$E_u(t)$ \& $E_B(t)$], evaluate the distribution of kinetic and magnetic energy across different scales [$E_u(k)$ \&  $E_B(k)$], determine the coherence length scale [$l_u(t)$ \& $l_B(t)$], examine the probability distribution function (PDF) of velocity and magnetic field fluctuations, evaluate the PDF of alignment angle between different fields, calculate the Skewness and Kurtosis [$S(f)$ \& $K(f)$], determine the structure functions of different orders [$S^p(l)$], and evaluate the hyper-flatness [$\mathcal{F}^6 (l)$].

The organization of the paper is as follows. In Sec. II we present about the dynamic equations. Our numerical solver and simulation details are described in Sec. III. The initial conditions, parameter details are shown in Sec. IV. Section V is dedicated to the simulation results on induction dynamo action that we obtained from our code and finally the summary and conclusions are listed in Sec. VI.

\section{Governing Equations}\label{Equations}
The governing equations for the single fluid MHD plasma are as follows,
\begin{eqnarray}
	&& \label{density} \frac{\partial \rho}{\partial t} + \vec{\nabla} \cdot \left(\rho \vec{u}\right) = 0\\
	&& \frac{\partial (\rho \vec{u})}{\partial t} + \vec{\nabla} \cdot \left[ \rho \vec{u} \otimes \vec{u} + \left(P + \frac{B^2}{2}\right){\bf{I}} - \vec{B}\otimes\vec{B} \right]\nonumber \\
	&& \label{velocity} ~~~~~~~~~ = \frac{1}{R_e} \nabla^2 \vec{u} + \vec{f}\\
	&& \label{EOS} P = C_s^2 \rho \\
	&& \label{Bfield} \frac{\partial \vec{B}}{\partial t} + \vec{\nabla} \cdot \left( \vec{u} \otimes \vec{B} - \vec{B} \otimes \vec{u}\right) = \frac{1}{R_m} \nabla^2 \vec{B}
\end{eqnarray}
for the above said system of equations, $\rho$, $\vec{u}$, $P$ and $\vec{B}$ represent the density, velocity, kinetic pressure and magnetic fields respectively.  $\vec{f}$ is the external drive available in the system. All quantities are appropriately normalised as discussed below. Eqn. \ref{density} to Eqn. \ref{Bfield} are solved by GMHD3D suite \cite{GTC, Biswas_FEC, Shishir_CF:2024} of codes using high performance computing. GMHD3D suite also provides a choice between utilizing the energy equation or the equation of state. For all cases studied here, we have used equation of state (see Eq. \ref{EOS}).

We define Alfven speed as, $V_A=\frac{u_0}{M_A}$, here $M_A$ is the Alfven Mach number of the plasma flow and $u_0$ is a typical velocity scale. Sound speed of the fluid is defined as $C_s = \frac{u_0}{M_s}$, where $M_s$ is the sonic Mach number of the fluid flow \textcolor{black}{and the dynamic sound speed $C_s$ contains the inherent information regarding the temperature of the system}. The initial magnetic field present in the plasma is calculated from relation $B_0 = V_A\sqrt{\rho_0}$, $\rho_0$ is the initial density of the flow. Thus in our simulations, time is normalized to Alfven times as $t = t_0*t'$, $t_0 = \frac{L}{V_A}$ and length to a typical characteristic length scale L.

The dimensionless numbers are defined as, $R_e = \frac{u_0L}{\mu}$, $R_m = \frac{u_0L}{\eta}$ , here $R_e$ and $R_m$ are the kinetic Reynolds number and magnetic Reynolds number, $\mu$ \& $\eta$ are the kinematic viscosity and magnetic diffusivity. Magnetic Prandtl number is also be defined as, $P_M = \frac{R_m}{R_e}$. The symbol '$\otimes$' represents the dyadic between two vector quantities.

\label{equations}

For solving the above set of equations at high grid resolution, we have developed a suite of GPU codes namely GMHD3D \cite{GTC, Biswas_FEC, Shishir_CF:2024}, which is briefly described in the following Section.

\section{\textbf{Simulation Details: \textit{GMHD3D} Solver}}
In this Section, we discuss the details of the numerical solver  along with the  benchmarking of the solver carried out by us.
In order to study the plasma dynamics governed by MHD equations described above, we have recently upgraded an already existing well bench-marked single GPU MHD solver \cite{rupak_thesis:2019}, developed in house at Institute for Plasma Research to multi-node, multi-card (multi-GPU) architecture for better performance \cite{Biswas_FEC, GTC, Shishir_CF:2024}. This newly upgraded GPU based magnetohydrodynamic solver (\textit{GMHD3D}) is now capable of handling very large grid sizes. \textit{GMHD3D} is a multi-node, multi-card, three dimensional (3D), weakly compressible, pseudo-spectral, visco-resistive solver \cite{Biswas_FEC, GTC, Shishir_CF:2024}. This suite (GMHD3D) includes both 2-dimensional and 3-dimensional HydroDynamic (HD) and MagnetoHydrodynamic (MHD) solvers. It uses pseudo-spectral technique to simulate the dynamics of 3D magnetohydrodynamic plasma in a cartesian box with periodic boundary condition. By this technique one  calculates the spatial derivative to evaluate non-linear term in governing equations with a standard $\frac{2}{3}$ de-aliasing rule \cite{dealiasing:1971}. OpenACC FFT library (AccFFT library \cite{Accfftw:2016}) is used to perform Fourier transform and Adams-bashforth time solver, for time integration. For 3D iso-surface visualization, an open source Python based data converter to VTK (Visualization Tool kit) by ``PyEVTK'' \cite{VTK} is developed, which converts ASCII data to VTK binary format. After dumping the state data files to VTK, an open source visualization softwares, VisIt 3.1.2 \cite{visit} and Paraview \cite{paraview} is used to visualize the data.
 Further, several other benchmarking studies have been performed at different grid resolutions \cite{Biswas_FEC, GTC, Shishir_CF:2024}. As will be discussed in the coming Section,  numerical simulations reported here are performed in $256^3$ grid sizes.

As discussed in the Introduction, to study the self-consistent dynamo action, an accurate selection of ``driving'' field is crucial, which we discuss in the Section to follow.   
\section{Initial Condition}

 Recently Yoshida and Morrison \cite{EPI2D:2017} (YM) proposed a new intermediate class of flow, which may be regarded as a topological bridge between quasi-2D  and 3D flow classes.  The flow is formulated as follows:
 \begin{equation} \label{base flow}
 	\vec{u}_b = u_0 \alpha    \vec{u}_+ + u_0 \beta    \vec{u}_-
 \end{equation}
 with
 \begin{align} 
 	\vec{u}_+ &= \begin{bmatrix} 
 		B sin(k_0y) - C cos(k_0z) \\
 		0\\
 		A sin(k_0x) \\
 	\end{bmatrix}
 \end{align}
 and 
 \begin{align} 
 	\vec{u}_- &= \begin{bmatrix} 
 		0\\
 		C sin(k_0z) - A cos(k_0x)\\
 		-B cos(k_0y) \\
 	\end{bmatrix}
 \end{align}
 so that,
 \begin{equation}\label{Yoshida_flow}
 	\begin{aligned}
 		u_x &= \alpha u_0 [ B \sin(k_0y) - C \cos(k_0z) ]\\
 		u_y &= \beta u_0 [ C \sin(k_0z) - A \cos(k_0x) ]\\
 		u_z &= u_0 [ \alpha A \sin(k_0x) - \beta B \cos(k_0y) ]
 	\end{aligned}
 \end{equation}
 where \textcolor{black}{${\vec u}_b$ represents the base flow}, \textcolor{black}{$k_0$}, $u_0$, $\alpha, \beta$, A, B and C are  arbitrary real constants. \textcolor{black}{For the present study, we consider the value of $\alpha$, A, B and C to be unity.}

 The variation of $\beta$ value in YM flow leads to new classes of base flows. For example, for \textcolor{black}{$\beta = 0$}, Yoshida et al. \cite{EPI2D:2017} classify this flow as EPI-2D flow which is given by :
 \begin{equation}\label{EPI2D}
 	\textcolor{black}{
 		\begin{aligned}
 			u_x &=  u_0 [ \sin(k_0y) - \cos(k_0z) ]\\
 			u_y &= 0\\
 			u_z &=  u_0[ \sin(k_0x)]
 		\end{aligned}
 	}
 \end{equation}
 \textcolor{black}{This flow (i.e, Eq. \ref{EPI2D}) is dependent on all the 3 spatial coordinates (i.e, $x, y, z$), whereas only two flow components are nonzero. Thus EPI-2D flow is quasi-2D in nature.}  
 
 As can be expected, for $\beta = 1$ Eq. \ref{Yoshida_flow} becomes the well known Arnold–Beltrami–Childress flow [ABC] like flow, 
 \begin{equation}\label{ABC_like}
 	\textcolor{black}{
 		\begin{aligned}
 			u_x &= u_0 [ \sin(k_0y) -  \cos(k_0z) ]\\
 			u_y &=  u_0 [  \sin(k_0z) -  \cos(k_0x) ]\\
 			u_z &=  u_0 [  \sin(k_0x) -  \cos(k_0y) ]
 		\end{aligned}
 	}
 \end{equation}
 As $\beta$ is varied from $0$ to $1.0$, a whole set of intermediate class of flows emerge, such that a normalized fluid helicity is exactly $0.0$ for $\beta = 0$ and is maximum for $\beta =1.0$ (i.e, ABC-like flows) \cite{Biswas_Scripta:2023}. The variation of $\beta$ value clearly leads to two distinguishable class viz helical ($\beta > 0$) and non-helical ($\beta=0$) class of base flows \cite{Biswas_Scripta:2023}.  
 
 The external forcing function $\vec{f}$ in Equation \ref{velocity} introduces both kinetic energy and kinetic helicity. This study utilized various sets of forcing functions to produce dynamos. The class of YM flows, denoted as $\vec{f}$, is utilized in the following manner:
 
\begin{eqnarray*}
\vec{f} =  && f_0 \{[\sin(k_fy) -  \cos(k_fz) ]\hat{x} \\ 
	 && + \beta [ \sin(k_fz) -  \cos(k_fx) ] \hat{y}\\
	 && + [ \sin(k_fx) - \beta  \cos(k_fy) ] \hat{z}\}
\end{eqnarray*}

where $f_0$ is the forcing or drive amplitude and $k_f$ is the forcing or drive length scale.

The helical form of this forcing function is determined by the parameter $\beta$ by the equation $\langle\vec{f} \cdot (\vec{\nabla} \times \vec{f})\rangle = - \beta k_f \langle\vec{f} \cdot \vec{f}\rangle$. Here $\langle...\rangle$ represents the ensemble average. Based on the preceding discussion, it is determined that when the value of $\beta$ is non zero, the forcing is helical in character (i.e, $\langle\vec{f} \cdot (\vec{\nabla} \times \vec{f})\rangle \neq 0$). Consequently, the generated dynamo is classified as helical dynamo. For $\beta = 0$ (i.e, $\langle\vec{f} \cdot (\vec{\nabla} \times \vec{f})\rangle = 0$), the dynamos are classified as non-helical dynamos.

We look at the impact of forcing at the largest feasible scale, specifically when $k_f = 1$. The magnitude of the forcing is given by $f_0$, which is decided by the average force balance.




 A recent study \cite{Biswas_Scripta:2023} based on kinematic model has demonstrated that a non-helical quasi-2-dimensional EPI2D flow by itself cannot generate fast dynamo action because it lacks sufficient stretching ability. The influence of ``magnetic back-reaction'' has not been address for this flow drive.  
 We have used a random perturbation as the starting seed magnetic field for our numerical studies. We have conducted numerical studies using a uniform magnetic field and observed that the dynamo's features are mostly unaffected by the initial specifications of the B-field in both scenarios. For the rest of the discussion, we will provide results derived from random perturbations as an initial magnetic field. 
 
We conduct our simulations using the provided initial conditions along with additional parameter spaces.

 \subsection{Parameter Details}
 For the class of YM flow profile, we evolve the set of equations mentioned in Section \ref{Equations} in a triply periodic box of length $L_x = L_y = L_z = 2\pi$, using time stepping $(dt) = 10^{-4}$. In our present investigation, the grid resolution is considered as $256^3$. We have conducted grid size scaling studies and determined that a grid resolution of $256^3$ is sufficient for the current investigation. Our investigation has identified that growth rates and spectral convergence remain consistent at a grid resolution of $256^3$. Consequently, we have presented datum with $256^3$ resolution. In light of these initial conditions and parameter spaces, we present the results of our numerical simulation.


\section{Simulation Results}
\begin{figure*}
	\begin{subfigure}{0.49\textwidth}
		\centering
		\includegraphics[scale=0.55]{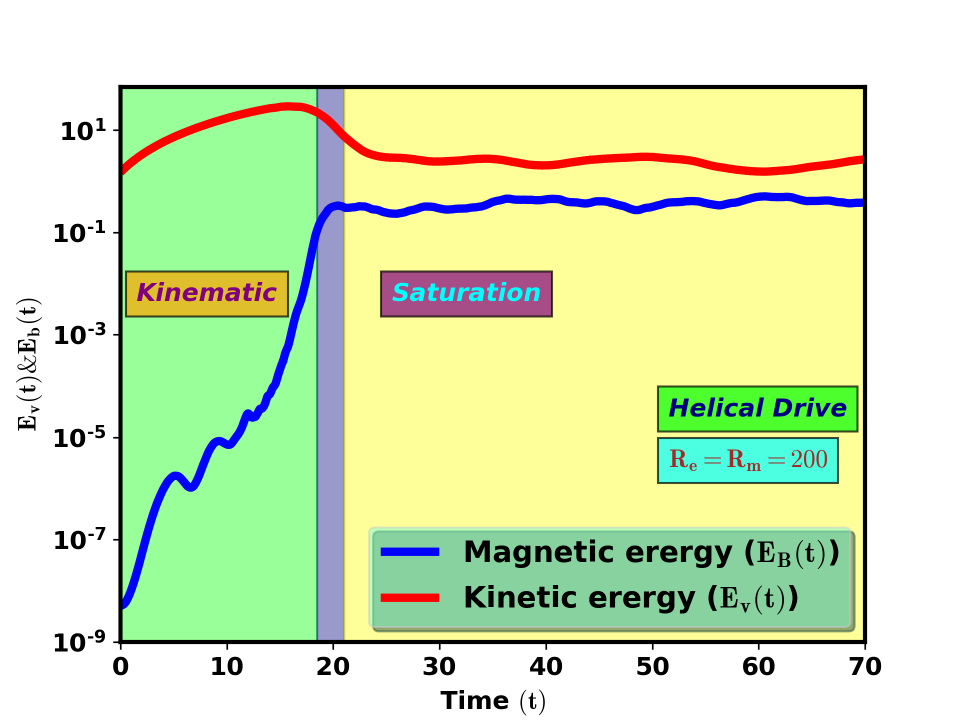}
		\caption{}
			\label{Sc Dyanmo basic}
	\end{subfigure}
	\begin{subfigure}{0.49\textwidth}
		\centering
		\includegraphics[scale=0.55]{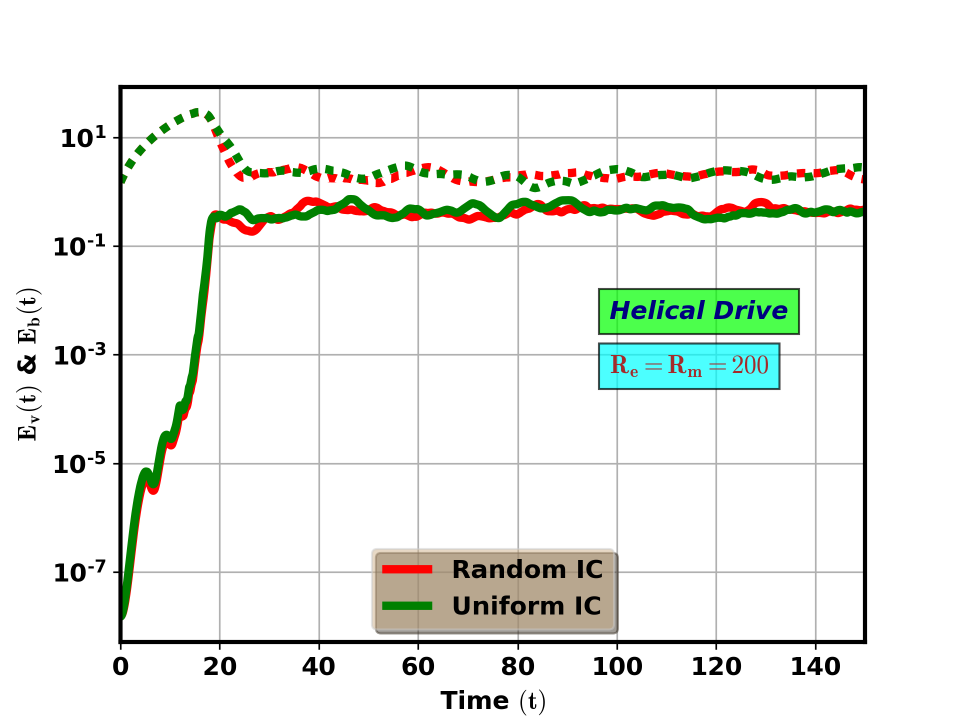}
		\caption{}
			\label{Initial condition indenpency}
	\end{subfigure}
	\caption{(a) Temporal evolution of kinetic ($E_v = \frac{1}{2} \int_{V} (u_x^2 + u_y^2 + u_z^2) dx dy dz$) and magnetic ($E_B = \frac{1}{2} \int_{V} (B_x^2 + B_y^2 + B_z^2) dx dy dz$) energies respectively. In the kinematic stage (light green shaded area), magnetic energy increases exponentially as a result of dynamo action. As the magnetic field increases, the dynamo goes through a transitional phase (light gray shaded area) before reaching a stable saturated state (light yellow shaded area) caused by magnetic feedback. This model is referred to as the ``Self-Consistent dynamo'' model. (b) The temporal evolution of kinetic and magnetic energy under two distinct initial conditions-uniform (green) and random seed (red). The magnetic and kinetic energy evolution is observed to be the same in both scenarios.} 
\end{figure*}

Dynamo action refers to the systematic and sustained generation of a magnetic field through the nonlinear transfer of energy from kinetic to magnetic mode by stretching, twisting, and folding magnetic field lines. First, we will discuss our simulation results using the ABC flow drive, also known as the helical drive. Fig. \ref{Sc Dyanmo basic} shows that a robust magnetic field is produced from an initial magnetic field via the dynamo process, which then begins to impact the structure of the velocity field. Therefore, the velocity field has been altered and is no longer the ABC field. A ``back reaction'' occurs when the magnetic field receives feedback from the velocity field and vice versa. This mutually reinforcing connection impacts the dynamics of the magnetic field and results in non-linear saturation. Three regions are distinguished: the kinematic regime where the magnetic field increases linearly up to a certain point, followed by a short transition region where the magnetic influence on the velocity field is significant enough to modify its form, and finally leading to the non-linear saturation region (Refer to Fig. \ref{Sc Dyanmo basic}). This model is referred to as the ``Self-Consistent Dynamo'' model.

In order to illustrate the concept of initial condition independence, our simulation is initiated under two distinct initial conditions. Both uniform and random seed magnetic fields have been used as initial conditions. It can be seen from Fig. \ref{Initial condition indenpency} that the magnetic energy increases exponentially with a very similar growth rate during the kinematic stage and reaches a statistically steady state (non-linear saturated stage) with the same value for both the initial condition. Therefore, any memory of the original seed field is observed to be absent. After demonstrating that the time evolution of the magnetic field is not influenced by the nature of seed magnetic field, we investigate how the initial seed magnetic field impacts the structural dynamics of the magnetic field. 

\begin{figure*}
	\centering
	\centering
	\begin{turn}{90} 
		\large{\textbf{\textcolor{blue}{Uniform}}}
	\end{turn}
	\begin{subfigure}{0.32\textwidth}
		\centering
		\includegraphics[scale=0.08780]{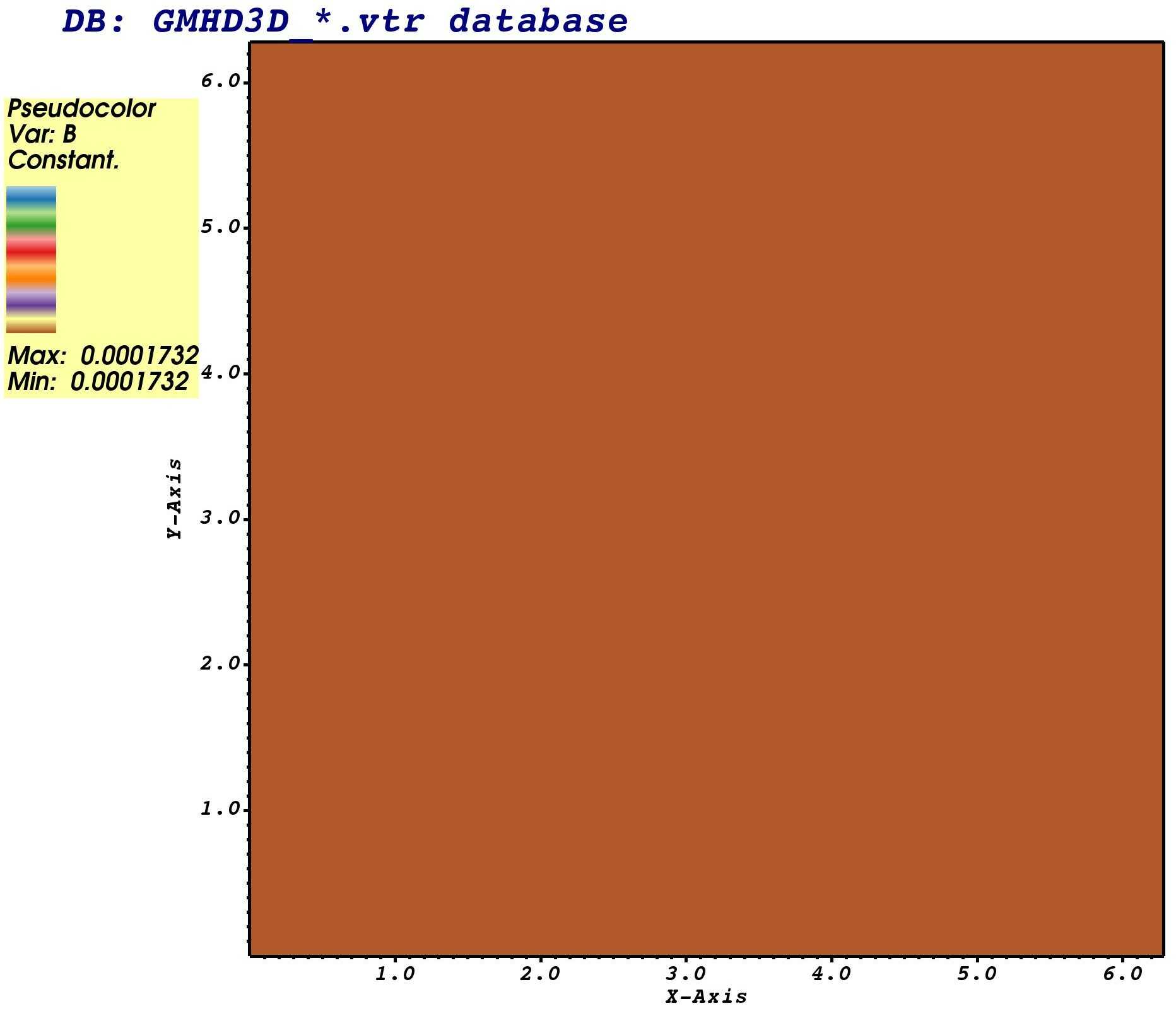}
		\caption{Time = 0.0}
	\end{subfigure}
	\begin{subfigure}{0.32\textwidth}
		\centering
		\includegraphics[scale=0.08780]{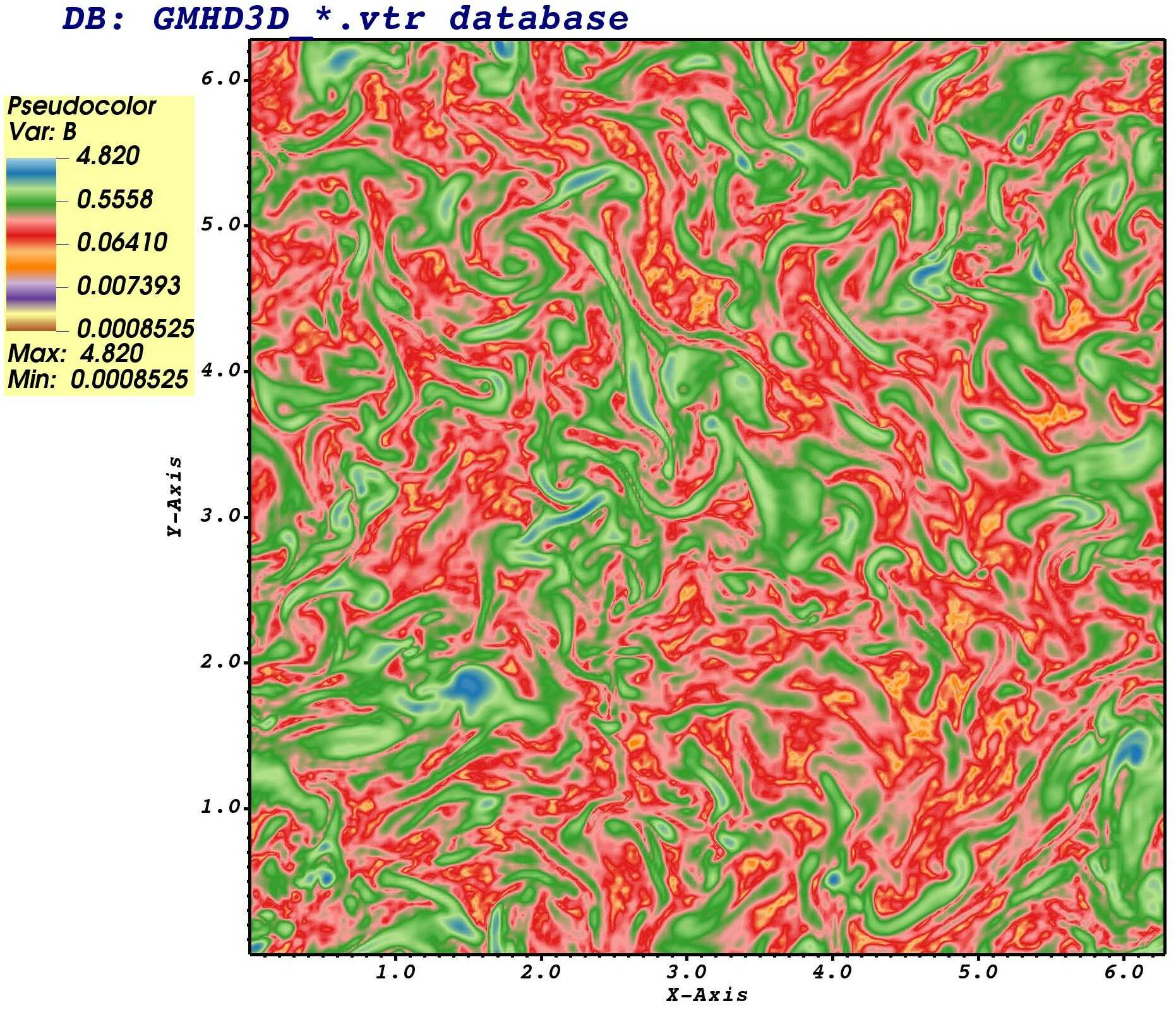}
		\caption{Time = 15.0}
	\end{subfigure}
	\begin{subfigure}{0.32\textwidth}
		\centering
		\includegraphics[scale=0.08780]{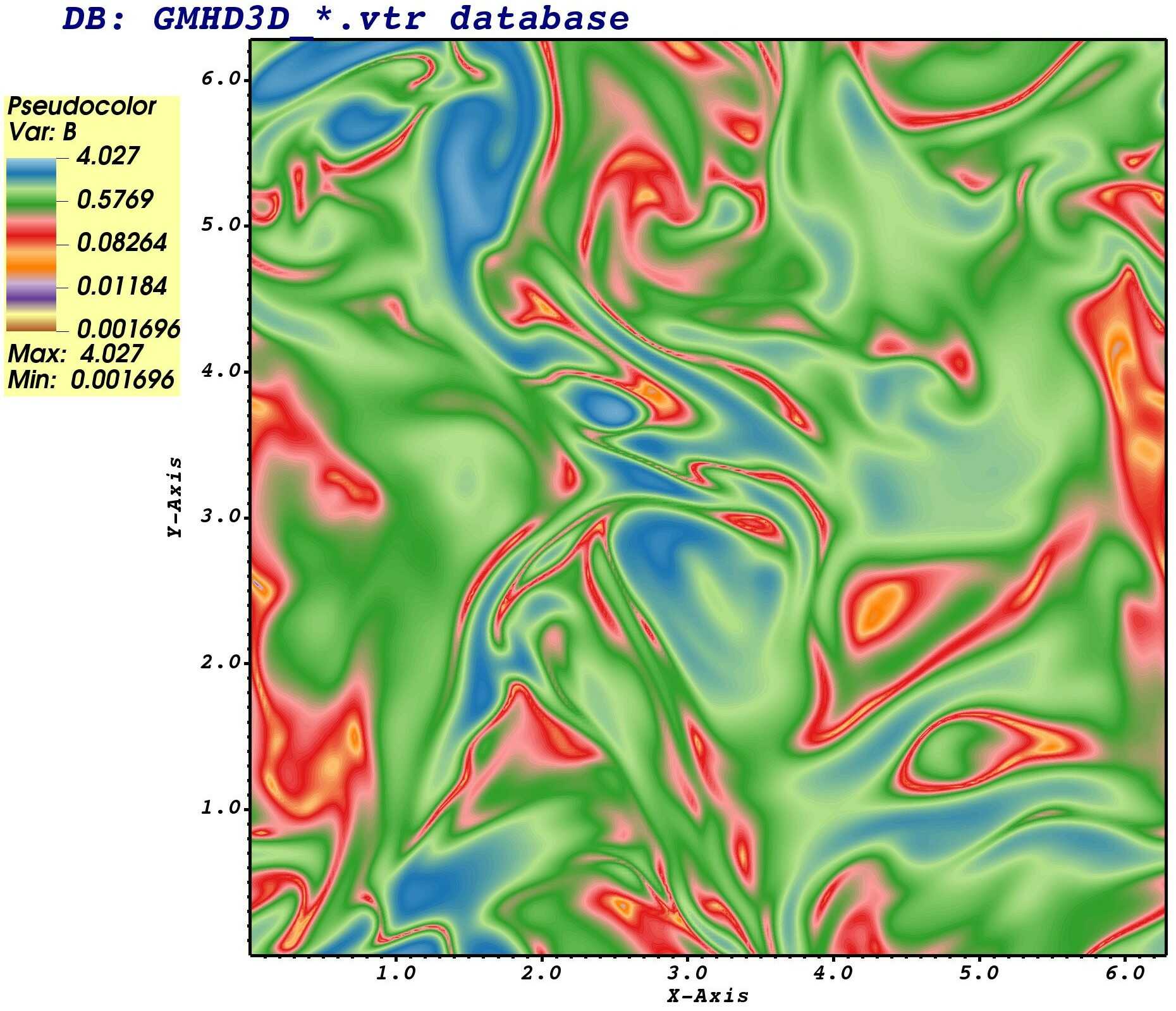}
		\caption{Time = 100.0}
	\end{subfigure}
	\begin{turn}{90} 
		\large{\textbf{\textcolor{blue}{Random}}}
	\end{turn}
	\begin{subfigure}{0.32\textwidth}
		\centering
		\includegraphics[scale=0.08780]{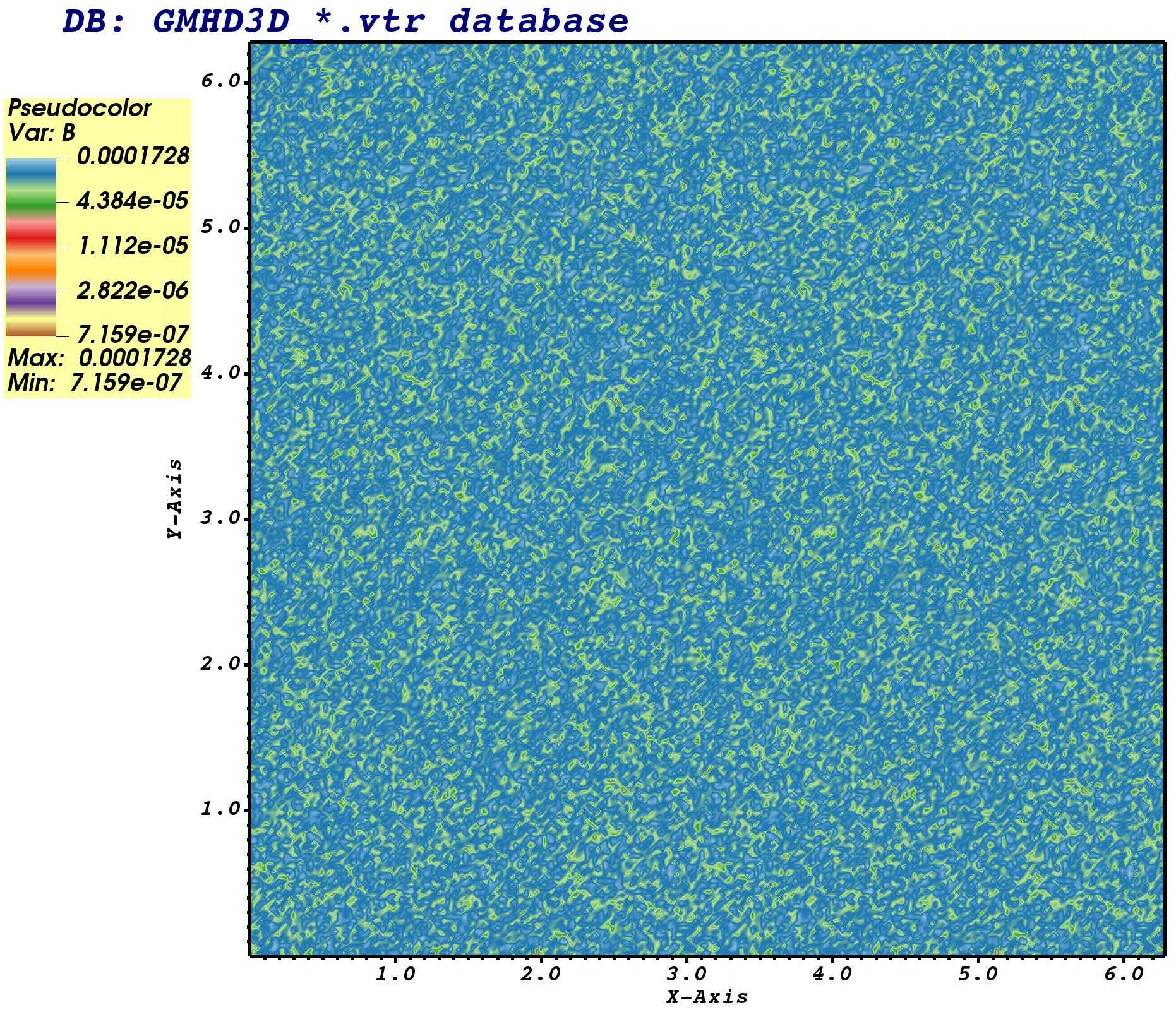}
		\caption{Time = 0.0}
	\end{subfigure}
	\begin{subfigure}{0.32\textwidth}
		\centering
		\includegraphics[scale=0.08780]{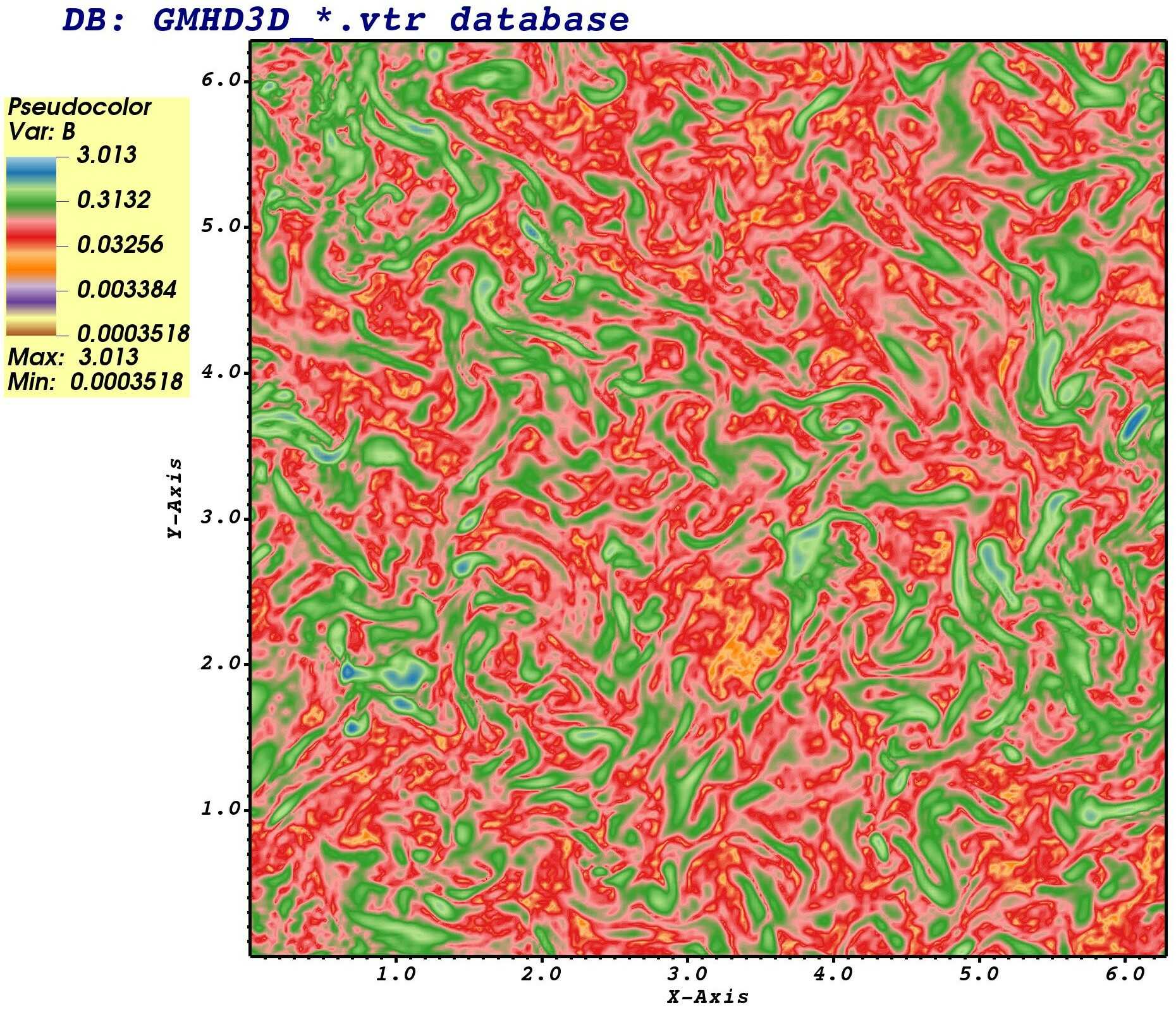}
		\caption{Time = 15.0}
	\end{subfigure}
	\begin{subfigure}{0.32\textwidth}
		\centering
		\includegraphics[scale=0.08780]{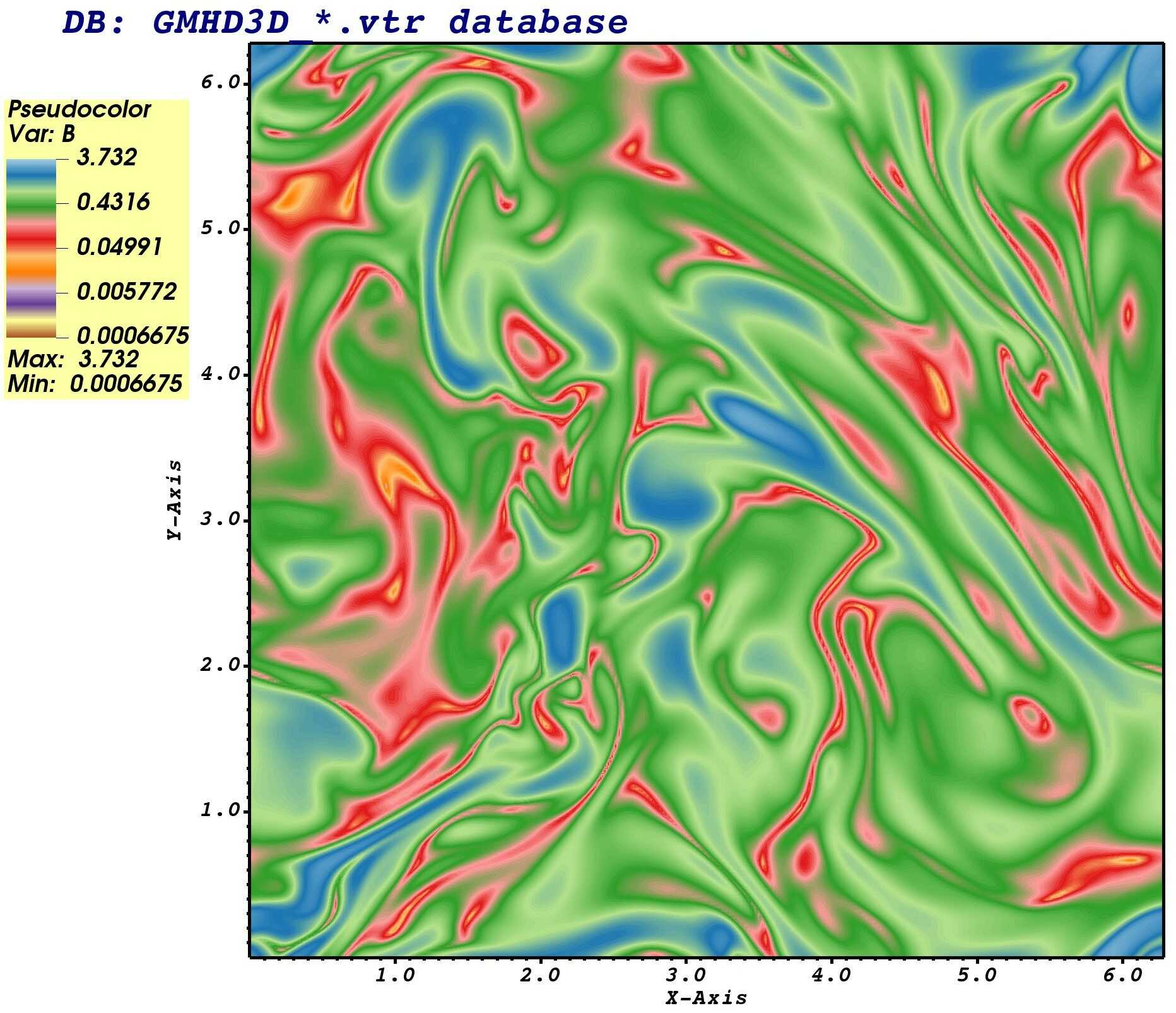}
		\caption{Time = 100.0}
	\end{subfigure}
		\caption{A two-dimensional slice ($X-Y$ plane at $Z = 0$) of the magnetic field in the initial (shown in first column), kinematic (shown in second column), and saturated (shown in third column) stages for two distinct seed fields: uniform (shown in first row) and random (shown in second row). Qualitatively speaking, the fields appear fairly similar in both the kinematic and saturation stages.} 
	\label{uniform and random initial field}
\end{figure*}

We provide a cross-section (X-Y plane view  at $Z = 0$) of the magnetic field's time evolution for both scenarios in Figure \ref{uniform and random initial field}. It is evident from Figure \ref{uniform and random initial field} that regardless of the initial condition (uniform or random seed), the magnetic field structures in the kinematic stage (represented in the middle column of Fig. \ref{uniform and random initial field}) and self-consistent stage (last column of Fig. \ref{uniform and random initial field}) are analogues if the drive is identical (helical ABC drive in this case). It is not dependent on the initial seed magnetic field.


In order to gain a more comprehensive understanding of the similarity between magnetic field structures, we have developed a number of diagnostic techniques. One of them is the time variation of the normalized magnetic field coherence length ($\frac{l_B}{L}$) as described in Ref. \cite{Seta_MNRAS:2020}.  The magnetic field coherence length, which is normalized to the size of the numerical domain $L$, is defined as:

\begin{equation}\label{lb definition}
	\frac{l_B}{L} = \frac{\int k^{-1} B(k) dk}{\int B(k) dk}
\end{equation}

\begin{figure*}
	\begin{subfigure}{0.32\textwidth}
		\centering
		\includegraphics[scale=0.36]{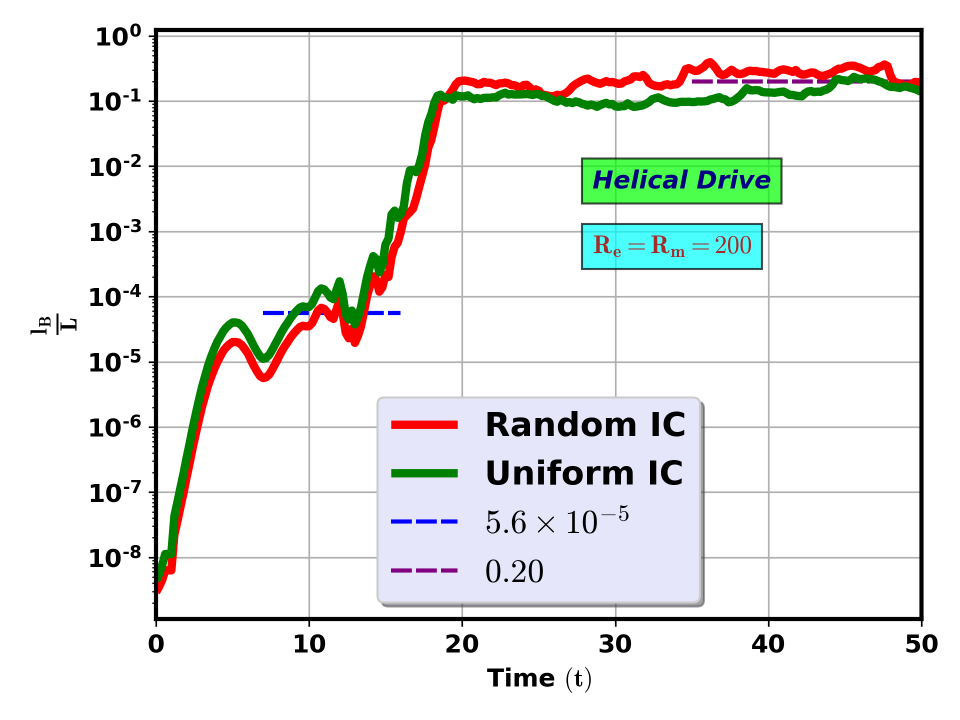}
		\caption{}
		\label{Coherence Length Initial Independency}
	\end{subfigure}
	\begin{subfigure}{0.32\textwidth}
		\centering
		\includegraphics[scale=0.36]{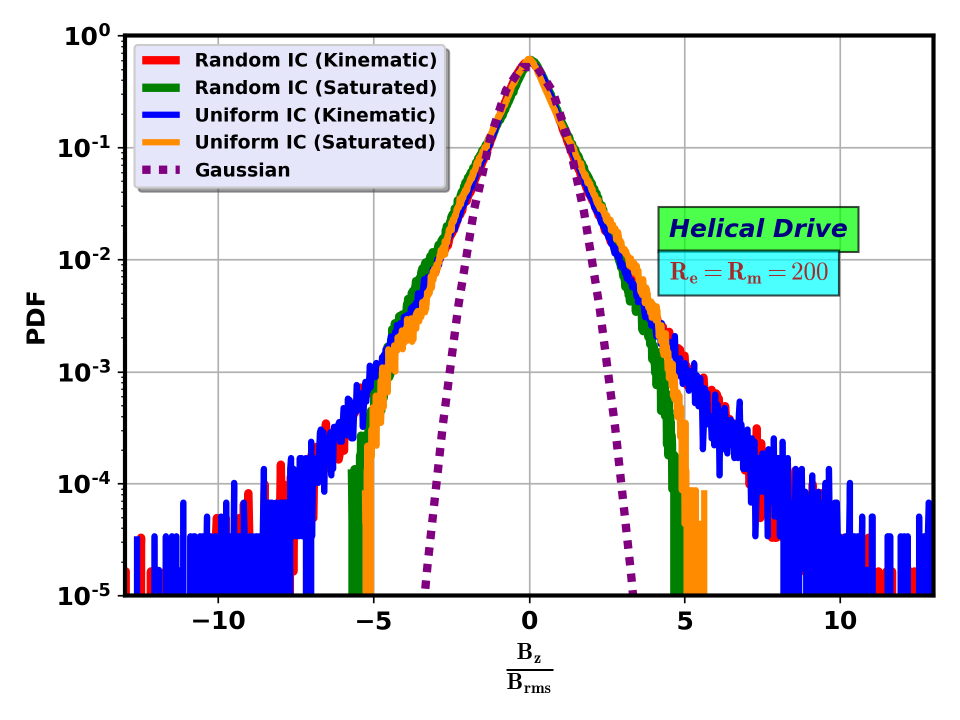}
		\caption{}
		\label{PDF Bz Initial Independency}
	\end{subfigure}
	\begin{subfigure}{0.32\textwidth}
		\centering
		\includegraphics[scale=0.36]{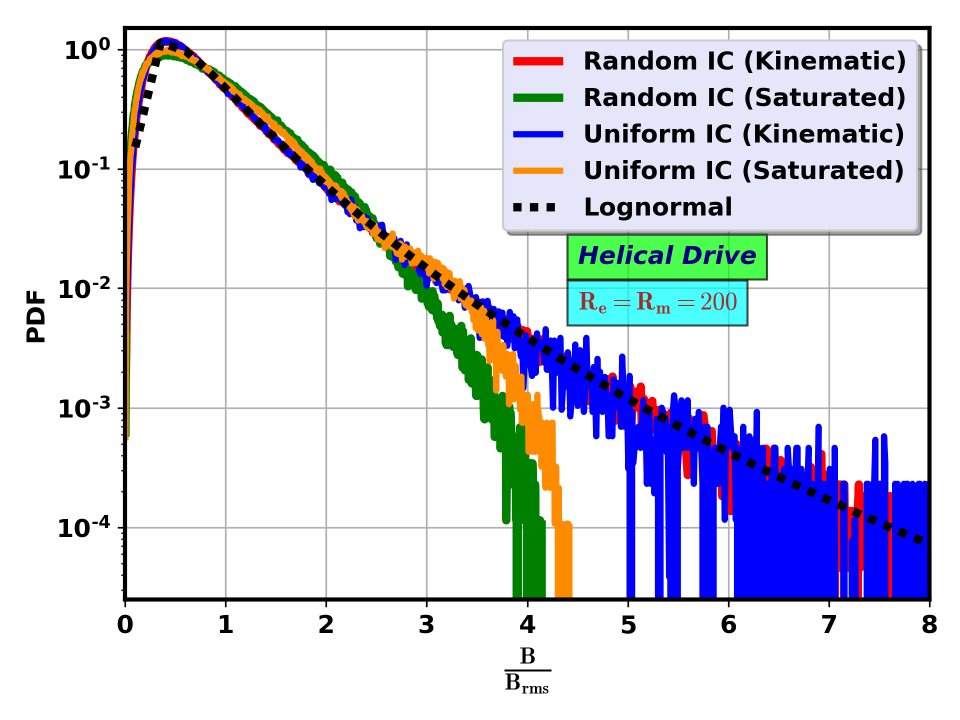}
		\caption{}
		\label{PDF B Initial Independency}
	\end{subfigure}
	\caption{(a) The magnetic field coherence length ($\frac{l_B}{L}$), normalized to the size of the numerical domain ($\frac{l_B}{L} = \frac{\int k^{-1} B(k) dk}{\int B(k) dk}$) as a function of time $t$ . (b \& c) The probability density functions (PDFs) for the normalized $z$-component of the magnetic field, $\frac{B_z}{B_{rms}}$, and the magnetic field strength, $\frac{B}{B_{rms}}$, for two distinct starting seed magnetic fields in both the kinematic and saturation stages. Here $B$ is defined as $\sqrt{B_x^2 + B_y^2 + B_z^2}$ and $B_{rms}$ is defined as $\sqrt{\langle B^2 \rangle}$.} 
\end{figure*}

An estimation has been obtained for various time snapshots and initial conditions, as illustrated in Figure \ref{Coherence Length Initial Independency}. It is clear that the normalized magnetic field coherence length is greater in the saturated phases compared to the kinematic stage, and it remains unaffected by the initial condition (See Figure \ref{Coherence Length Initial Independency}).


We have computed the normalized probability density functions (PDFs) for the $z$-component of the magnetic field $\frac{B_z}{B_{rms}}$ and the magnetic field strength $\frac{B}{B_{rms}}$ (where $B$ is defined as $\sqrt{B_x^2 + B_y^2 + B_z^2}$ and $B_{rms}$ is defined as $\sqrt{\langle B^2 \rangle}$) at both the kinematic and saturated stages, considering two distinct initial conditions.  During the kinematic stage, the magnetic field exhibits more intermittency, even when accounting for statistical fluctuations. This is evident from the longer tails observed for larger values of $\frac{B_z}{B_{rms}}$ and $\frac{B}{B_{rms}}$ in Figures \ref{PDF Bz Initial Independency} and \ref{PDF B Initial Independency}, respectively. The extended tail is a clear indicator of intermittency \cite{Intermittency_Aditya:1992}. 


The probability density function (PDF) of the normalized magnetic field strength ($\frac{B}{B_{rms}}$) during the kinematic stage has a log-normal distribution (See Fig. \ref{PDF B Initial Independency}). Furthermore, it is observed that the tails of these distributions are longer when compared to those of the saturated magnetic field, as depicted in Figure \ref{PDF B Initial Independency}. The saturated stage is characterized by a higher degree of Gaussian distribution or volume filling compared to the kinematic stage (refer to Figure \ref{PDF Bz Initial Independency}). Nevertheless, it has been determined that the PDFs exhibit statistical similarity regardless of the two distinct initial conditions of the magnetic fields. 
Therefore, it can be inferred that the initial configuration of the seed magnetic field has no impact on the statistical structure of the field in the subsequent kinematic and saturation stages of the small-scale dynamo.



\begin{figure*}
	\begin{subfigure}{0.49\textwidth}
		\centering
		\includegraphics[scale=0.55]{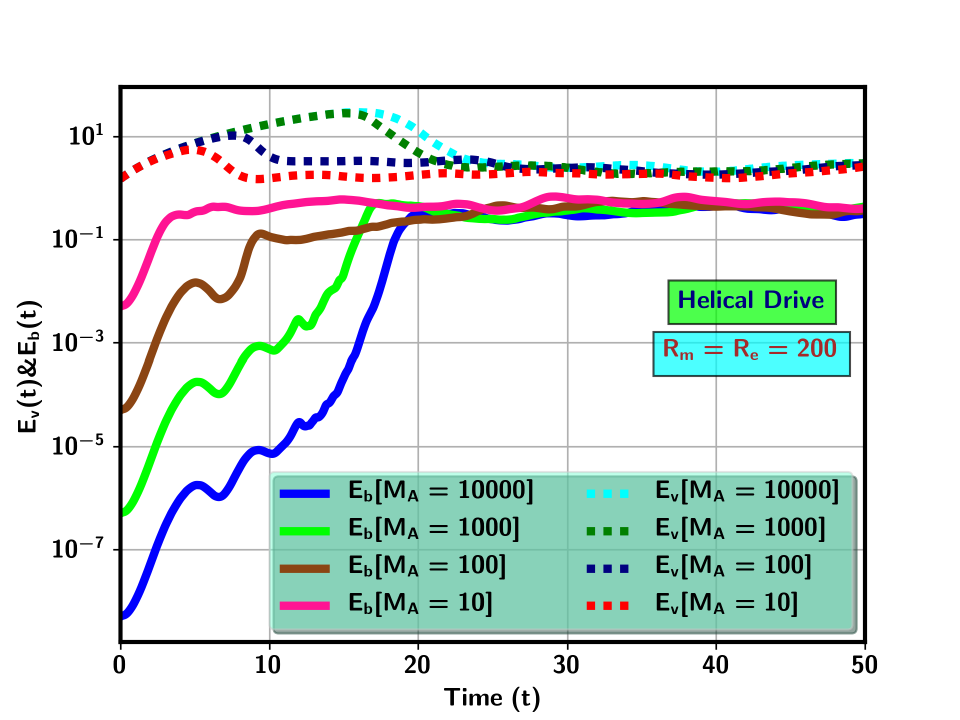}
		\caption{}
			\label{helical Ma}
	\end{subfigure}
	\begin{subfigure}{0.49\textwidth}
		\centering
		\includegraphics[scale=0.55]{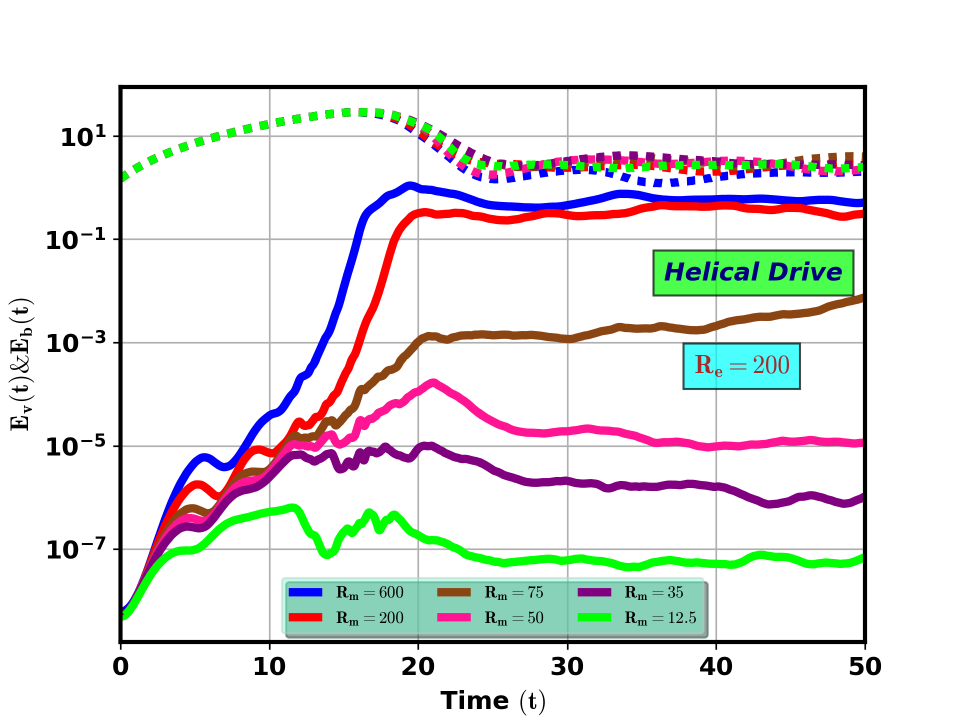}
		\caption{}
			\label{helical Rm}
	\end{subfigure}
		\caption{(a) Alfven speed influences helically forced dynamo action. Dynamo action is more pronounced in super Alfvenic regimes when driven by helical forces. (b) This study also examines the impact of magnetic Prandtl number ($P_m = \frac{R_m}{R_e}$) on helical dynamo action while keeping the kinetic Reynolds number ($R_e$) constant. Helical dynamo activity is inhibited at lower values of the magnetic Prandtl number ($P_m$). Here the dotted lines indicate the kinetic energy.} 
	\label{ABC type}
\end{figure*}

After verifying the independence of properties of dynamo on the nature of the initial seed magnetic field, we have investigated the impact of Alfven speed on dynamo instability. We conduct our simulation by varying the Alfven Mach numbers (i.e., $M_A = 10000, 1000, 100, 10$), while keeping all other parameters constant. For the two scenarios, when $M_A$ equals $10000$ and $1000$, three unique slopes have been identified as shown in Figure \ref{helical Ma}. The magnetic energy ($E_B = \frac{1}{2} \int_{V} (B_x^2 + B_y^2 + B_z^2) dx dy dz$) initially grows exponentially over time. After undergoing several orders of magnitude amplification, the exponent of increment decreases for $M_A = 1000$ and $10000$. Subsequently, it abruptly begins to rise with a greater exponent (refer to Fig. \ref{helical Ma}). It is interesting to note that the initial growth rate of magnetic energy is the same for both scenarios, but they diverge at a later stage. Figure \ref{helical Ma} shows that the saturation values of kinetic and magnetic energies are the same for all situations, regardless of the Alfven speed. Based on the aforementioned observations, it can be deduced that the dynamo effect becomes more feasible in super-Alfvenic regimes (refer to Fig. \ref{helical Ma}) when the velocity field is helically forced, irrespective of the seed magnetic field. The dynamo action is quite strong, resulting in a final magnetic energy that is comparable in magnitude to the kinetic energy.


Studying the impact of magnetic Prandtl number ($P_m$) on self-consistent dynamo activity is crucial to the overall understanding of the dynamo problem.  We study the impact of $P_m$ by keeping the kinetic Reynolds number ($R_e$) constant at a specific value and conduct numerical experiments for various magnetic Reynolds numbers ($R_m$). Fig. \ref{helical Rm} shows that the increase rate of magnetic energy decreases when the magnetic Prandtl number ($P_m$) decreases. Therefore, it is determined that the dynamo action is inhibited at lower $P_m$ values when a helical drive is present (Refer to Appendix \ref{Appen B} for more details).


\begin{figure*}
	\centering
	\begin{turn}{90} 
		\large{\textbf{\textcolor{blue}{\doublebox{Kinematic}}}}
	\end{turn}
	\begin{subfigure}{0.31\textwidth}
		\centering
		\includegraphics[scale=0.05580]{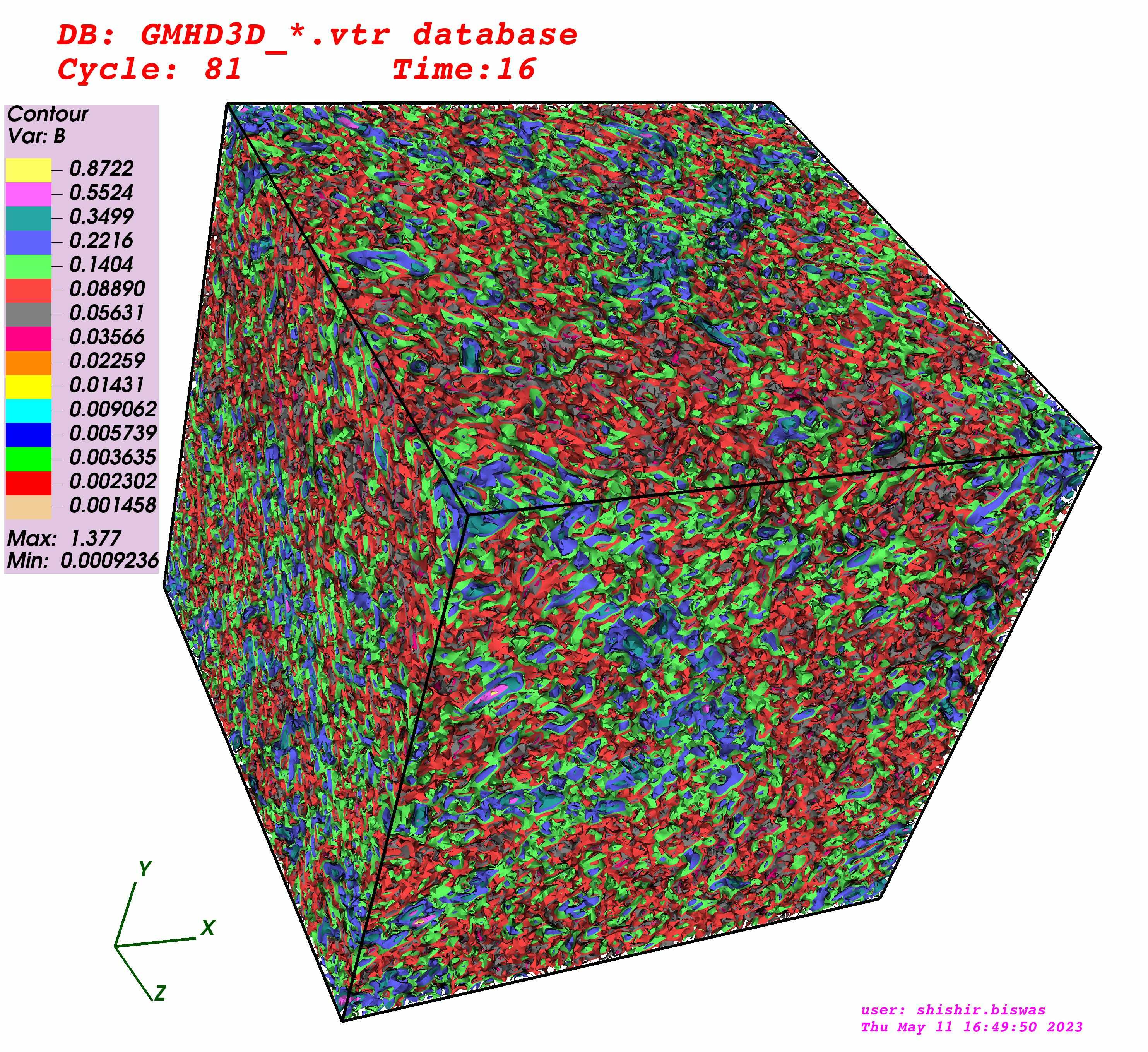}
		\caption{$P_m = 2.0$}
	\end{subfigure}
	\begin{subfigure}{0.31\textwidth}
		\centering
		\includegraphics[scale=0.05580]{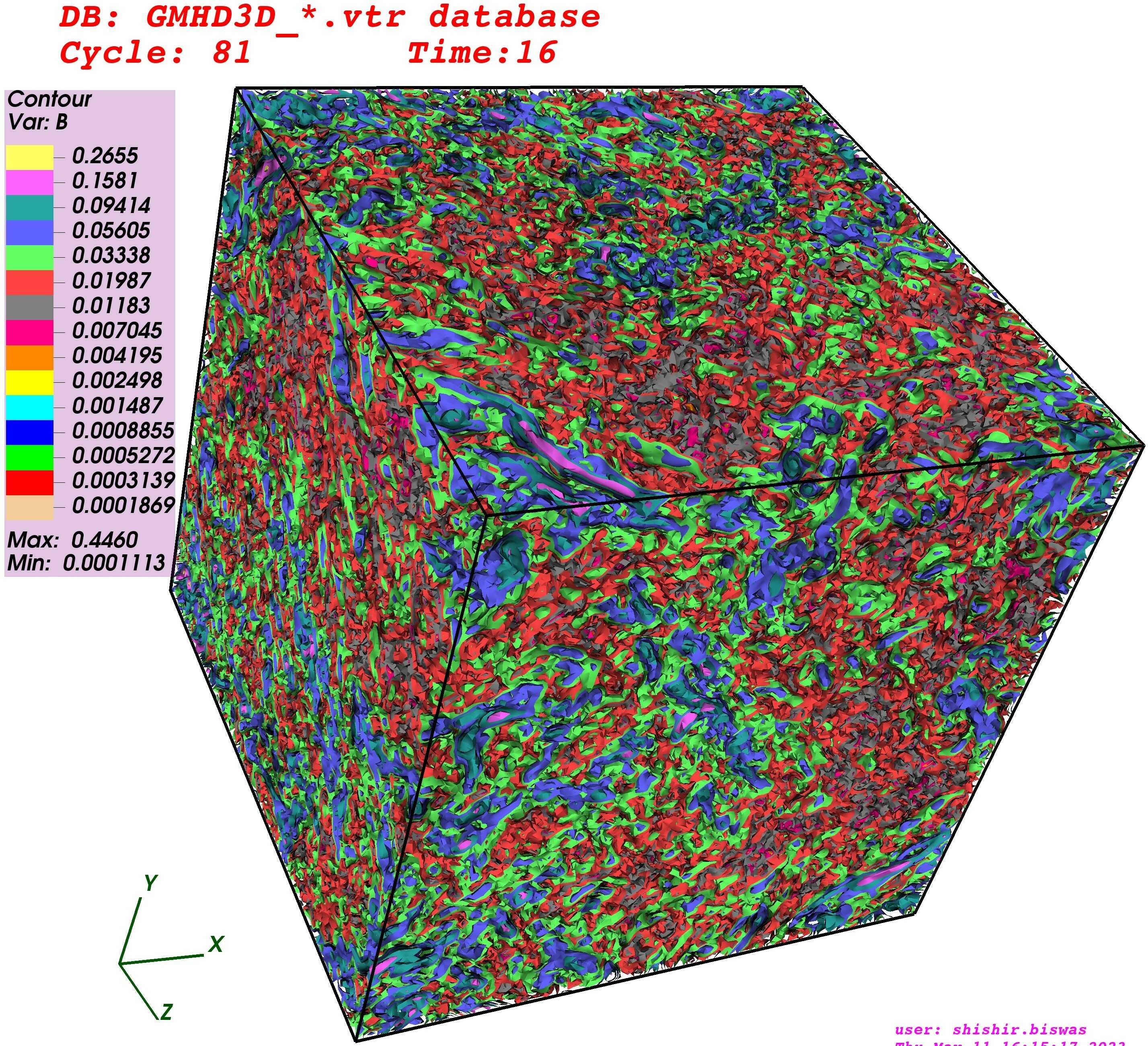}
		\caption{$P_m = 1.0$}
	\end{subfigure}
	\begin{subfigure}{0.31\textwidth}
		\centering
		\includegraphics[scale=0.05580]{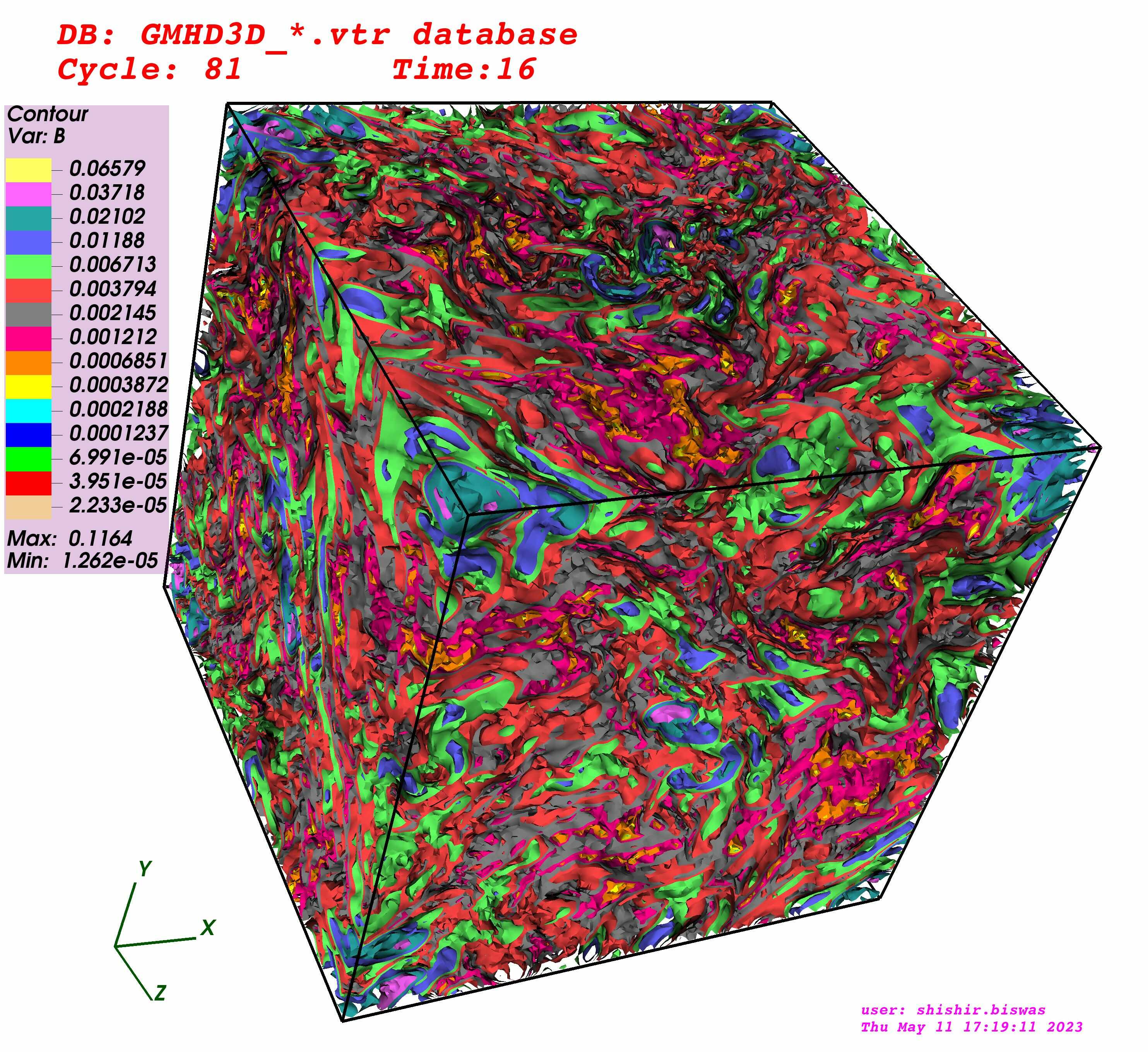}
		\caption{$P_m = 0.25$}
	\end{subfigure}
	\begin{turn}{90} 
		\large{\textbf{\textcolor{blue}{\doublebox{Saturated}}}}
	\end{turn}
	\begin{subfigure}{0.31\textwidth}
		\centering
		\includegraphics[scale=0.05580]{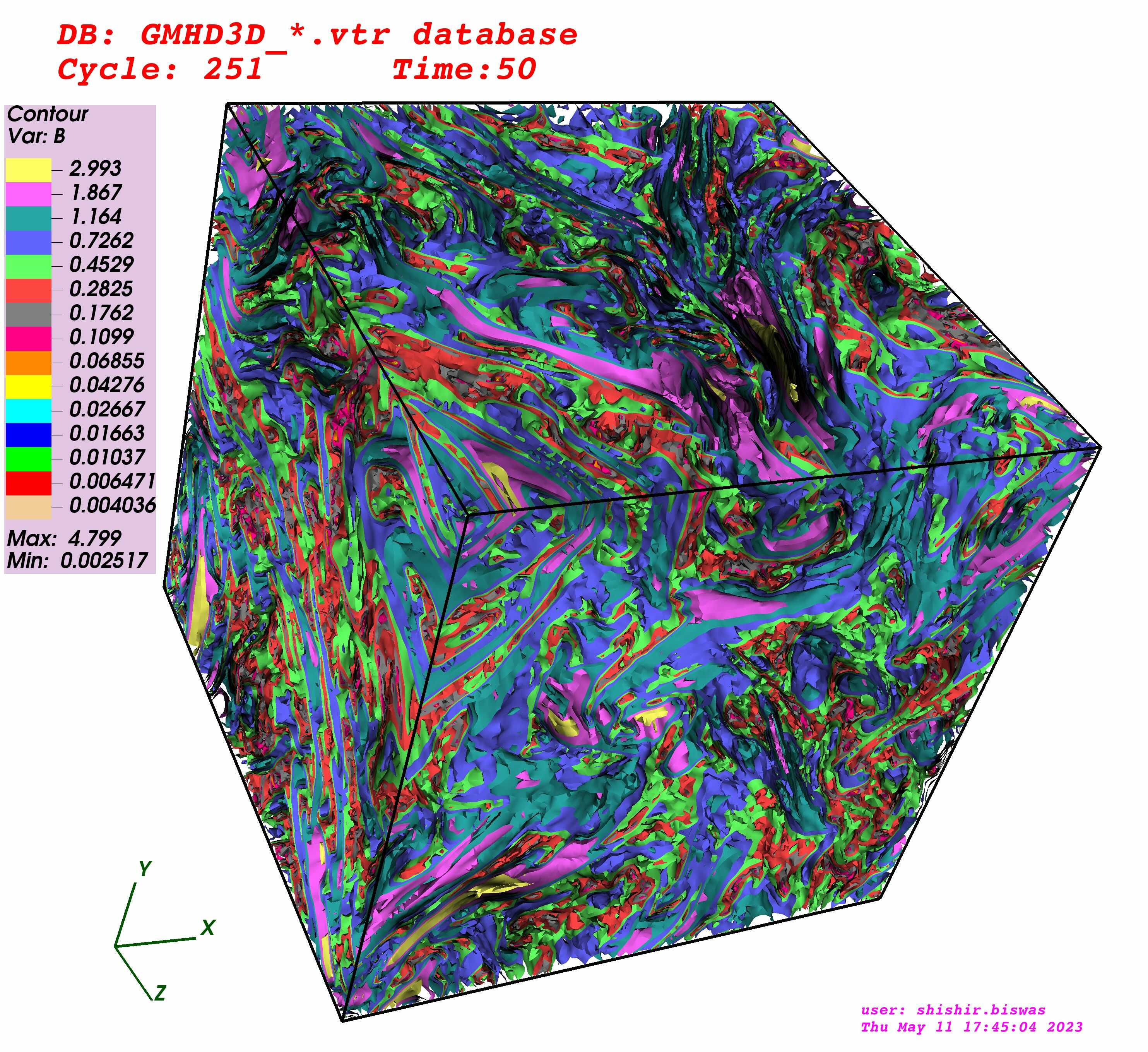}
		\caption{$P_m = 2.0$}
	\end{subfigure}
	\begin{subfigure}{0.31\textwidth}
		\centering
		\includegraphics[scale=0.05580]{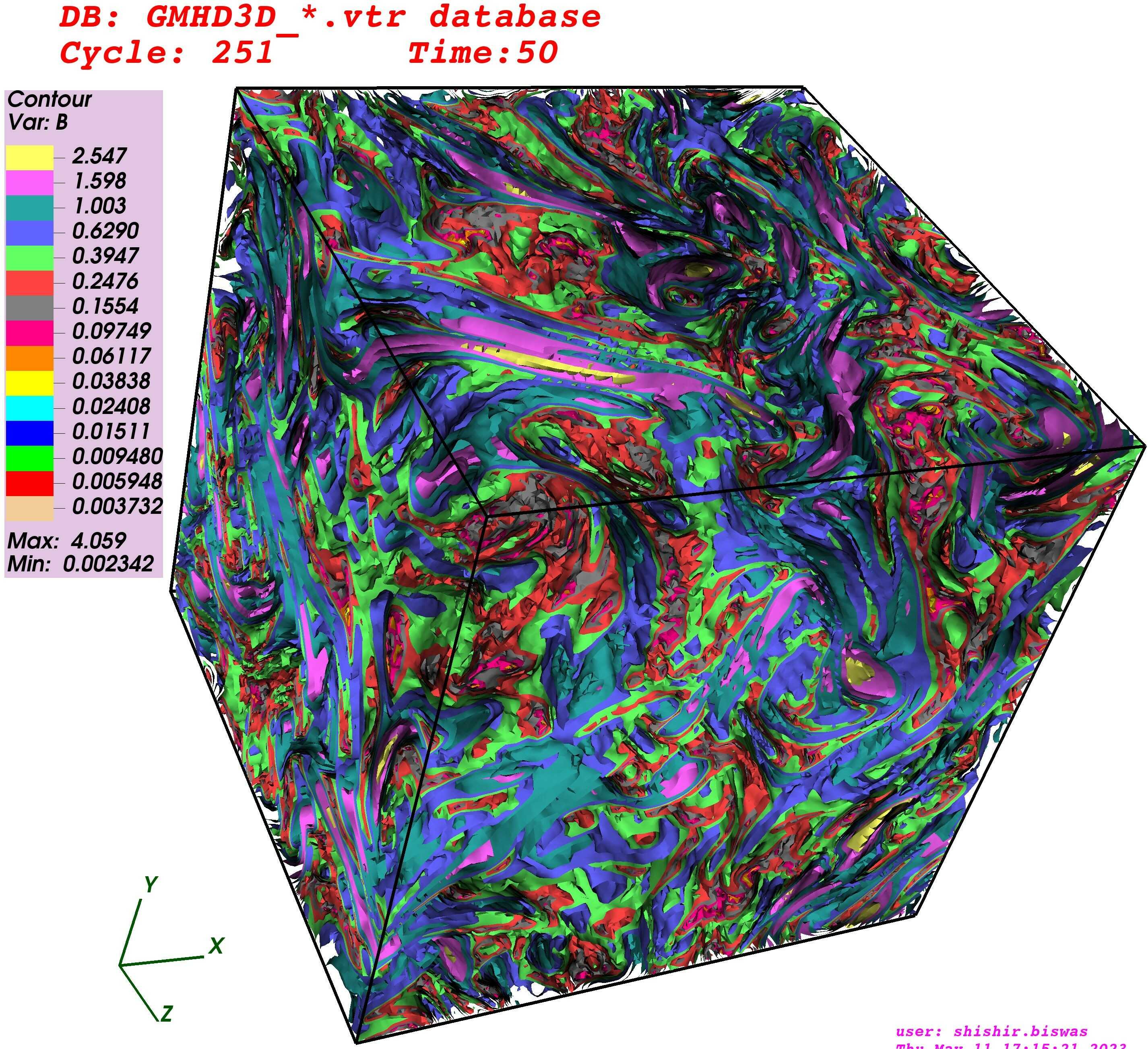}
		\caption{$P_m = 1.0$}
	\end{subfigure}
	\begin{subfigure}{0.31\textwidth}
		\centering
		\includegraphics[scale=0.05580]{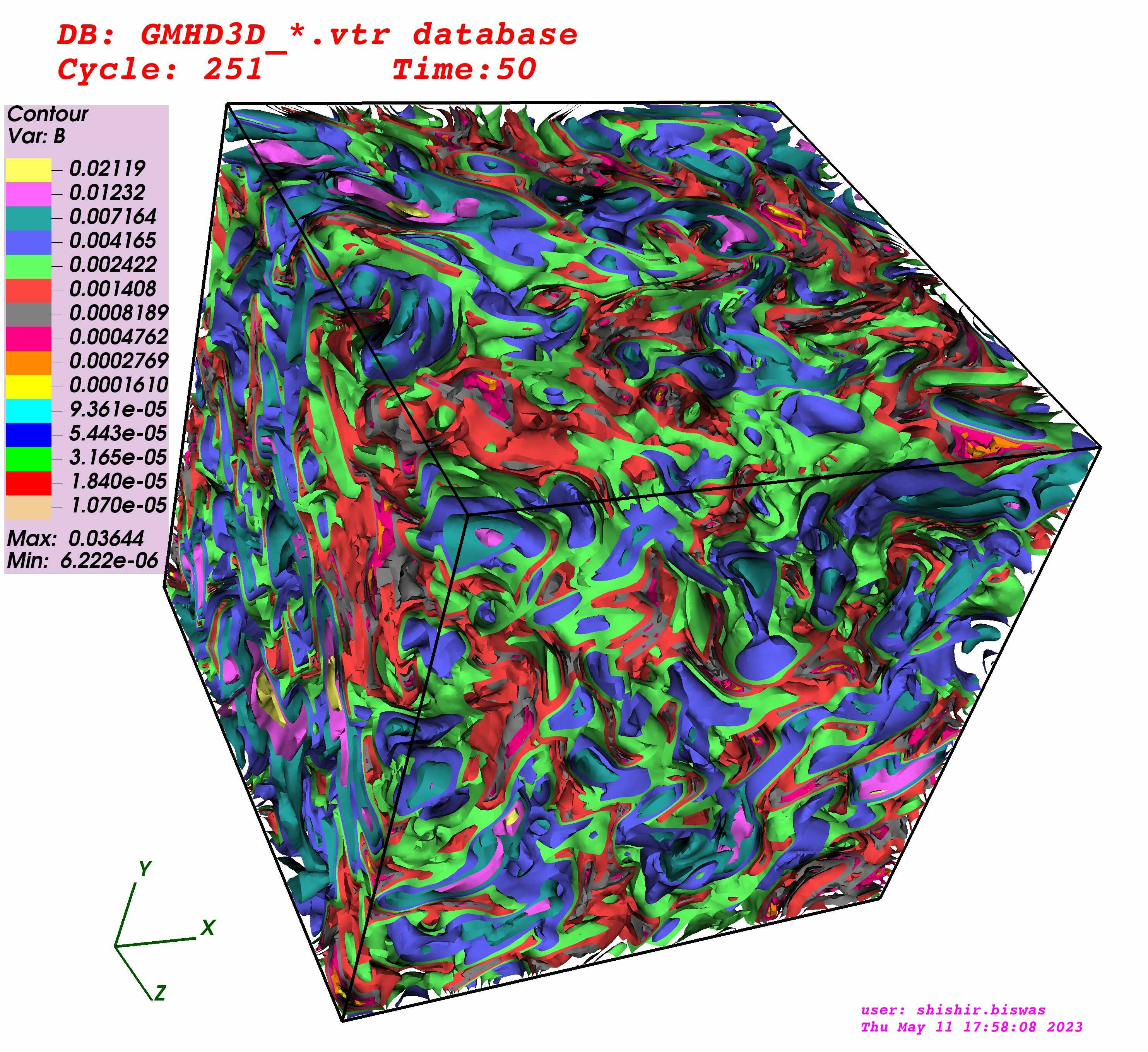}
		\caption{$P_m = 0.25$}
	\end{subfigure}
		\caption{ Visualization of three-dimensional magnetic field iso-surfaces for a helical dynamo in the kinematic stage (top row) and saturated stage (bottom row) at magnetic Prandtl numbers of (a \& d) $2$, (b \& e) $1$, and (c \& f) $0.25$. The magnetic fields in the saturated stages exhibit structures larger than those in the kinematic stages. The magnetic field structures increase in size when the magnetic Prandtl number ($P_m$) decreases in both the kinematic and saturation stages. Visualization is conducted using a logarithmic scale.} 
	\label{helical drive IsoB}
\end{figure*}

Following an investigation regarding the temporal evolution of magnetic energy across distinct $P_m$ limits, we have also investigated the dynamics of the magnetic field iso-surface (IsoB) at various $P_m$ ranges. Figure \ref{helical drive IsoB} illustrates the iso-surfaces of magnetic fields during the kinematic and saturation stages for various $P_m$ values. The magnetic field iso-surfaces exhibit smaller scale structure during the kinematic stage and larger structures at the saturated stage, as shown in Figure \ref{helical drive IsoB}. The magnetic field iso-surfaces' structure increases in size when $P_m$ decreases throughout both the kinematic and saturation periods, as seen in Fig. \ref{helical drive IsoB}. A lower value of $P_m$ at a fixed kinetic Reynolds number ($R_e$) leads to a high resistive plasma framework, which suppresses magnetic energy fluctuations, resulting in the formation of these larger scale structures.


\begin{figure*}
		\centering
	\begin{subfigure}{0.32\textwidth}
		\centering
		\includegraphics[scale=0.38]{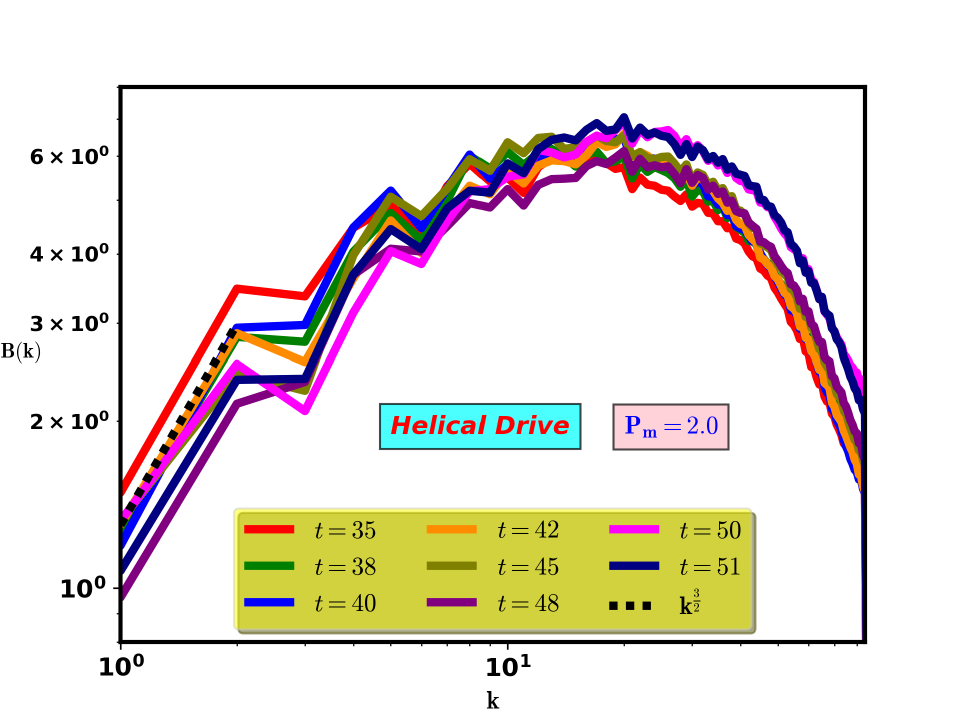}
		\caption{}
	\end{subfigure}
	\begin{subfigure}{0.32\textwidth}
		\centering
		\includegraphics[scale=0.38]{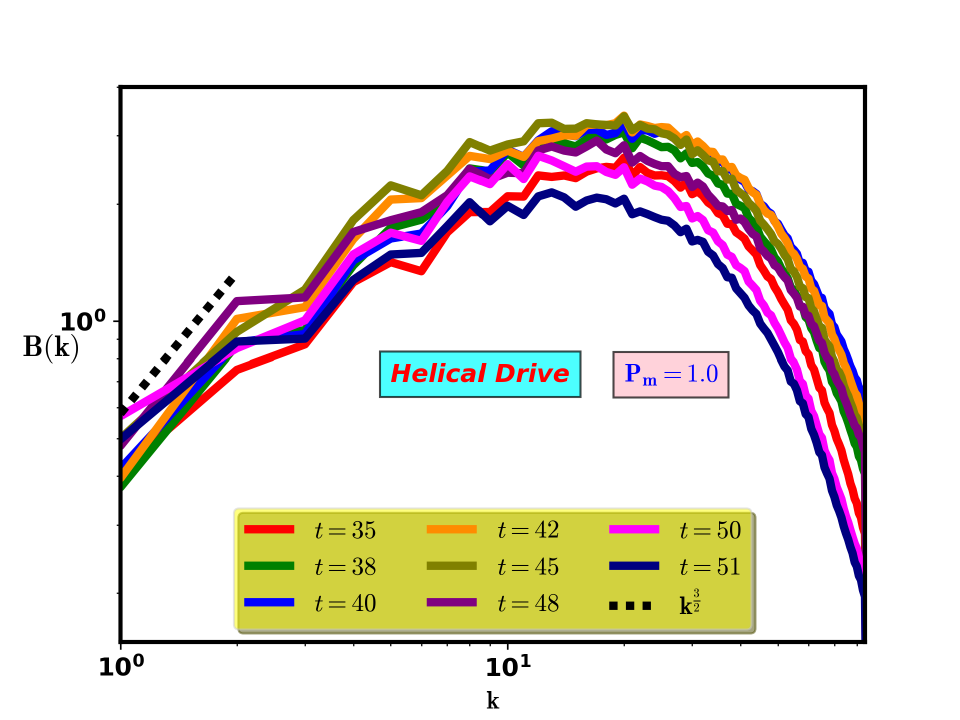}
		\caption{}
	\end{subfigure}
	\begin{subfigure}{0.32\textwidth}
		\centering
		\includegraphics[scale=0.38]{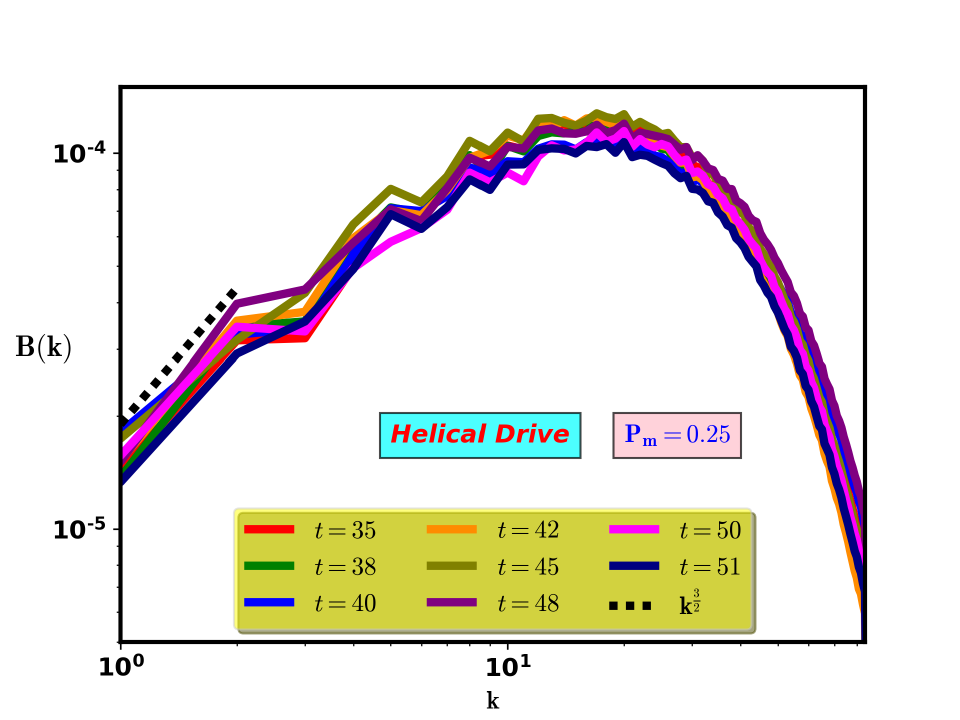}
		\caption{}
	\end{subfigure}
		\caption{ Determination of the spectral density of magnetic energy $B(k)$ for the following $P_m$ values: (a) $2.0$, (b) $1.0$, and (c) $0.25$, such that $\int |B(k, t)|^2 dk$ is the total energy at time t and $k = \sqrt{k_x^2 + k_y^2 + k_z^2}$. The dynamos are small-scale dynamos (SSD) and are independent of $P_m$. The magnetic field exhibits a well-established Kazantsev $k^\frac{3}{2}$ scaling \cite{Kazantsev:1968}.} 
\label{helical specta}
\end{figure*}

An additional noteworthy observation is that, relative to the flow drive's length scale, the magnetic field iso-surfaces are predominantly composed of small-scale structures for various $P_m$ value ranges, as illustrated in Fig. \ref{helical drive IsoB}. To verify this further, the spectral characteristics of the magnetic fields are examined. We have computed the magnetic energy spectral density $B(k)$ for a range of $P_m$ values, namely $P_m = 0.25, 1.0$, and $2.0$, such that $\int |B(k, t)|^2 dk$ is the total energy at time t and $k = \sqrt{k_x^2 + k_y^2 + k_z^2}$. Our spectral analysis shows that most of the power is concentrated on higher modes, indicating smaller length scales, in all situations (refer to Fig. \ref{helical specta}). Our spectral analysis indicates that the dynamos are small scale dynamos (SSD). Furthermore, it has been determined that the magnetic field exhibits a Kazantsev spectrum \cite{Kazantsev:1968} with a scaling of $B(k) \propto k^\frac{3}{2}$ which is unaffected by the parameter $P_m$ (refer to Figure \ref{helical specta}).


\begin{figure*}
	\begin{subfigure}{0.49\textwidth}
		\centering
		\includegraphics[scale=0.49]{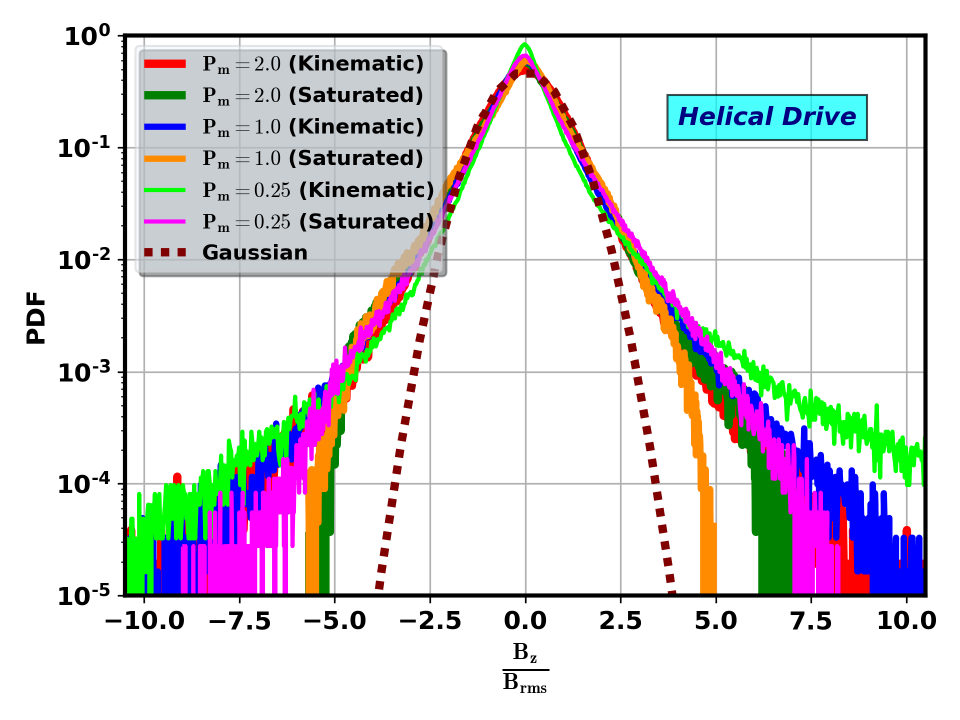}
		\caption{}
		\label{PDF Bz Helical}
	\end{subfigure}
	\begin{subfigure}{0.49\textwidth}
		\centering
		\includegraphics[scale=0.49]{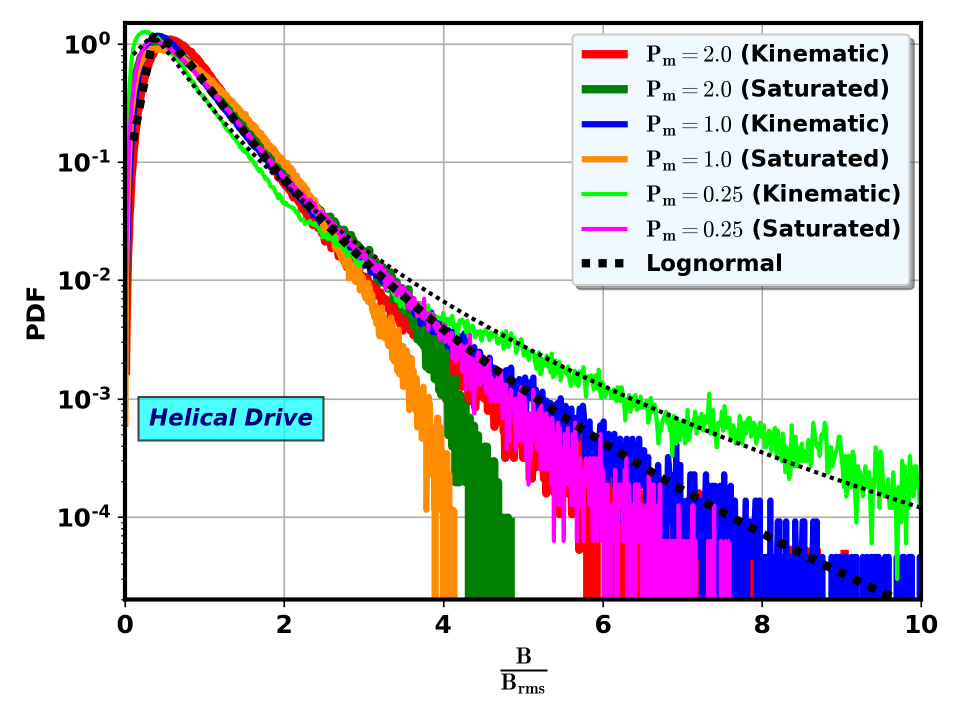}
		\caption{}
		\label{PDF B Helical}
	\end{subfigure}
	\caption{The probability density functions (PDFs) for the (a) normalized $z$-component of the magnetic field, $\frac{B_z}{B_{rms}}$, and (b) the magnetic field strength, $\frac{B}{B_{rms}}$, are determined for a helically driven dynamo in both the kinematic and saturated stages. Here $B$ is defined as $\sqrt{B_x^2 + B_y^2 + B_z^2}$ and $B_{rms}$ is defined as $\sqrt{\langle B^2 \rangle}$.} 
	\label{PDF Helical Dynamo}
\end{figure*}

PDFs of normalized single components of the magnetic field $\frac{B_z}{B_{rms}}$ and magnetic field strength $\frac{B}{B_{rms}}$ in the kinematic and saturated dynamo stages for various $P_m$ have also been computed (albeit with the z-component being the only one displayed here; we have calculated PDFs for other components as well and determined that they are identical). Figure \ref{PDF Bz Helical} clearly indicates that the probability density function (PDF) of the magnetic field is not Gaussian. Furthermore, it has been observed that the magnetic field structure during the kinematic stage exhibits greater intermittency compared to the self-consistent stage. The validity of this observation has been confirmed by testing with various $P_m$ values, and consistent results have been obtained. This intermittency is rationalized by the extended tails illustrated in Figures \ref{PDF B Helical} and \ref{PDF Bz Helical}. It is worth mentioning that the probability density function (PDF) of the kinematic stage follows a log-normal distribution, as seen in Figure \ref{PDF B Helical}. 


During both the kinematic and saturated stages, the magnetic field exhibits characteristics of typical intermittency. However, as the field approaches saturation, the degree of intermittency reduces. These observations align with previous studies conducted using varying parameter spaces and starting conditions. This ensures that the aforementioned discovery remains unaffected by minor adjustments to the model configuration and parameters.


\begin{figure}
	\begin{subfigure}{0.49\textwidth}
		\centering
		\includegraphics[scale=0.49]{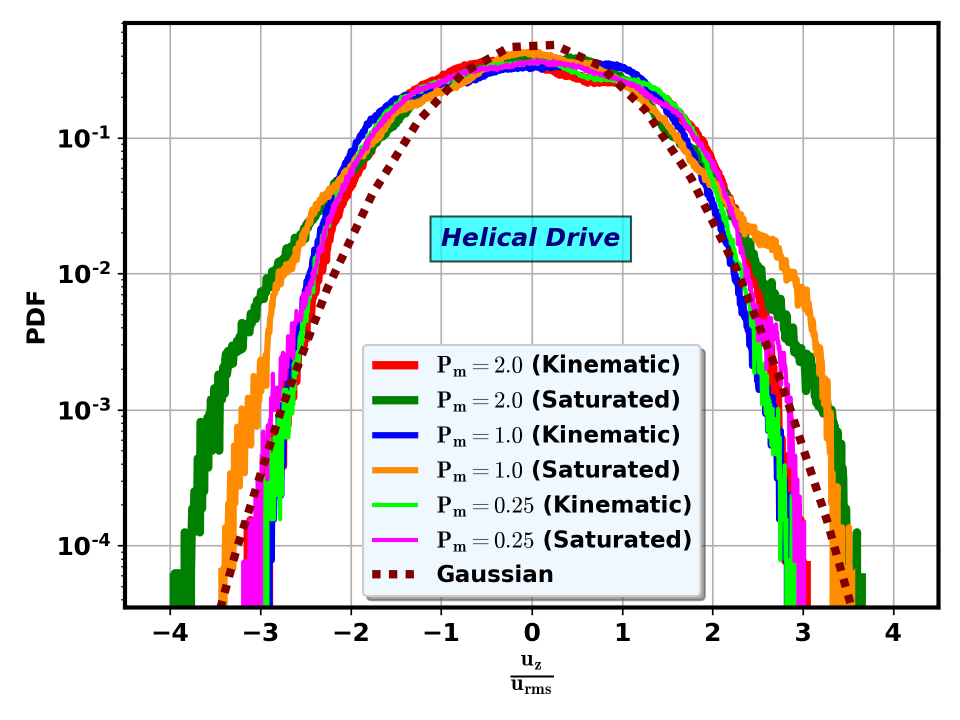}
	\end{subfigure}
	\caption{The probability density functions (PDFs) for the normalized $z$-component of the velocity field $\frac{u_z}{u_{rms}}$ and for a helical-driven dynamo at various values of $P_m$ in both the kinematic and saturated phases are presented. The PDFs exhibit a close resemblance to a Gaussian distribution. Here $u_{rms}$ is defined as $\sqrt{\langle u_x^2 + u_y^2 + u_z^2 \rangle}$.}	
		\label{PDF Vz Helical}
\end{figure}

In addition, we have calculated the probability density function (PDF) of a specific component of the velocity field, $\frac{u_z}{u_{rms}}$ (where $u_{rms}$ is defined as $\sqrt{\langle u_x^2 + u_y^2 + u_z^2 \rangle}$), during both the kinematic and saturated dynamo stages. The results of our numerical study for various $P_m$ values are presented in Figure \ref{PDF Vz Helical}. Based on Figure \ref{PDF Vz Helical}, it can be observed that the probability density functions (PDFs) have a close resemblance to Gaussian distributions in both the kinematic and saturation stages. This pattern holds true for the various $P_m$ instances investigated in this study. Thus, despite the non-Gaussian nature of the magnetic PDFs, the velocity field PDFs are nearly Gaussian. For homogeneous turbulence, this Gaussian characteristic holds true in general.


The previous discussion already demonstrated that the probability density functions (PDFs) of the magnetic field exhibit a non-Gaussian distribution, while the velocity field shows a close approximation to Gaussianity. A deviation from Gaussianity in a distribution can be characterized by two important parameters: Skewness and Kurtosis. In the present work, Kurtosis is defined statistically as follows,


\begin{equation}
	K(f) = \frac{1}{N} \sum_{1}^{N} \left(\frac{f_i - \bar{f}}{\sigma}\right)^4 - 3
\end{equation}
Here, $f_i$ is the statistical quantity of interest, $N$ is the number of grid points, and $\sigma$ represents the standard deviation.
Similarly, the Skewness of a distribution can be defined as follows:
\begin{equation}
	S(f) = \frac{1}{N} \sum_{1}^{N} \left(\frac{f_i - \bar{f}}{\sigma}\right)^3.
\end{equation}

Based on these definitions, the values of $K(f)$ and $S(f)$ are both 0 for a distribution that is entirely Gaussian. We have computed the Skewness and Kurtosis for both the velocity and magnetic field. The specific information can be found in Table \ref{Kurtosis Helical} and Table \ref{Skewness Helical}.

\begin{table*}
	\centering
	\textcolor{black}{
		\begin{tabular}{ |c|c|c|c|c|c|c| }
			\hline
			\textbf{$P_m$} & \textbf{$\mathbf{K (B_z)}$ [Kin]} & \textbf{$\mathbf{K (B_z)}$ [Sat]} & \textbf{$\mathbf{K (B)}$ [Kin]} & \textbf{$\mathbf{K (B)}$ [Sat]} & \textbf{$\mathbf{K (u_z)}$ [Kin]} & \textbf{$\mathbf{K (u_z)}$ [Sat]}\\
			\hline
			2 & 3.560 &	2.783 & 8.777 &	3.397 & -0.709 &	-0.109\\
			\hline
			1 & 5.935 &	3.260 &	11.480 &	2.875 & -0.768 & -0.069\\
			\hline
			0.25 & 19.087 &	4.208 &	34.944 &	6.568 & -0.751 & -0.645\\
			\hline
		\end{tabular}
	}
	\caption{ Kurtosis $K$ for the velocity and magnetic field distribution in the case of helical dynamo. High Kurtosis value for the magnetic field in the kinematic stage indicates existence of intermittency (see Figure \ref{PDF Helical Dynamo}) . However, the Kurtosis value for the velocity field is close to $0$ (less than 1), indicating a high degree of Gaussianity (Refer to Fig. \ref{PDF Vz Helical})}.	
	\label{Kurtosis Helical}
\end{table*} 

Based on the calculated value, it is observed that the Kurtosis and Skewness of the magnetic field are larger in the kinematic stage compared to the saturated stage. This suggests that the kinematic stage is more intermittent. We have computed the Kurtosis and Skewness for the velocity field, and it is evident that both values are close to 0 (less than 1), indicating a clear sign of Gaussian distribution.

\begin{table*}
	\centering
	\textcolor{black}{
		\begin{tabular}{ |c|c|c|c|c| }
			\hline
			\textbf{$P_m$} &  \textbf{$\mathbf{S (B)}$ [Kin]} & \textbf{$\mathbf{S (B)}$ [Sat]} & \textbf{$\mathbf{S (u_z)}$ [Kin]} & \textbf{$\mathbf{S (u_z)}$ [Sat]}\\
			\hline
			2 & 2.111 &	1.586 & 0.096 &	-0.152 \\
			\hline
			1 & 2.517 &	1.470 & -0.173 & -0.0264 \\
			\hline
			0.25 & 4.427 &	2.024 & -0.024 & -0.002\\
			\hline
		\end{tabular}
	}
	\caption{ The Skewness $S$ of the magnetic and velocity field distribution for the helically driven scenario. A high skewness value observed for the magnetic field during the kinematic stage indicates the presence of intermittency, as seen in Figure \ref{PDF Helical Dynamo}. However, the Skewness of the velocity field is close to 0 (less than 1), indicating a strong indication of Gaussian distribution (see Figure \ref{PDF Vz Helical}). A positive Skewness indicates that the distribution is skewed towards the right from the origin, whereas a negative Skewness indicates that the distribution is skewed towards the left from the origin.}
	\label{Skewness Helical}
\end{table*}

To understand the key mechanism behind the saturation of helical dynamo, we have calculated PDF of these following quantities:
\begin{eqnarray}
	\cos(\theta)_{u,B} = \frac{\vec{u}\cdot\vec{B}}{|\vec{u}||\vec{B}|}\\
	\cos(\theta)_{j,B} = \frac{\vec{j}\cdot\vec{B}}{|\vec{j}||\vec{B}|}\\
	\cos(\theta)_{\omega, u} = \frac{\vec{\omega}\cdot\vec{u}}{|\vec{\omega}||\vec{u}|}
\end{eqnarray}
The variables $\vec{u}$, $\vec{B}$, $\vec{j}$, and $\vec{\omega}$ represent velocity, magnetic field, current density, and vorticity, respectively.

\begin{figure*}
	\centering
	\begin{subfigure}{0.32\textwidth}
		\centering
		\includegraphics[scale=0.36]{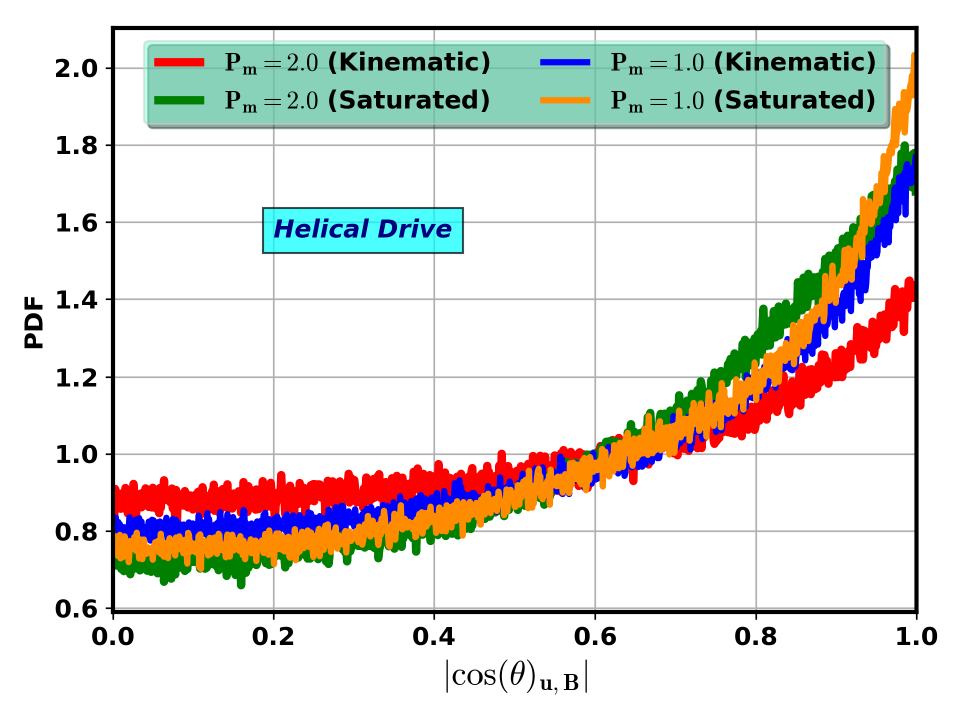}
		\caption{}
		\label{Cos uB}
	\end{subfigure}
	\begin{subfigure}{0.32\textwidth}
		\centering
		\includegraphics[scale=0.36]{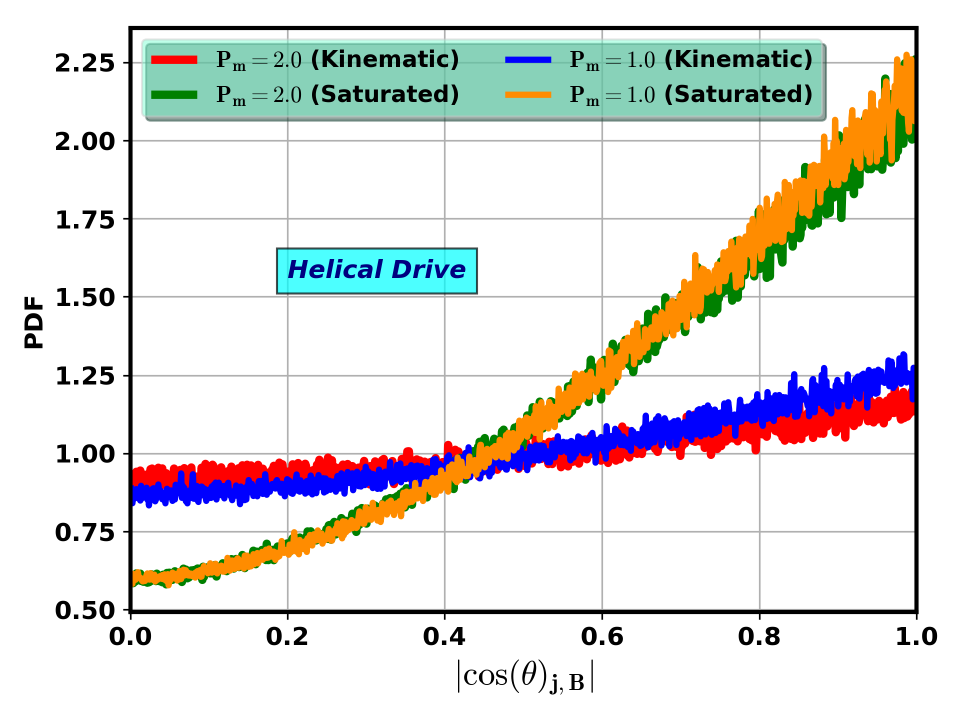}
		\caption{}
		\label{Cos jB}
	\end{subfigure}
	\begin{subfigure}{0.32\textwidth}
		\centering
		\includegraphics[scale=0.36]{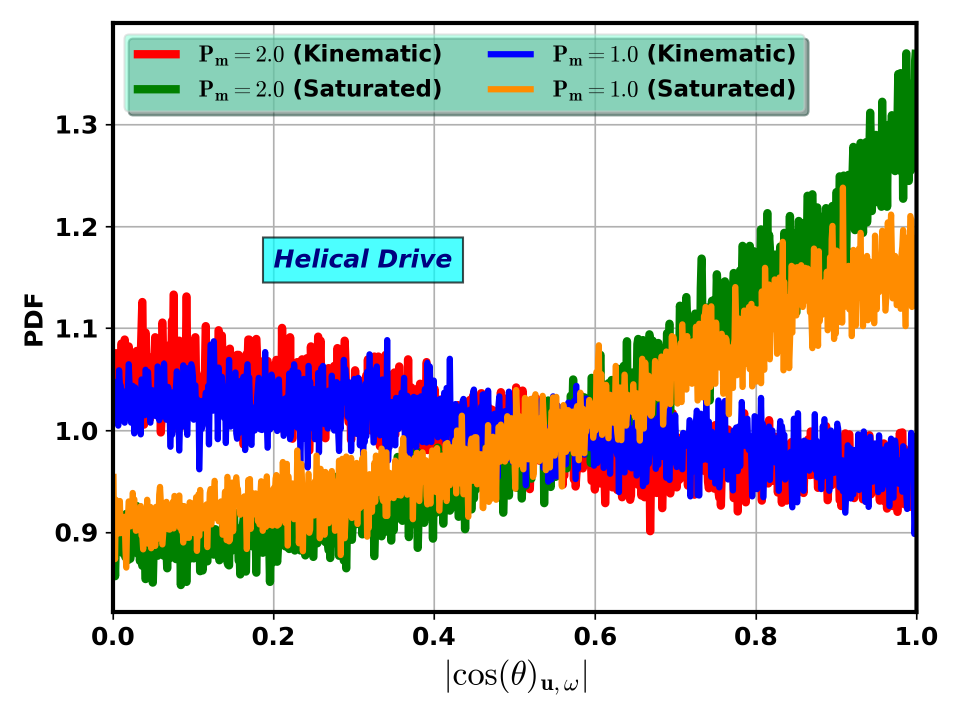}
		\caption{}
		\label{Cos omegaB}
	\end{subfigure}
	\caption{ PDFs are provided for the absolute value of the cosine of the angle between (a) velocity and magnetic field ($|\cos(\theta)_{u,B}|$), (b) current density and magnetic field ($|\cos(\theta)_{j,B}|$), and (c) the velocity and vorticity ($|\cos(\theta)_{u,\omega}|$) for various values of $P_m$ in the presence of helical drive.}
	
\end{figure*}

Based on our numerical analysis, it has been determined that the probability of the cosine of the angle between velocity and magnetic field, denoted as $|\cos(\theta)_{u,B}|$, is significantly greater during the saturated stage compared to the kinematic stage. A stronger alignment between $\vec{u}$ and $\vec{B}$ results in a drop in the induction term $\vec{\nabla} \times (\vec{u} \times \vec{B})$, leading to a reduction in the amplification of the magnetic field. This relationship is illustrated in Figure \ref{Cos uB}. Additionally, it has been documented that the probability the cosine angle formed by the current density and magnetic field ($|\cos(\theta)_{j,B}|$) is significantly greater by an order of magnitude during the saturated stage. Therefore, when the alignment between the $\vec{j}$ and $\vec{B}$ increases, the Lorentz force (or the effect of the velocity field on the flow, represented by $\vec{j} \times \vec{B}$) decreases. This means that the field approaches a force-free state when it reaches saturation (refer to Figure \ref{Cos jB}). In other words, a higher degree of alignment between $\vec{u}$ and $\vec{B}$ results in a reduced efficiency of magnetic induction. Similarly, a greater degree of alignment between vector $\vec{j}$ and $\vec{B}$ causes a decrease in the Lorentz force, meaning that the field becomes more force-free.


We have computed the probability density function (PDF) of the absolute value of the cosine of the angle $\theta$ between the velocity and vorticity field, denoted as $|\cos(\theta)_{u, \omega}|$.  The amplification of the vorticity field is dependent upon the angle formed by the vectors $\vec{u}$ and $\omega$. In the saturated stage, it is noticed that these two fields are aligned with each other (refer to Figure \ref{Cos omegaB}), similar to the alignment of $\vec{u}$ and $\vec{B}$, and $\vec{j}$ and $\vec{B}$. 

Structure function of various orders is an additional essential quantity that must be computed in this context. The structure function of field $\vec{f}$ with an order of $p$ is formally defined as:

\begin{equation}
	S^p(l) = \langle\left[(\vec{f} (\vec{r} + \vec{l}) - \vec{f} (\vec{r}))\cdot\left(\frac{\vec{l}}{l}\right)\right]^p\rangle
\end{equation} 
where $\langle....\rangle$ represents the ensemble average, and $\vec{l}$ is used to measure the separation distance between two points in the volume.

For extremely high Reynolds numbers, Kolmogorov \cite{kolmogorov:1941} determined that the structure function should exhibit the following scaling relationship: 

\begin{equation}
	S^p(l) \sim l^{\frac{p}{3}}
\end{equation}

According to Kolmogorov's phenomenological theory \cite{kolmogorov:1941}, the second-order structure function scales as, 
\begin{equation}
	S^2(l) \sim l^{\frac{2}{3}}
\end{equation}
In addition, there exists a precise relationship for the third-order structure function in homogeneous, isotropic turbulence, denoted as, 

\begin{equation}
	S^3(l) = -\frac{4}{5} \epsilon l
\end{equation}
easily obtained by employing the K\'arm\'an-Howarth-Monin relation \cite{Karman_Howarth:1938} for the energy flux.
This relationship is alternatively referred to as the fourth-fifths law. Here $\epsilon$ denotes the energy-dissipation rate.

\begin{figure*}
	\centering
	\begin{subfigure}{0.32\textwidth}
		\centering
		\includegraphics[scale=0.36]{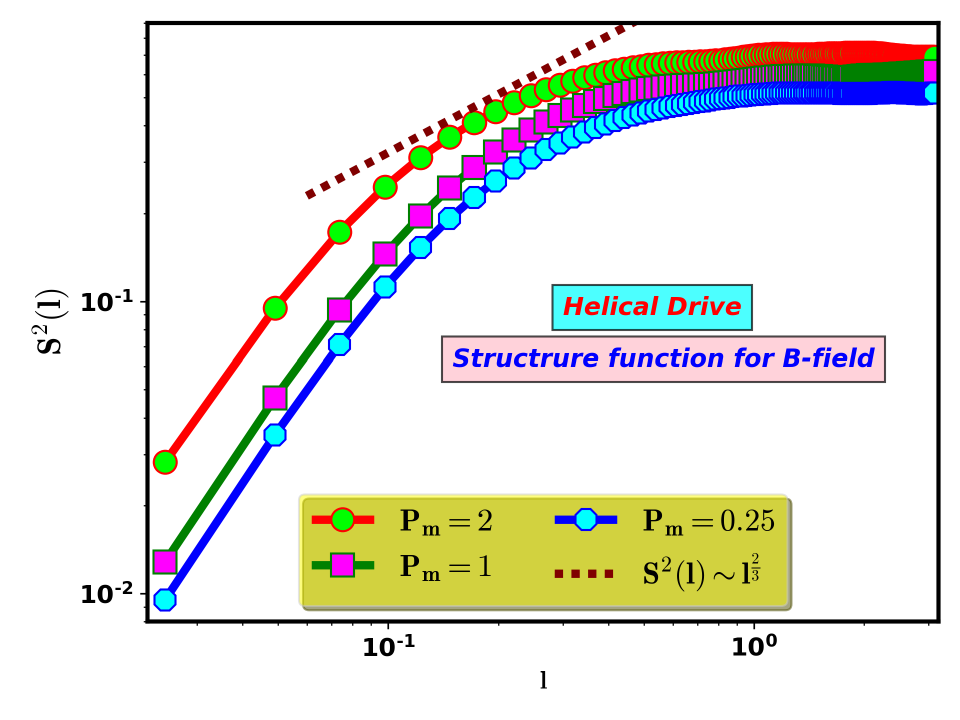}
		\caption{}
	\end{subfigure}
	\begin{subfigure}{0.32\textwidth}
		\centering
		\includegraphics[scale=0.36]{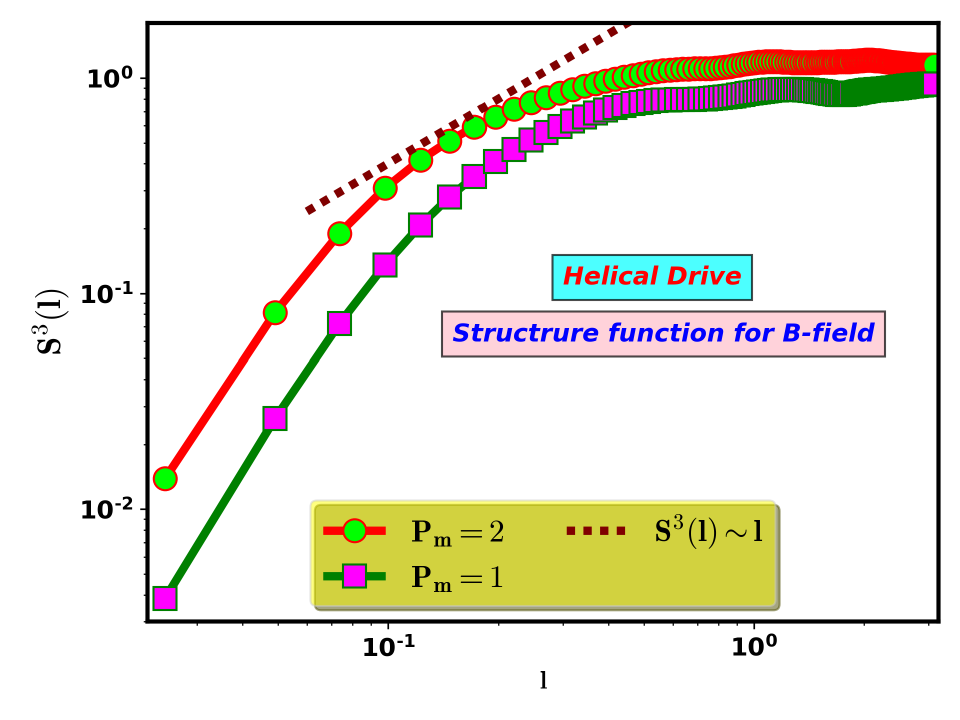}
		\caption{}
	\end{subfigure}
	\begin{subfigure}{0.32\textwidth}
		\centering
		\includegraphics[scale=0.36]{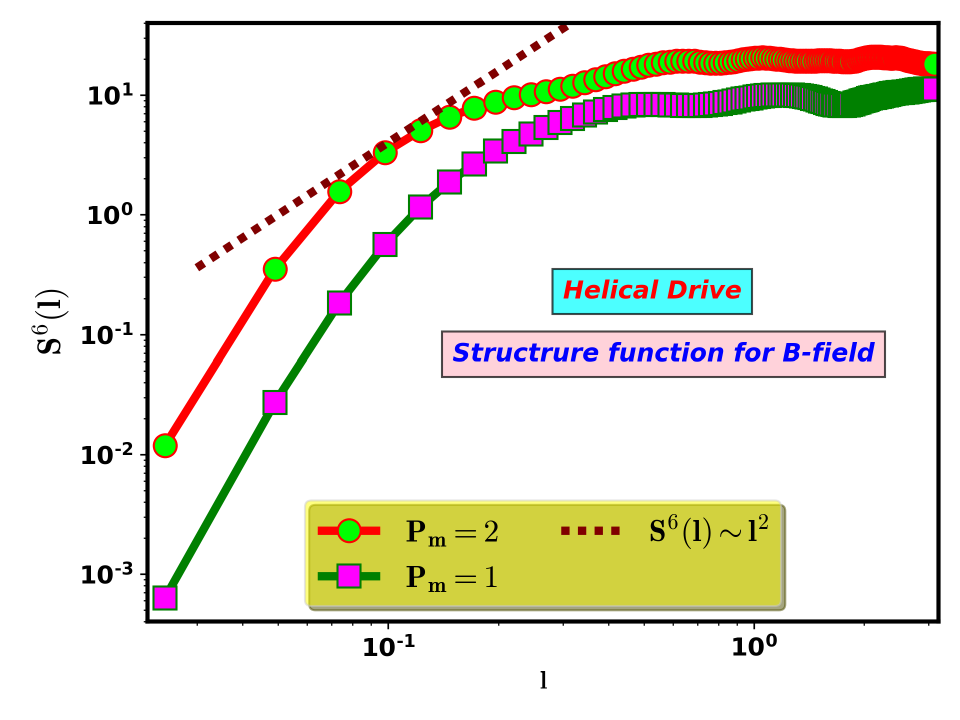}
		\caption{}
	\end{subfigure}
	\begin{subfigure}{0.32\textwidth}
		\centering
		\includegraphics[scale=0.36]{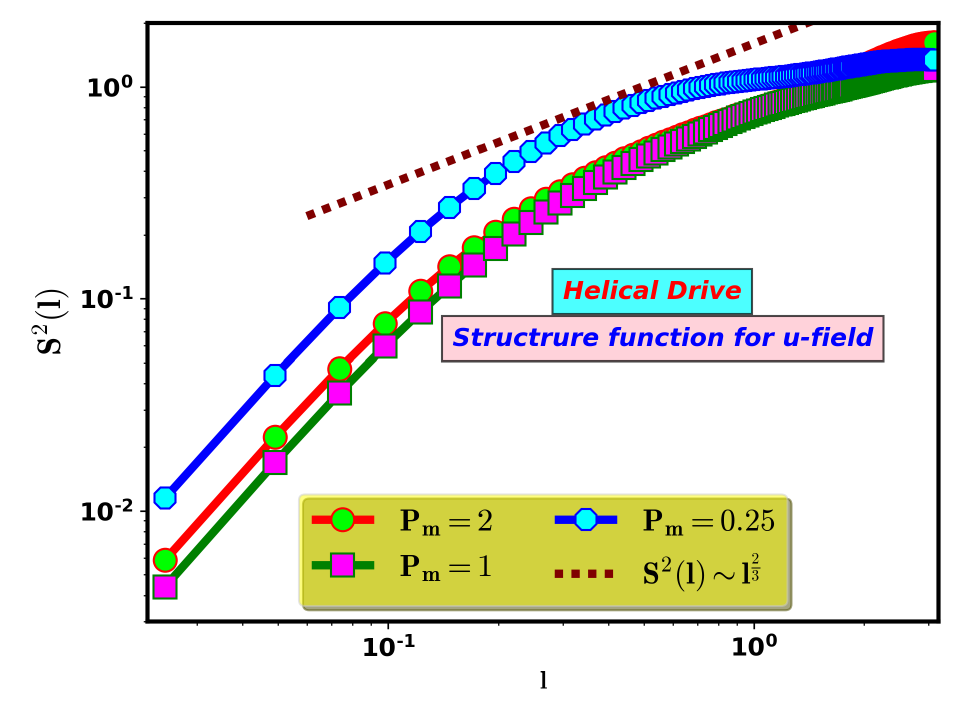}
		\caption{}
	\end{subfigure}
	\begin{subfigure}{0.32\textwidth}
		\centering
		\includegraphics[scale=0.36]{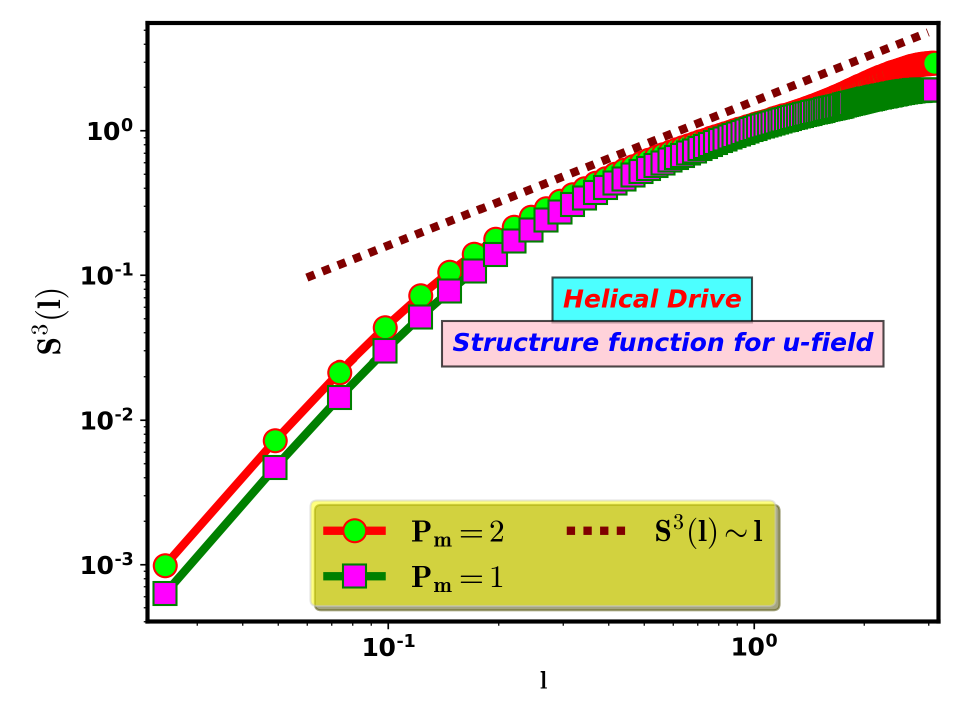}
		\caption{}
	\end{subfigure}
	\begin{subfigure}{0.32\textwidth}
		\centering
		\includegraphics[scale=0.36]{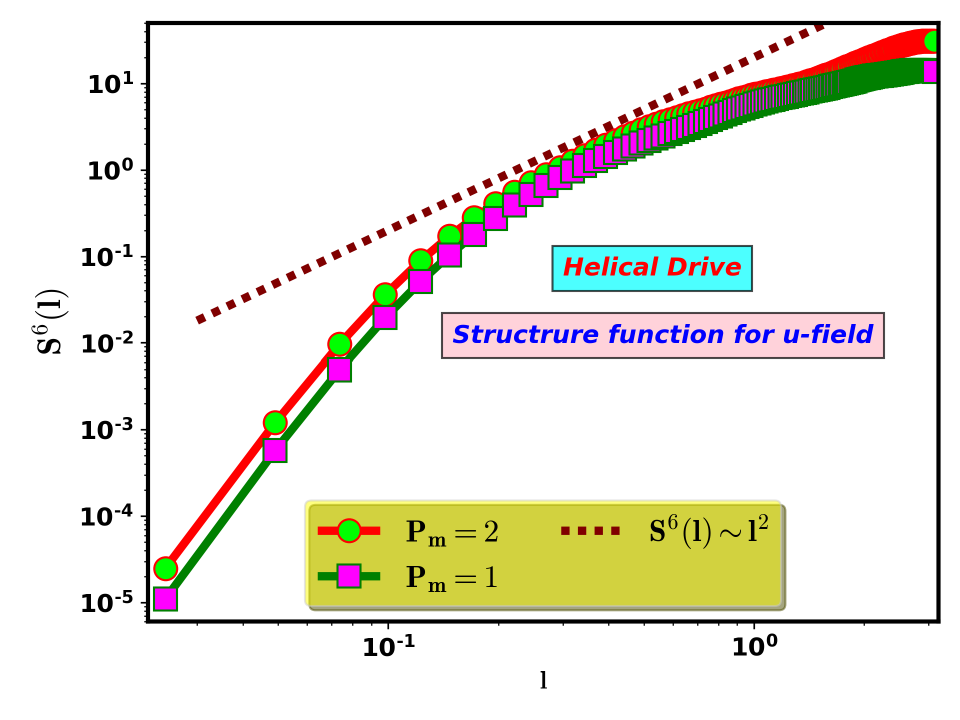}
		\caption{}
	\end{subfigure}
	\caption{The plots display the structure functions of three distinct orders, namely $p = 2, 3$, and $6$, as a function of $l$. These structure functions represent the velocity and magnetic field for various $P_m$ values in the context of a helical dynamo. The numerical analysis confirms that Kolmogrov's scalings \cite{kolmogorov:1941}, namely $S^2 (l) \sim l^\frac{2}{3}$, $S^3 (l) \sim l$, and $S^6 (l) \sim l^2$, are valid for both the velocity and magnetic field.}

	\label{Structure Function Helical}
\end{figure*}

We have computed the second-order ($S^2 (l)$), third-order ($S^3 (l)$), and sixth-order ($S^6 (l)$) structure functions of velocity and magnetic field for a helical dynamo. Figure \ref{Structure Function Helical} (top row) displays the second-order, third-order, and sixth-order structure functions for the magnetic field, whereas figure \ref{Structure Function Helical} (bottom row) showcases the corresponding structure functions for the velocity field. Our numerical calculation reveals that the predictions made by Kolmogorov \cite{kolmogorov:1941} are evident for both lower and higher order calculations (i.e, $p = 2, 3$, and $6$). Additionally, it is noteworthy that the magnetic Prandtl number ($P_m$) has minor influence on the dependency of $S^2 (l)$, $S^3 (l)$, and $S^6 (l)$. Nevertheless, the $P_m$ does not alter the slope of the structural functions.


The ratio of the sixth-order structure function to the second-order structure function for velocity and magnetic field serves as a reliable indicator of the intermittency present in turbulence. The concept being referred to is hyper-flatness, which has the following mathematical definition: 

\begin{equation}
	\mathcal{F}^6(l) = \frac{S^6 (l)}{(S^2 (l))^3}
\end{equation}

\begin{figure*}
	\begin{subfigure}{0.49\textwidth}
		\centering
		\includegraphics[scale=0.49]{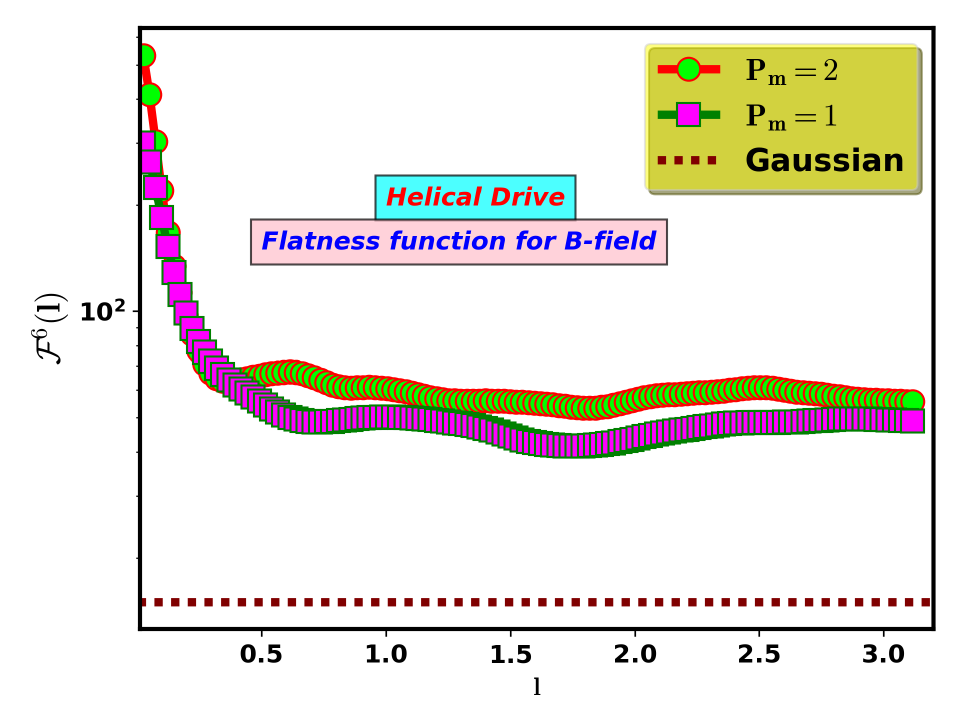}
		\caption{}
	\end{subfigure}
	\begin{subfigure}{0.49\textwidth}
		\centering
		\includegraphics[scale=0.49]{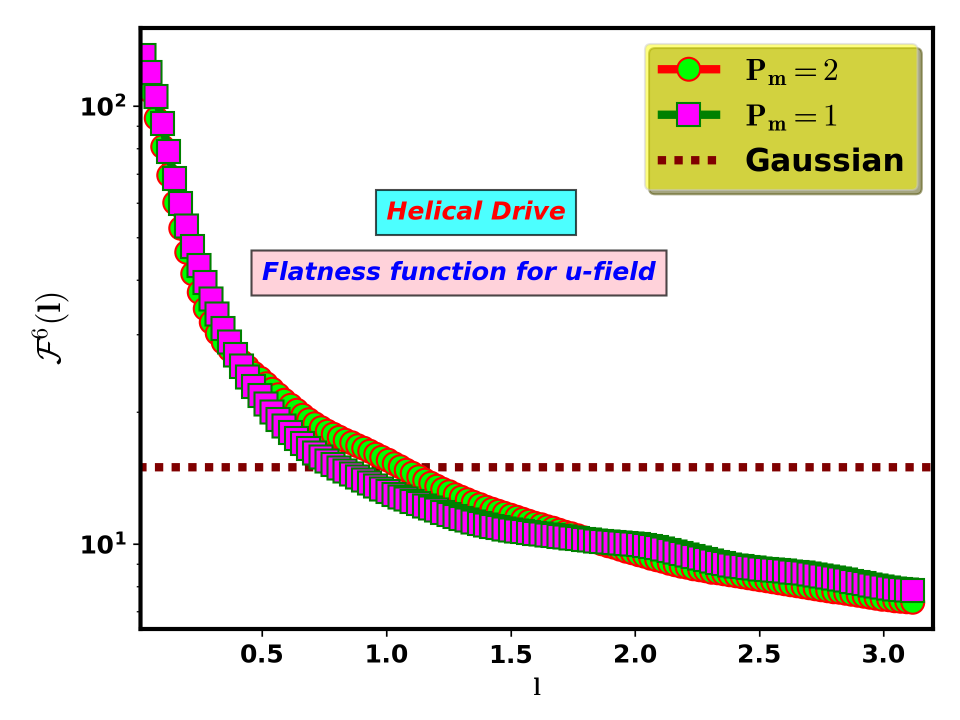}
		\caption{}
	\end{subfigure}
	\caption{ The hyper-flatness $\mathcal{F}^6(l)$ is calculated as a function of $l$ for two separate cases: (a) magnetic field and (b) velocity field. These calculations are performed at various values of $P_m$ for a helical dynamo.} 
	\label{Flatness Helical}
\end{figure*}

For the helical dynamo, we plot hyper-flatness ($\mathcal{F}^6(l)$) versus $l$ for the magnetic field and velocity field (refer to Fig. \ref{Flatness Helical}). This graphic demonstrates that the hyper-flatness ($\mathcal{F}^6(l)$), increases as the length scale, $l$, approaches zero ($l \to 0$). This suggests the presence of small-scale intermittency at the scales of dissipation range.


\begin{figure*}
	\begin{subfigure}{0.49\textwidth}
		\centering
		\includegraphics[scale=0.55]{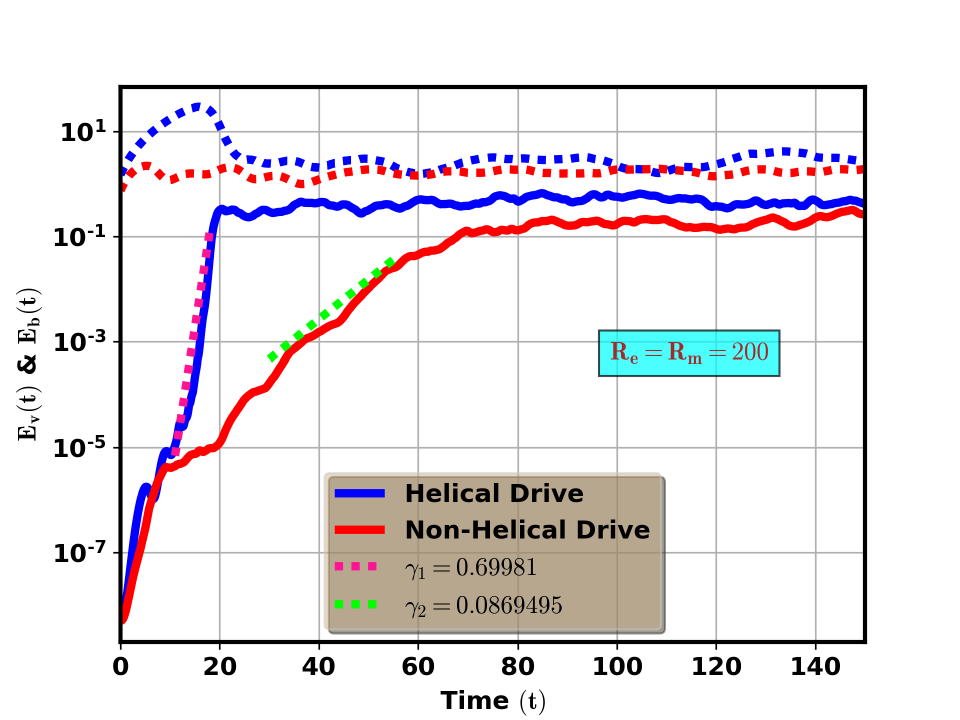}
		\caption{}
	\end{subfigure}
	\begin{subfigure}{0.49\textwidth}
		\centering
		\includegraphics[scale=0.55]{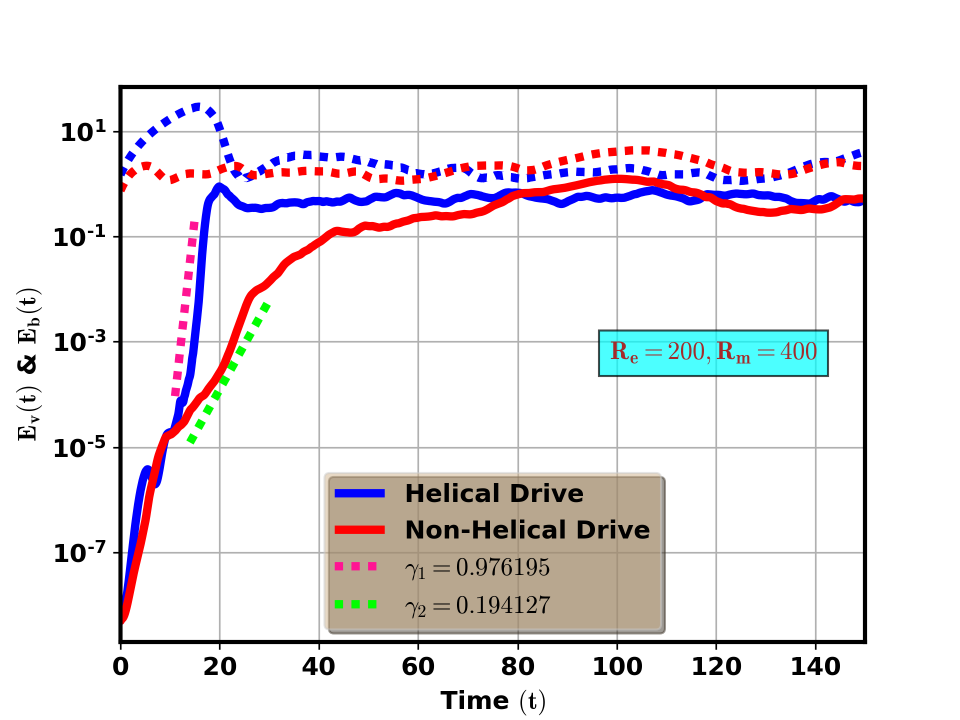}
		\caption{}
	\end{subfigure}
		\caption{Comparing helically driven dynamo with non-helically driven dynamo. The growth rate of the dynamo is significantly greater in the helically driven case compared to the non-helically driven condition, as determined by a linear fit. Therefore, the helical dynamo saturates more quickly because of the varying strength of the ``back reaction''. Numerical experiments were conducted for $P_m$ values of (a) $1.0$ and (b) $2.0$.} 
		\label{Compare helical vs nonhelical}
\end{figure*}

Following an extensive investigation of the influence of helical drive on the activity of self-consistent dynamos, the present study investigates the fate of self-consistent dynamos in the presence of non-helical drive. As previously mentioned, our drive is YM flow drive, and its helical character can be controlled using specific parameter spaces. Therefore, we begin by examining the independence of the initial condition when a non-helical drive is present (refer to Appendix \ref{Appen A} for further information). Additionally, we conduct a comparison of the time evolution of magnetic energy ($E_B = \frac{1}{2} \int_{V} (B_x^2 + B_y^2 + B_z^2) dx dy dz$) under two distinct drives: helical (ABC) and non-helical (EPI-2D). Fig. \ref{Compare helical vs nonhelical} shows that the magnetic energy growth rate ($\gamma = \frac{d}{dt}(\ln E_B(t))$) is significantly greater for the helically driven case compared to the non-helically driven case. This difference is quantified by a linear fit shown by the dotted lines, and the slopes of these fits are provided in the figure legend. In the kinematic stage, the impact of the magnetic field on the velocity field is minimal due to the small strength of the magnetic field. When the magnetic field reaches a certain strength that affects the flow, non-linearity develops earlier in the helical case, causing the exponential growth rate to slow down earlier compared to the non-helical case (as illustrated in Fig. \ref{Compare helical vs nonhelical}). Hence, the helically forced turbulent dynamo saturates earlier than the non-helically forced scenario as observed in Fig. \ref{Compare helical vs nonhelical}.


\begin{figure*}
	\begin{subfigure}{0.49\textwidth}
		\centering
		\includegraphics[scale=0.55]{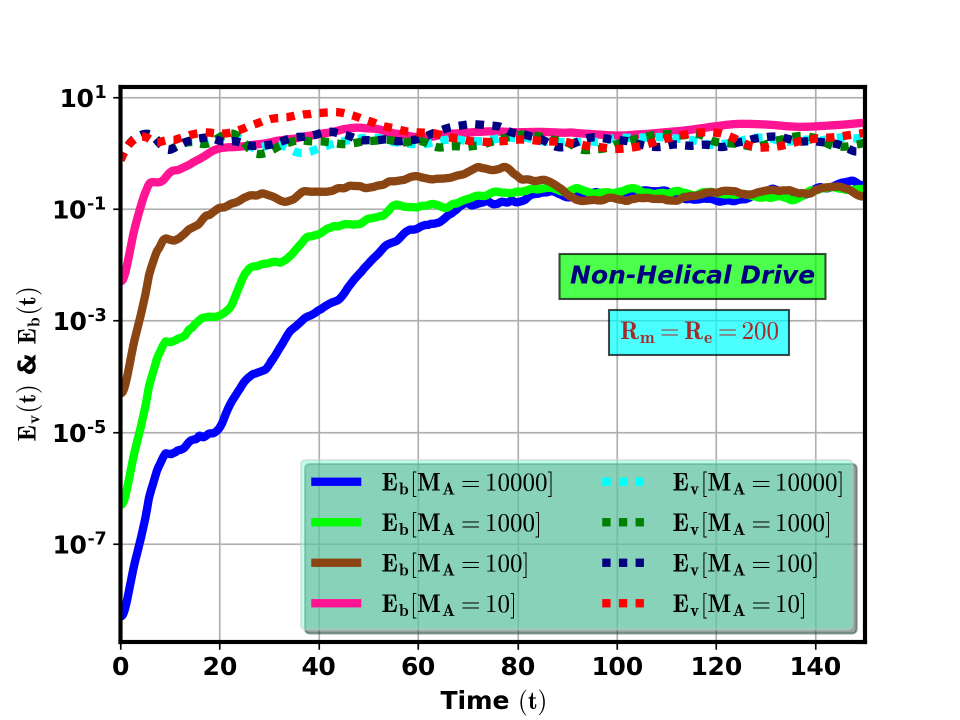}
		\caption{}
		\label{nonhelical Ma}
	\end{subfigure}
	\begin{subfigure}{0.49\textwidth}
		\centering
		\includegraphics[scale=0.55]{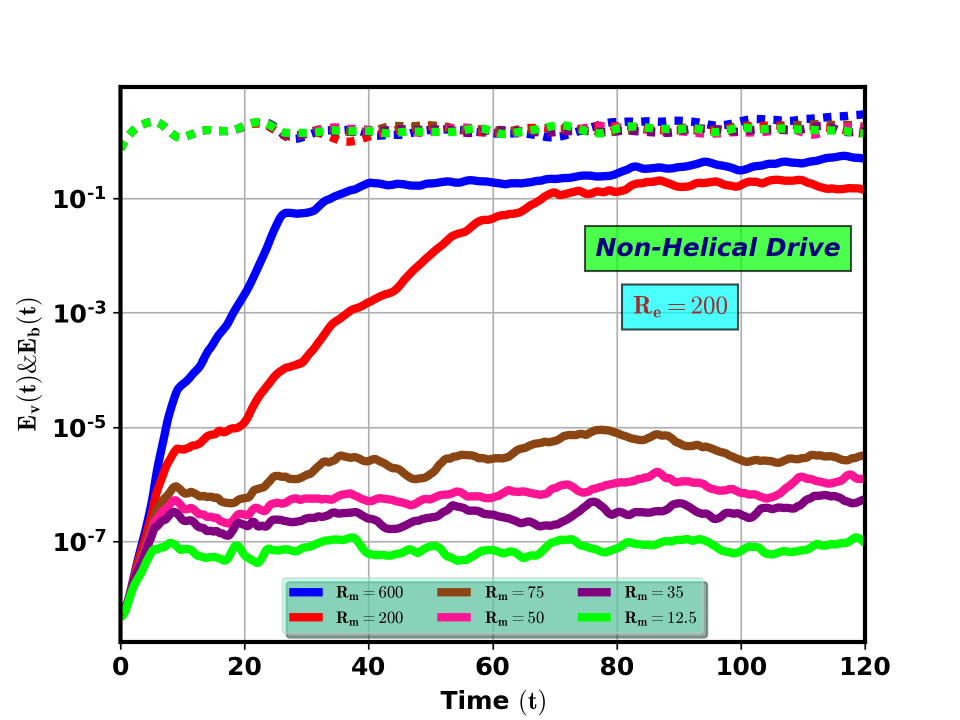}
		\caption{}
		\label{nonhelical Rm}
	\end{subfigure}
		\caption{(a) The non-helical forced dynamo action is being influenced by the Alfven speed. Dynamo activity is more pronounced in super Alfvenic regimes when driven by non-helical forces. (b) The impact of the magnetic Prandtl number ($P_m = \frac{R_m}{R_e}$) on the action of a non-helical dynamo is examined in this study, keeping the kinetic Reynolds number ($R_e$) constant. At lower $P_m$ limit, the non-helical dynamo activity is suppressed. Here the dotted lines
			indicate the kinetic energy.} 
\end{figure*}

Additionally, the impact of Alfven speed is examined in the case of non-helically driven dynamos. We have conducted simulations with various Alfven Mach numbers ($M_A = 10000, 1000, 100, 10$) and found that the most effective dynamo activity occurs in the super Alfvenic regime when the velocity field is driven by non-helical forcing (Refer to Fig. \ref{nonhelical Ma}). Additionally, it is noteworthy to mention that the saturation value of kinetic and magnetic energy remains constant regardless of the value of $M_A$; this suggests that it is not influenced by Alfven speed (refer to Figure \ref{nonhelical Ma}). This observation aligns with the previous helically forced instance.


Following this, the impact of the magnetic Prandtl number ($P_m$) on these non-helical dynamos is investigated. We maintain a constant kinetic Reynolds number ($R_e$) while adjusting the magnetic Reynolds number ($R_m$) to study the impact of $P_m$. Thus, we can also view this as a depiction of the magnetic Prandtl number ($P_m$) relationship. Figure  \ref{nonhelical Rm} indicates that the dynamo action is inhibited at lower values of magnetic Prandtl number ($P_m$) with a non-helical drive (refer to Appendix \ref{Appen B} for details), similar to the helical situations mentioned earlier. 


\begin{figure*}
	\centering
	\begin{turn}{90} 
		\large{\textbf{\textcolor{blue}{\doublebox{Kinematic}}}}
	\end{turn}
	\begin{subfigure}{0.31\textwidth}
		\centering
		\includegraphics[scale=0.05580]{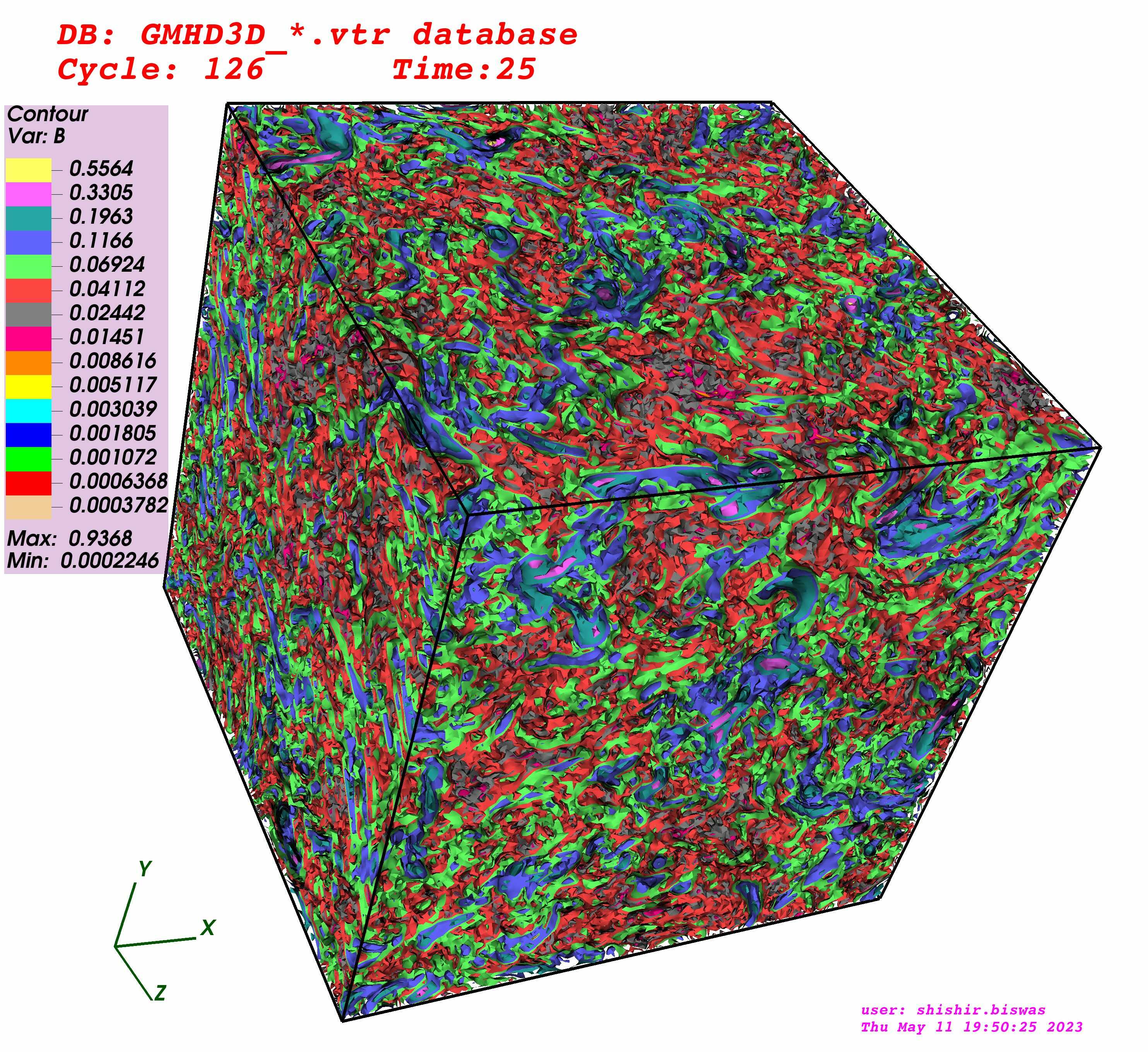}
		\caption{$P_m = 2.0$}
	\end{subfigure}
	\begin{subfigure}{0.31\textwidth}
		\centering
		\includegraphics[scale=0.05580]{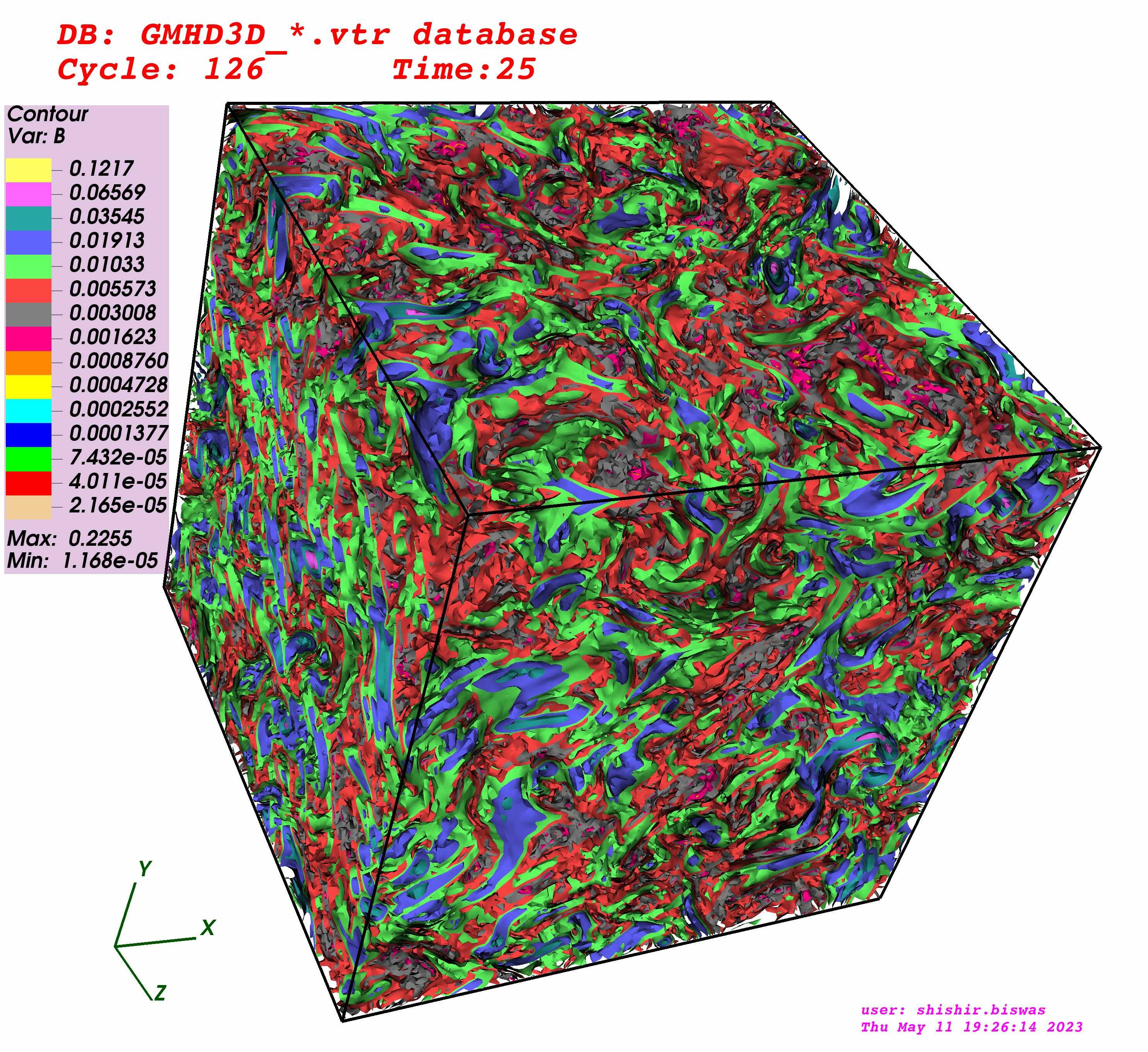}
		\caption{$P_m = 1.0$}
	\end{subfigure}
	\begin{subfigure}{0.31\textwidth}
		\centering
		\includegraphics[scale=0.05580]{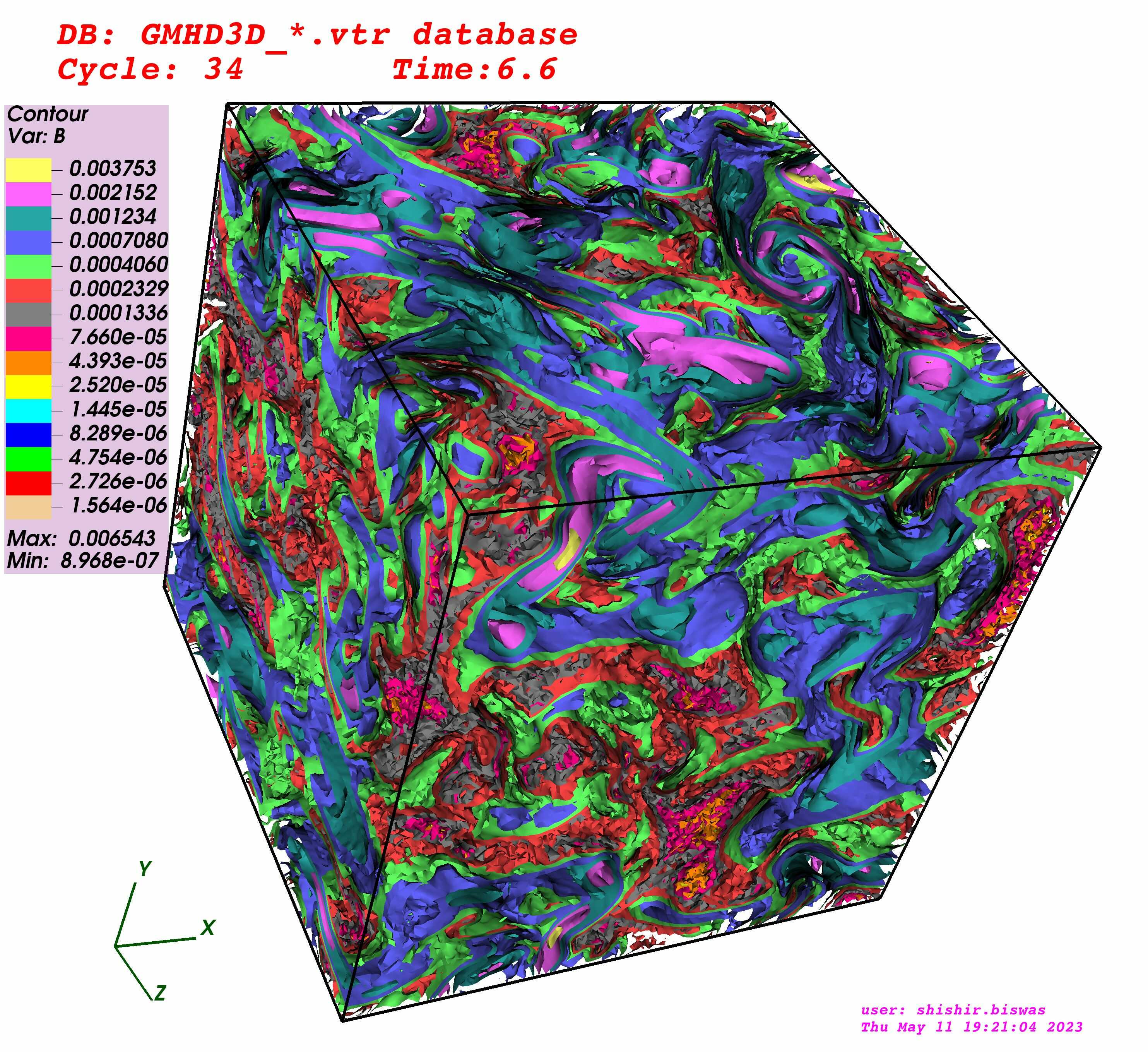}
		\caption{$P_m = 0.25$}
	\end{subfigure}
	\begin{turn}{90} 
		\large{\textbf{\textcolor{blue}{\doublebox{Saturated}}}}
	\end{turn}
	\begin{subfigure}{0.31\textwidth}
		\centering
		\includegraphics[scale=0.05580]{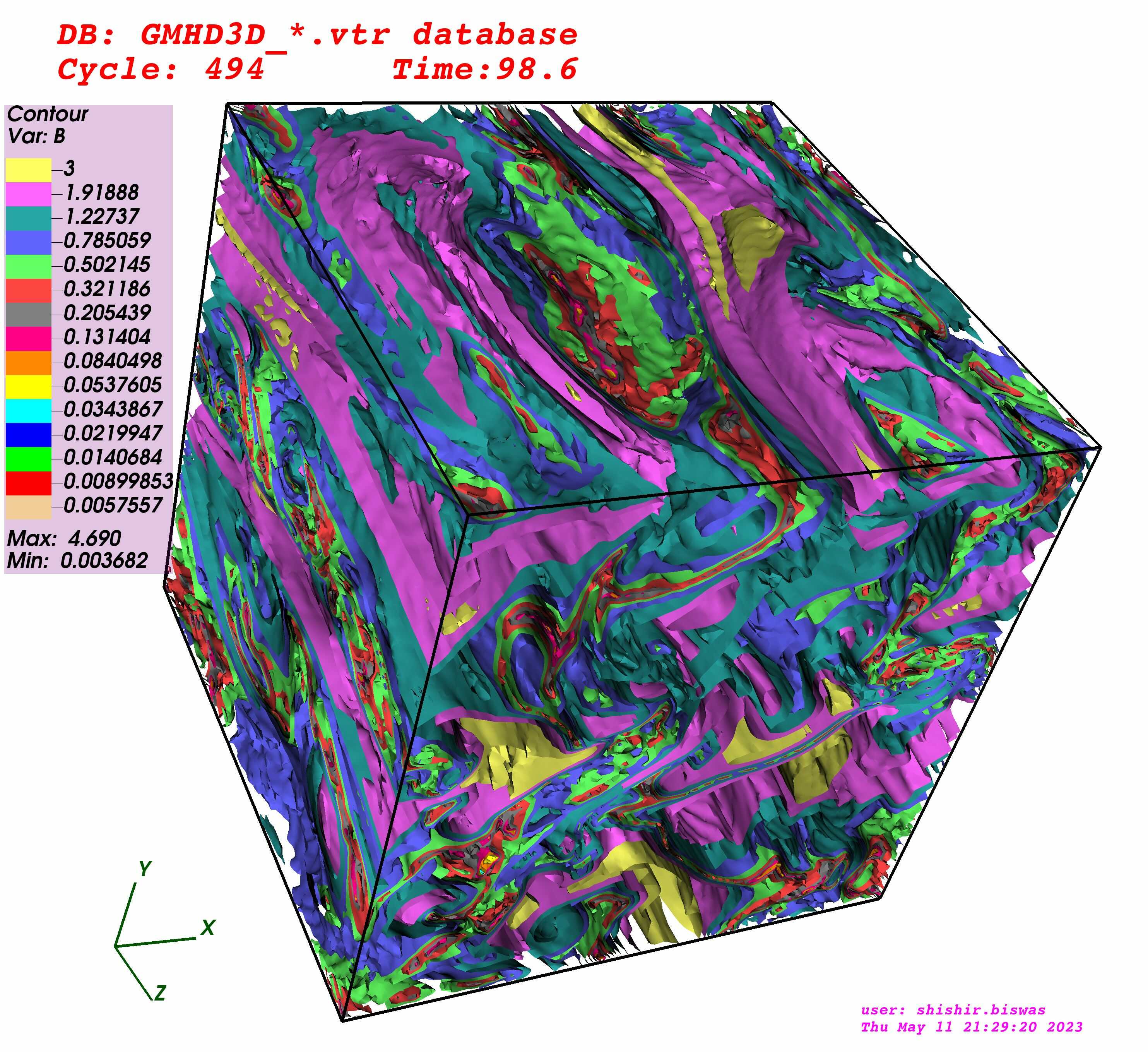}
		\caption{$P_m = 2.0$}
	\end{subfigure}
	\begin{subfigure}{0.31\textwidth}
		\centering
		\includegraphics[scale=0.05580]{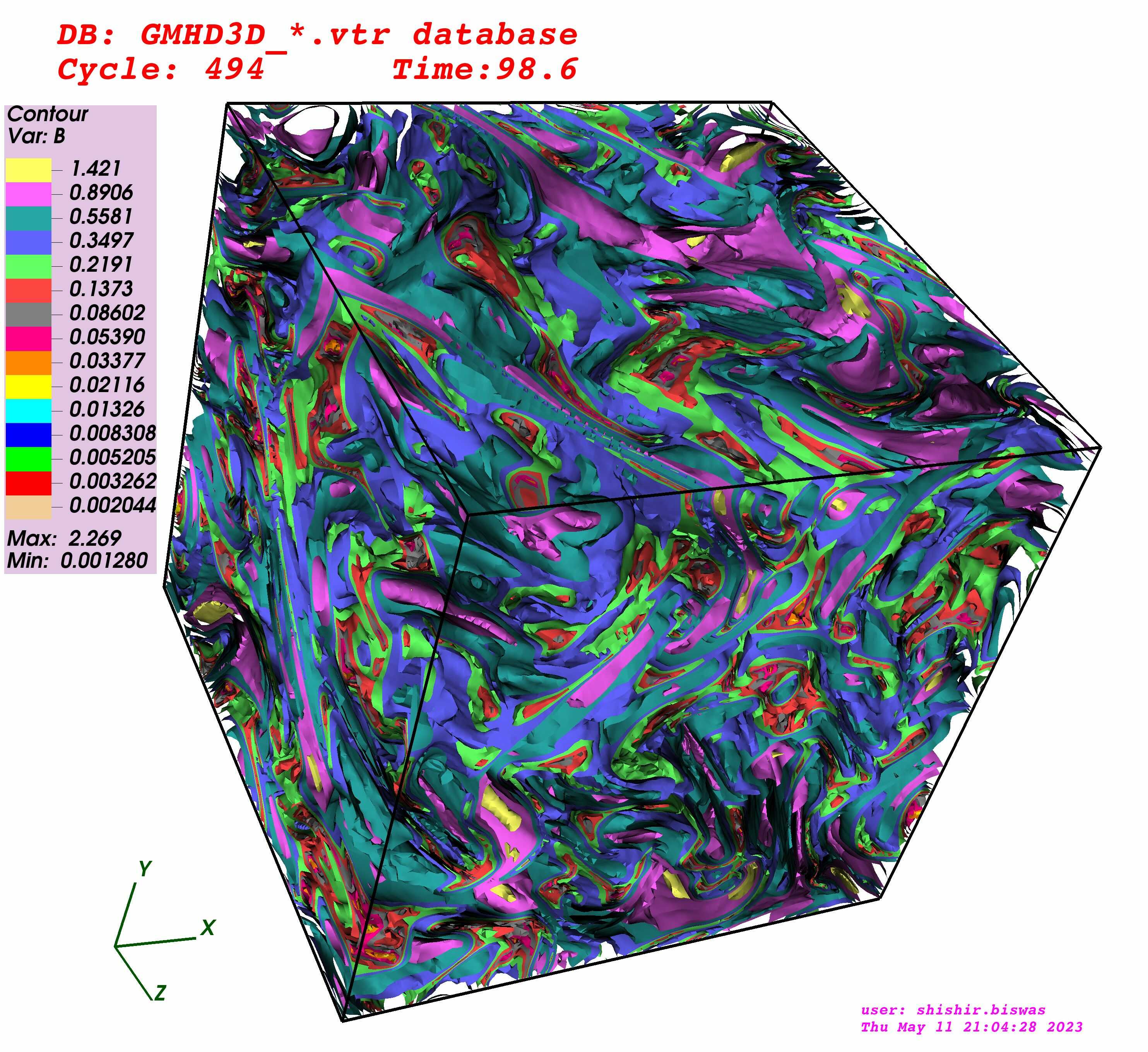}
		\caption{$P_m = 1.0$}
	\end{subfigure}
	\begin{subfigure}{0.31\textwidth}
		\centering
		\includegraphics[scale=0.05580]{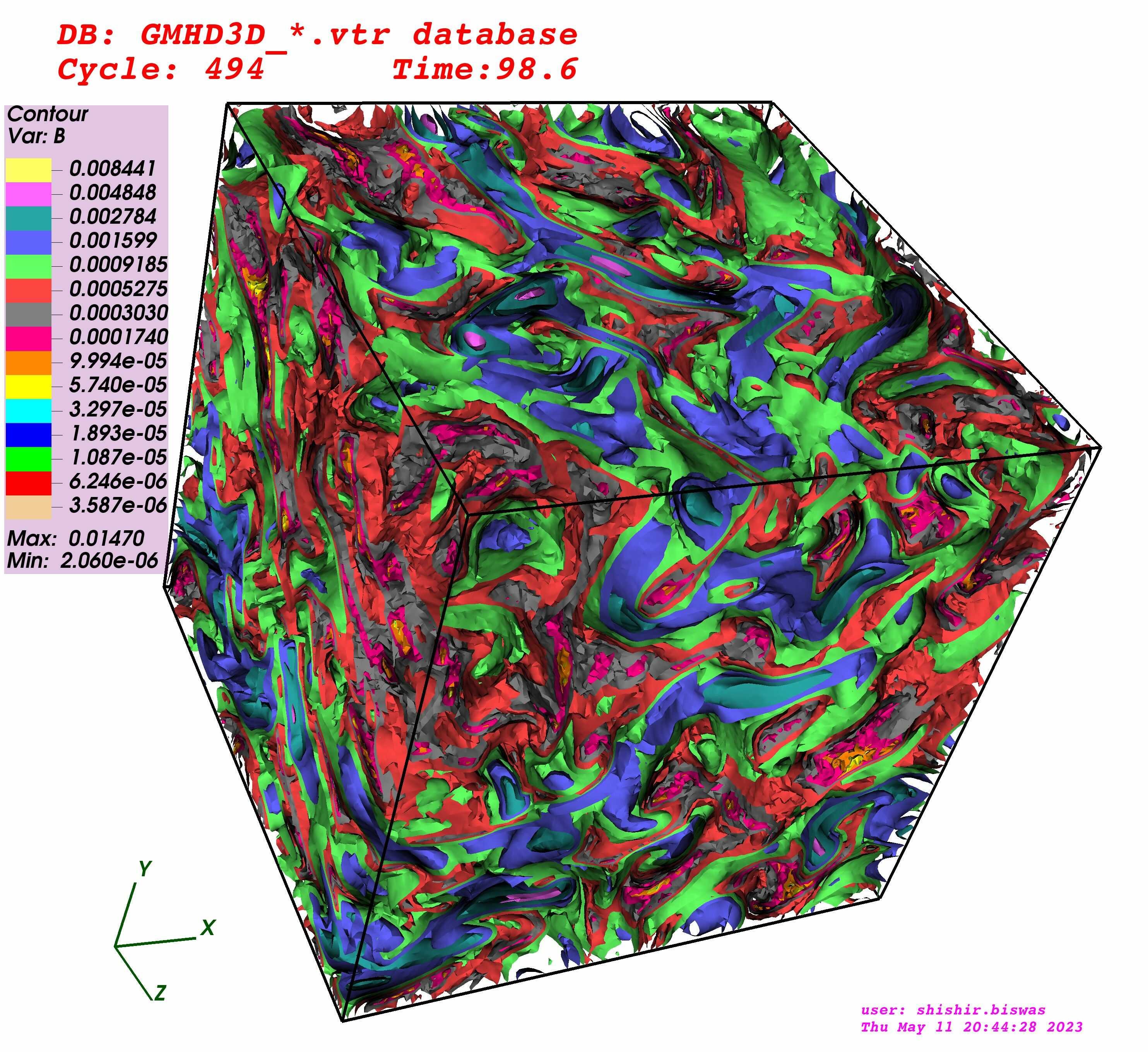}
		\caption{$P_m = 0.25$}
	\end{subfigure}
		\caption{ Visualization of three-dimensional magnetic field iso-surfaces (Iso-B surfaces) for a non-helical dynamo in the kinematic stage (top row) and saturated stage (bottom row) for (a \& d) $P_m = 2$, (b \& e) $P_m = 1$, and (c \& f) $P_m = 0.25$. The magnetic fields in the saturated stages exhibit structures that are larger than those in the kinematic stages. In the kinematic stage, the magnetic field structures increase in size as the magnetic Prandtl number ($P_m$) decreases. However, in the saturated stage, the opposite occurs, with larger structures forming as $P_m$ is increased. This observation is the complete opposite of the one made for the helical example. The log scale is utilized for visualization.} 
	\label{non helical drive IsoB}
\end{figure*}

Based on the preceding discussion, it has been determined that the time evolution of magnetic energy under helical or non-helical drive behaves identically across a range of $P_m$ values. The iso-surface dynamics of magnetic fields (IsoB) for non-helical dynamos at various $P_m$ values are now presented. The magnetic field in the kinematic stage exhibits smaller scale structures compared to the saturated phases, as observed in Fig. \ref{non helical drive IsoB}. For all values of $P_m$, the magnetic field structures in saturated phases appear visibly larger in size (Refer to Fig. \ref{non helical drive IsoB}). This observation is found to be identical as the helical dynamos addressed previously.


\begin{figure*}
	\centering
		\begin{turn}{90} 
			\large{\textbf{\textcolor{blue}{\doublebox{Saturated}}}}
		\end{turn}
	\begin{subfigure}{0.31\textwidth}
		\centering
		\includegraphics[scale=0.05580]{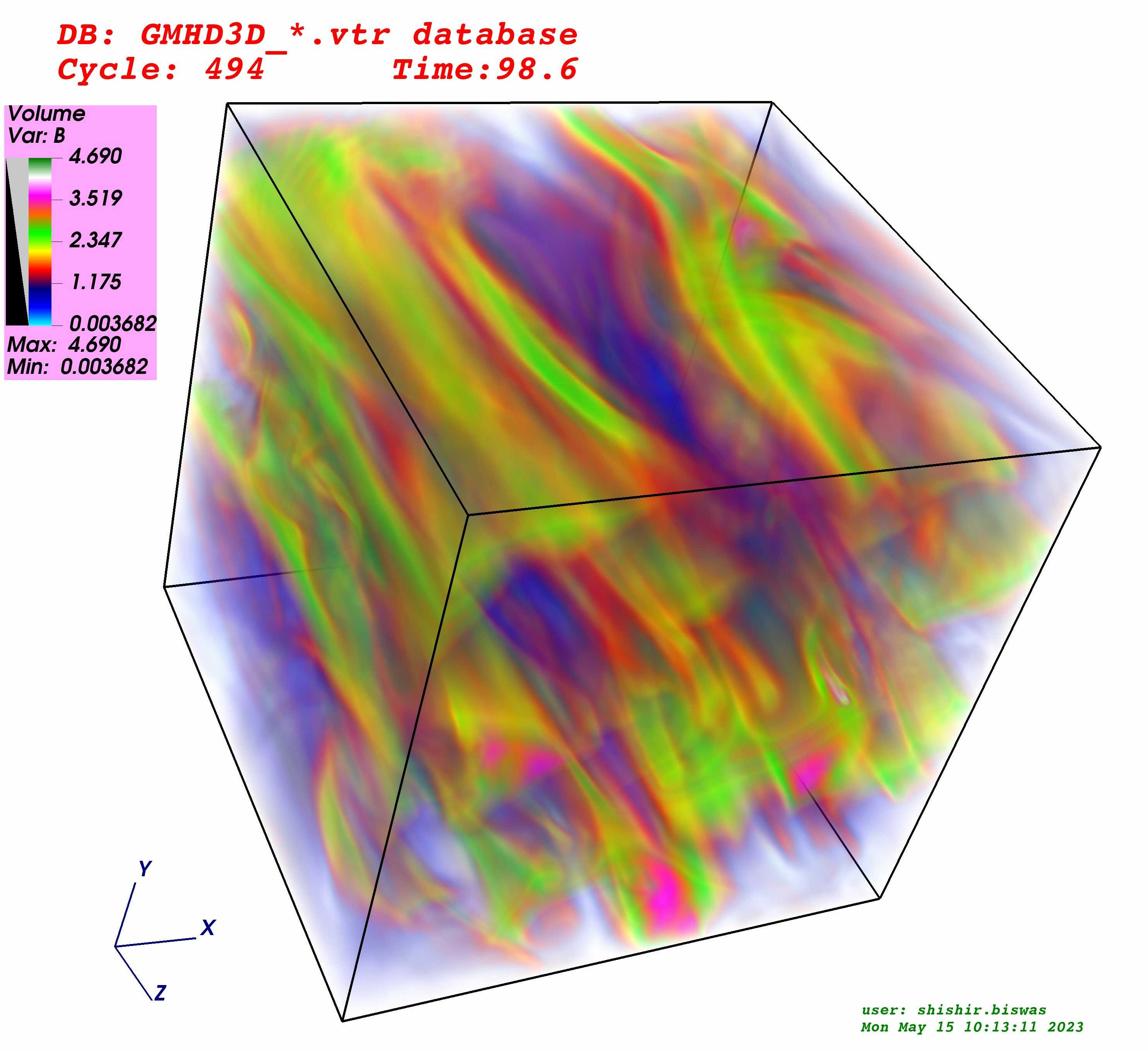}
		\caption{$P_m = 2.0$}
	\end{subfigure}
	\begin{subfigure}{0.31\textwidth}
		\centering
		\includegraphics[scale=0.05580]{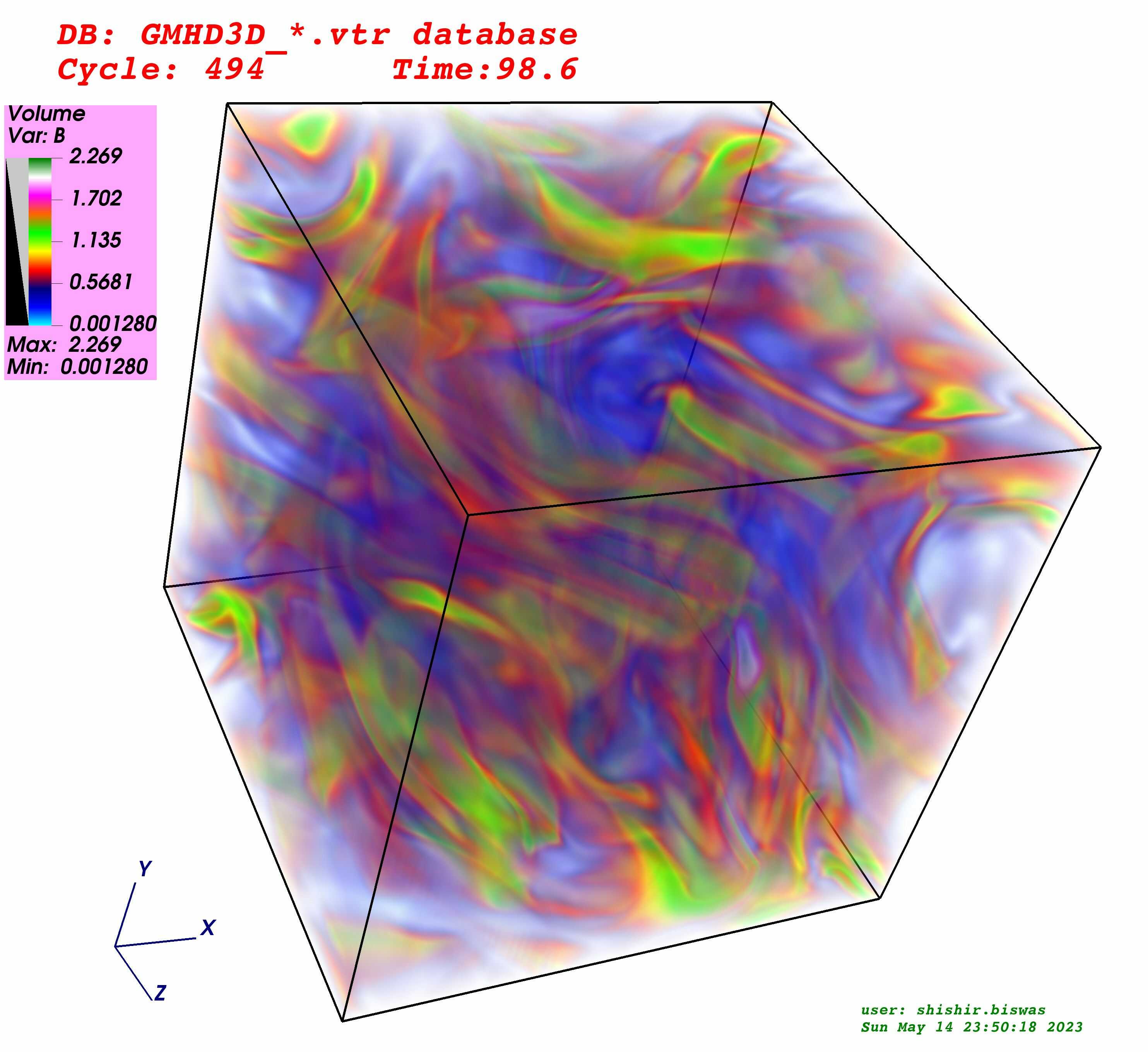}
		\caption{$P_m = 1.0$}
	\end{subfigure}
	\begin{subfigure}{0.31\textwidth}
		\centering
		\includegraphics[scale=0.05580]{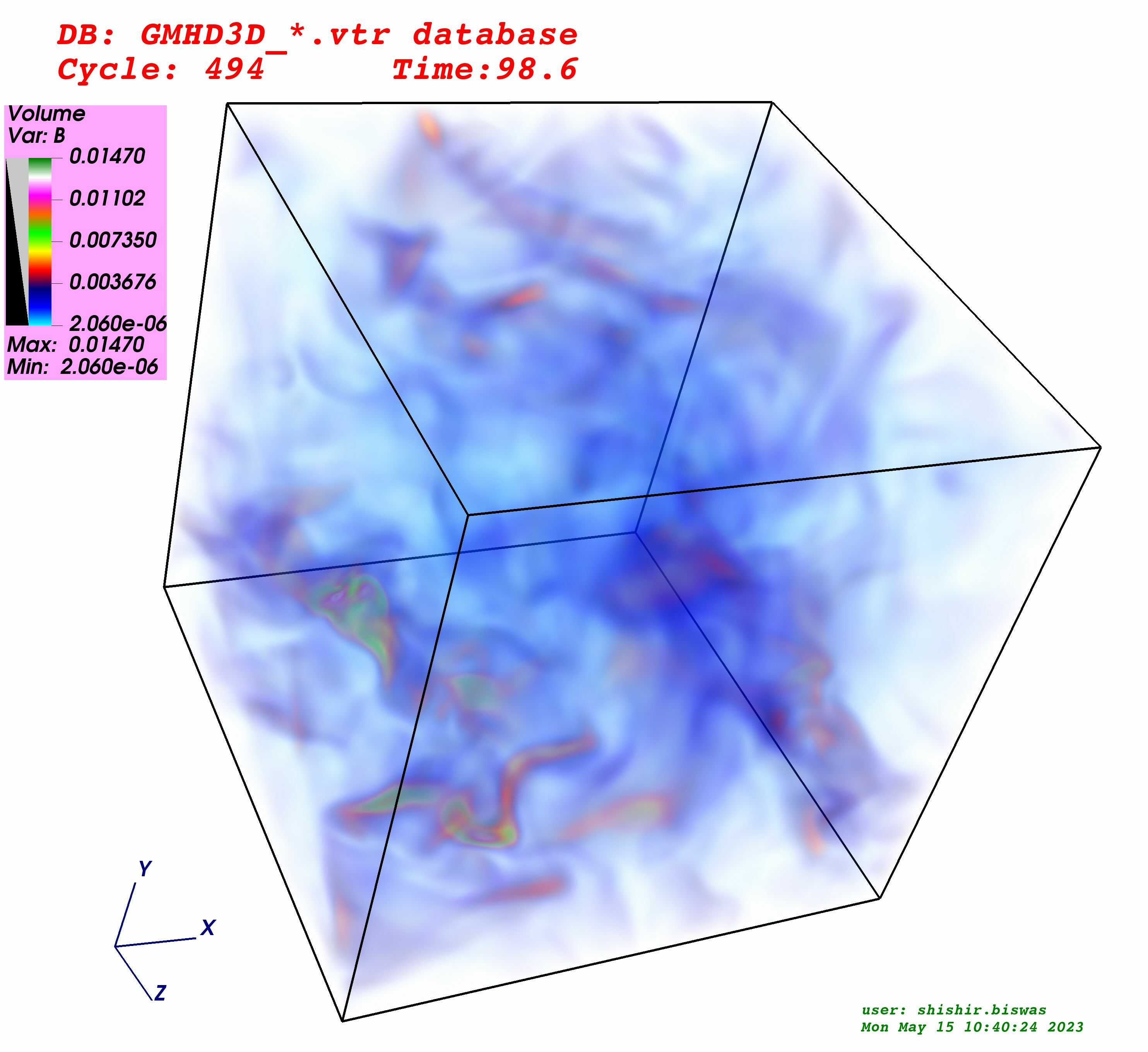}
		\caption{$P_m = 0.25$}
	\end{subfigure}
	\caption{Volume rendering visualization of magnetic field for non-helical  dynamo in saturated stage at (a) $P_m = 2$ , (b) $P_m = 1$ , and (c) $P_m = 0.25$. The magnetic field structures are getting larger as $P_m$ is increased in saturated stage. Using this volume rendering visualization technique larger scale structures are clearly identified for higher $P_m$ values. Visualization is done in linear scale.} 
		\label{non helical drive volume}
\end{figure*}

It is interesting to observe that the magnetic field iso-surface (IsoB) structure increases in kinematic stage as the $P_m$ value decreases (See Fig. \ref{non helical drive IsoB}). This is caused by the increase in magnetic resistivity, which subsequently minimizes smaller fluctuations in the kinematic phase of non-helical dynamos.  The magnetic field iso-surface structures exhibit a reversal nature with $P_m$ in the saturated stage. During the saturated stage, it is observed that the structures become increasingly larger as the parameter $P_m$ increases (refer to Fig. \ref{non helical drive IsoB}). We have implemented an alternative visualization tool to comprehend this unique behavior and do cross verification. We used volume metric rendering to depict the magnetic fields at saturation stage in a linear scale and observed the reversal nature more prominently (See Fig. \ref{non helical drive volume}). Figure \ref{non helical drive volume} clearly demonstrates that the magnetic field structures' scales increase when the magnetic Prandtl number ($P_m$) increases for non-helical dynamos. This is a noteworthy and diametrically opposed observation to the helical case.


\begin{figure*}
	\centering
	\begin{subfigure}{0.32\textwidth}
		\centering
		\includegraphics[scale=0.39]{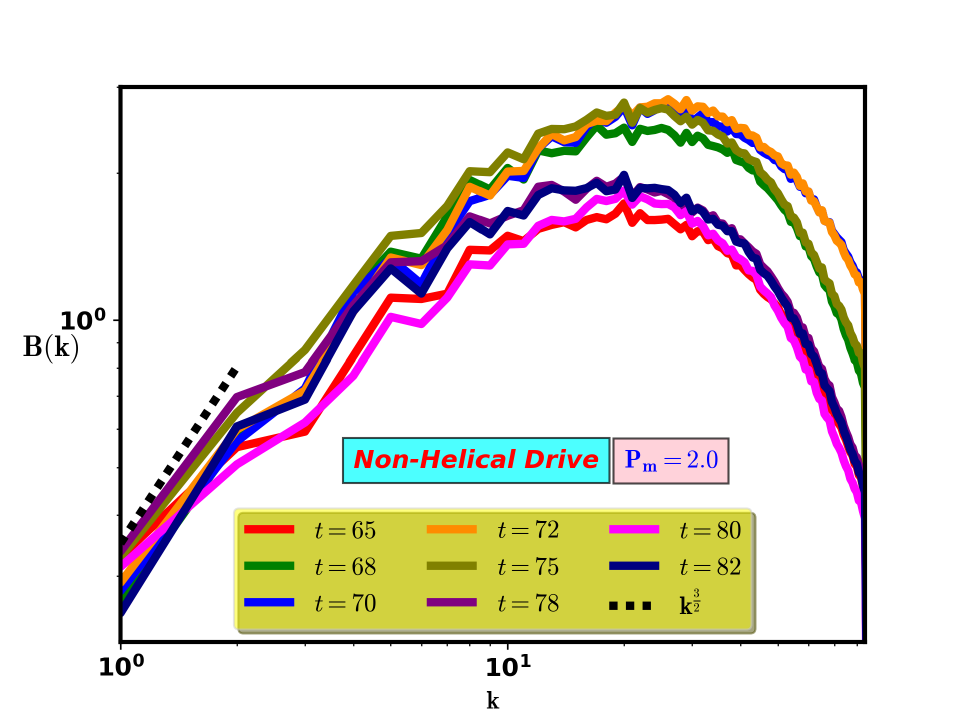}
		\caption{}
	\end{subfigure}
	\begin{subfigure}{0.32\textwidth}
		\centering
		\includegraphics[scale=0.39]{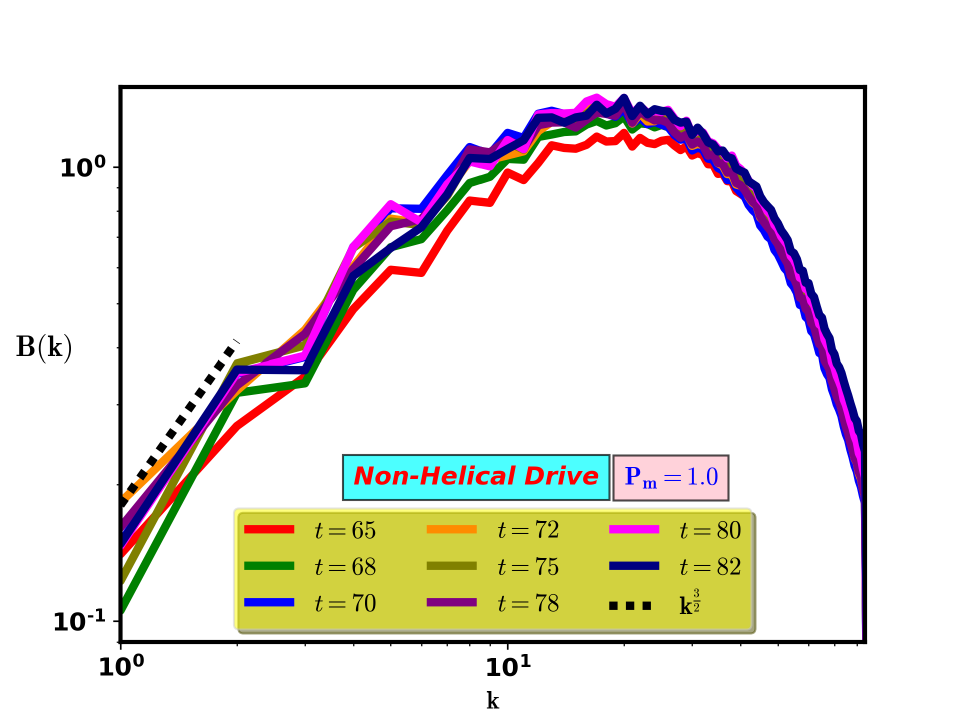}
		\caption{}
	\end{subfigure}
	\begin{subfigure}{0.32\textwidth}
		\centering
		\includegraphics[scale=0.39]{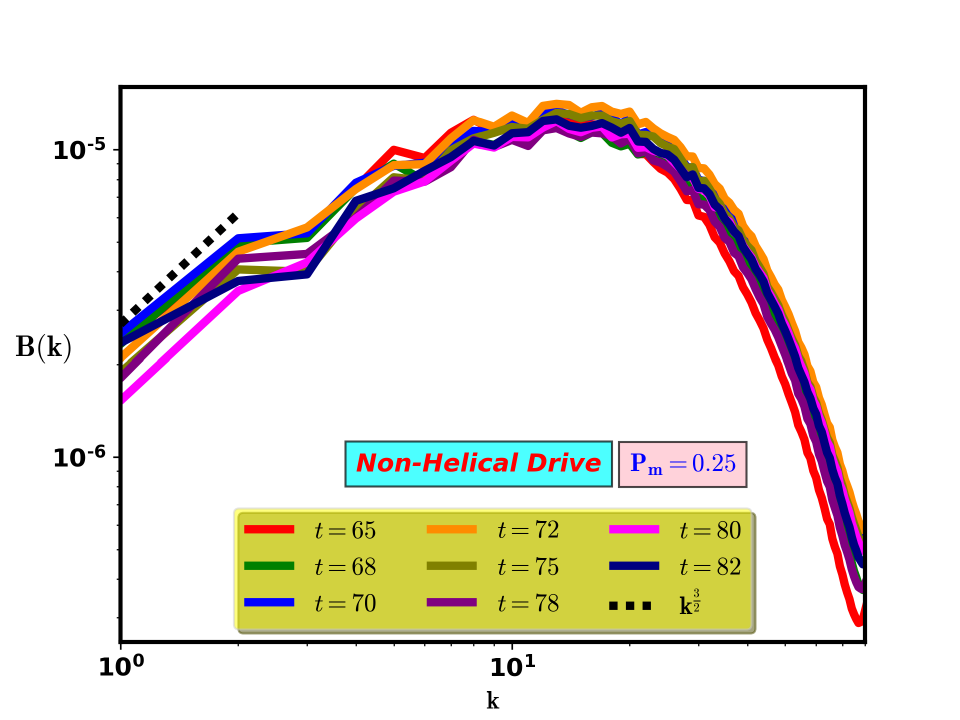}
		\caption{}
	\end{subfigure}
	\caption{ Determination of the spectral density of magnetic energy $B(k)$ for the following $P_m$ values: (a) $2.0$, (b) $1.0$, and (c) $0.25$, such that  $\int |B(k, t)|^2 dk$ is the total energy at time t and $k = \sqrt{k_x^2 + k_y^2 + k_z^2}$. The dynamos are unambiguously designated as small scale dynamos (SSD), which is not dependent on $P_m$. The magnetic field exhibits a well-established Kazantsev $k^\frac{3}{2}$ scaling \cite{Kazantsev:1968}.} 
	\label{non helical specta}
\end{figure*}

It can be observed from Figures \ref{non helical drive IsoB} and \ref{non helical drive volume} that the structures of the magnetic fields are predominantly composed of shorter scales. We have conducted spectral analysis to confirm this further. The spectral density of magnetic energy $B(k)$ is represented in Figure \ref{non helical specta} as a function of $k$ for $P_m$ values of $0.25$, $1.0$, and $2.0$, where $\int |B(k, t)|^2 dk$ represents the total energy at time t and $k = \sqrt{k_x^2 + k_y^2 + k_z^2}$. Our spectral estimate strongly indicates that the majority of power is concentrated at smaller length scales, namely on higher modes. Based on our comprehensive spectral analysis, it is obvious that these dynamos operate as small scale dynamos (SSD). The magnetic field precisely adheres to the Kazantsev $k^\frac{3}{2}$ spectral scaling \cite{Kazantsev:1968}, which remains consistent across a variety of $P_m$ values as seen in Fig. \ref{non helical specta}.


Therefore, based on our numerical analysis, it has been determined that the influence of both helical and non-helical drives results in the formation of small scale dynamos (SSD) with a substantial Kazantsev scaling of $k^\frac{3}{2}$ \cite{Kazantsev:1968}. Furthermore, our analysis demonstrates that this conclusion holds true across a variety of $P_m$ values.


\begin{figure}
	\centering
	\includegraphics[scale=0.5]{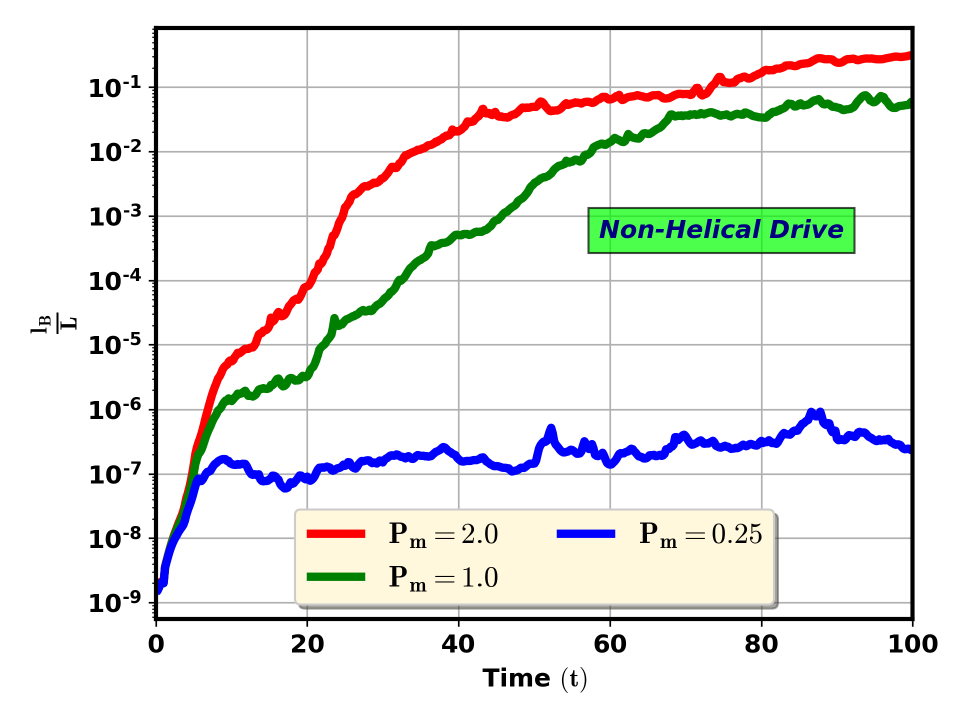}
	\caption{The magnetic field coherence length, normalized to the size of the numerical domain ($\frac{l_B}{L} = \frac{\int k^{-1} B(k) dk}{\int B(k) dk}$), has been computed for a non-helically driven dynamo for various values of magnetic Prandtl number ($P_m$). Refer to Fig. \ref{non helical drive IsoB} \& \ref{non helical drive volume} for comparison.} 
	\label{lb vs t Non-Helical}
\end{figure}

In order to have a clearer understanding of the dynamics of the magnetic field, we have computed the coherence length of the magnetic field as defined in Equation \ref{lb definition}. The normalized magnetic field coherence length ($\frac{l_B}{L}$) is found to be greater in the saturated stages than in the kinematic stage. Furthermore, the ratio of $l_B$ to $L$ is shown to rise over time as $P_m$ increases in the presence of a non-helical drive (refer to Figure \ref{lb vs t Non-Helical}). This result aligns with our findings about the dynamics of the magnetic field, as depicted in figures \ref{non helical drive IsoB} \& \ref{non helical drive volume}.


\begin{figure*}
	\begin{subfigure}{0.49\textwidth}
		\centering
		\includegraphics[scale=0.49]{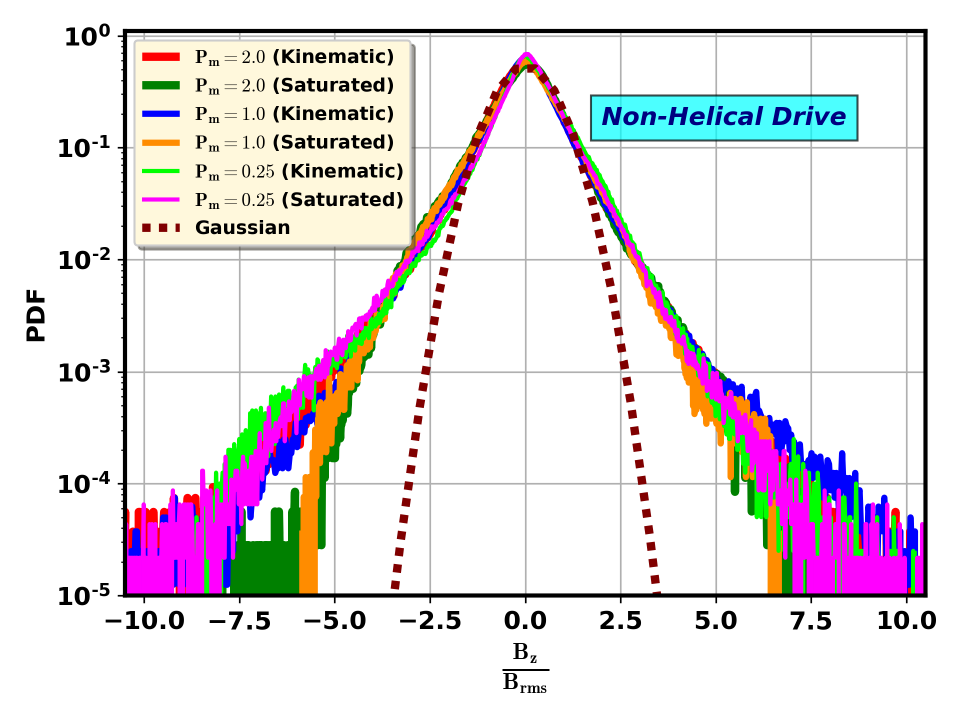}
		\caption{}
		\label{PDF Bz Non-Helical}
	\end{subfigure}
	\begin{subfigure}{0.49\textwidth}
		\centering
		\includegraphics[scale=0.49]{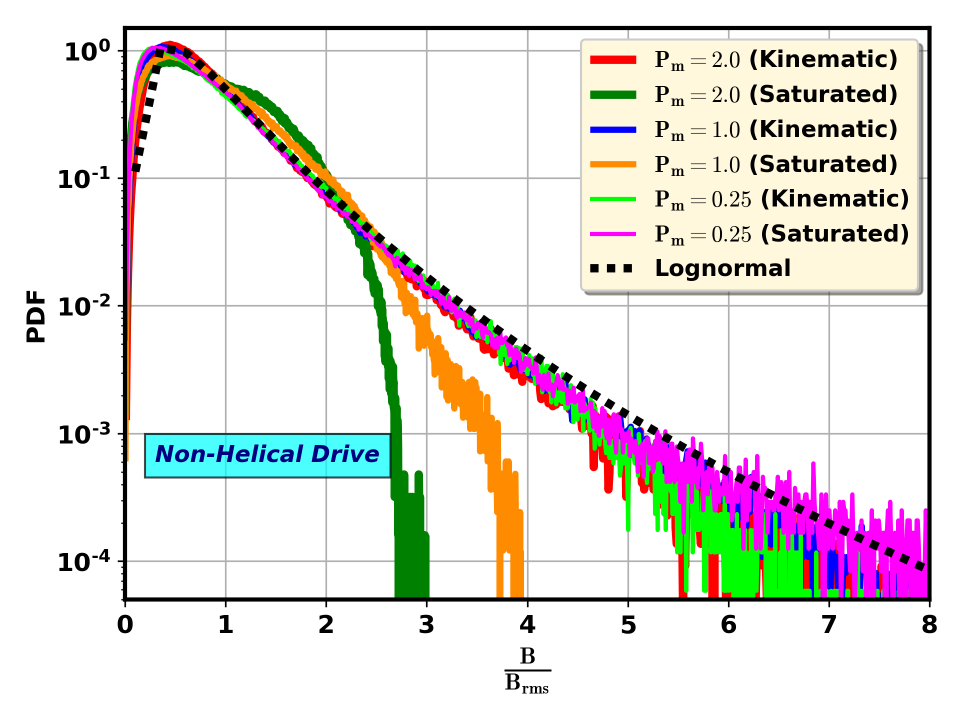}
		\caption{}
		\label{PDF B Non-Helical}
	\end{subfigure}
	\caption{(a) The probability density functions (PDFs) for the normalized $z$-component of the magnetic field $\frac{B_z}{B_{rms}}$ and (b) the magnetic field strength $\frac{B}{B_{rms}}$ are calculated for a non-helically driven dynamo in both the kinematic and saturated stages. Here $B$ is defined as $\sqrt{B_x^2 + B_y^2 + B_z^2}$ and $B_{rms}$ is defined as $\sqrt{\langle B^2 \rangle}$.}
	\label{PDF Non-Helical Dynamo}
\end{figure*}

We have computed the probability density functions (PDFs) of the normalized $z$-component of the magnetic field, denoted as $\frac{B_z}{B_{rms}}$, as well as the magnetic field strength, denoted as $\frac{B}{B_{rms}}$ (where $B$ is defined as $\sqrt{B_x^2 + B_y^2 + B_z^2}$ and $B_{rms}$ is defined as $\sqrt{\langle B^2 \rangle}$), during both the kinematic and saturated dynamo stages. We have performed these calculations for various values of the magnetic Prandtl number ($P_m$), similar to our previous analyses. Additionally, we have confirmed that the PDFs for the other components of the magnetic field exhibit identical behavior. Figure \ref{PDF Non-Helical Dynamo} unambiguously indicates that the distribution is non-Gaussian.  Additionally, it is observed that the kinematic stage exhibits a more intermittent distribution of the magnetic field compared to the self-consistent stage.  The intermittent nature of this phenomenon is supported by the presence of a larger tail, which conforms to a log-normal distribution, as depicted in Figures \ref{PDF B Non-Helical} \& \ref{PDF Bz Non-Helical}. 


It is crucial to note that as the value of $P_m$ increases, the coherence length scale ($\frac{l_B}{L}$) of the magnetic field in the saturated regime also increases (refer to Figure \ref{lb vs t Non-Helical}). This suggests that for higher $P_m$ cases, the structure should be more volume-filling or less intermittent, resulting in earlier truncation of PDF (PDF is truncated in earlier B values). This observation is precisely derived from the PDF computation, where the green curve is terminated at an earlier point  (See Fig. \ref{PDF Non-Helical Dynamo}).


\begin{figure*}
	\centering
	\begin{turn}{90} 
		\large{\textbf{\textcolor{blue}{\doublebox{Kinematic}}}}
	\end{turn}
	\begin{subfigure}{0.31\textwidth}
		\centering
		\includegraphics[scale=0.05580]{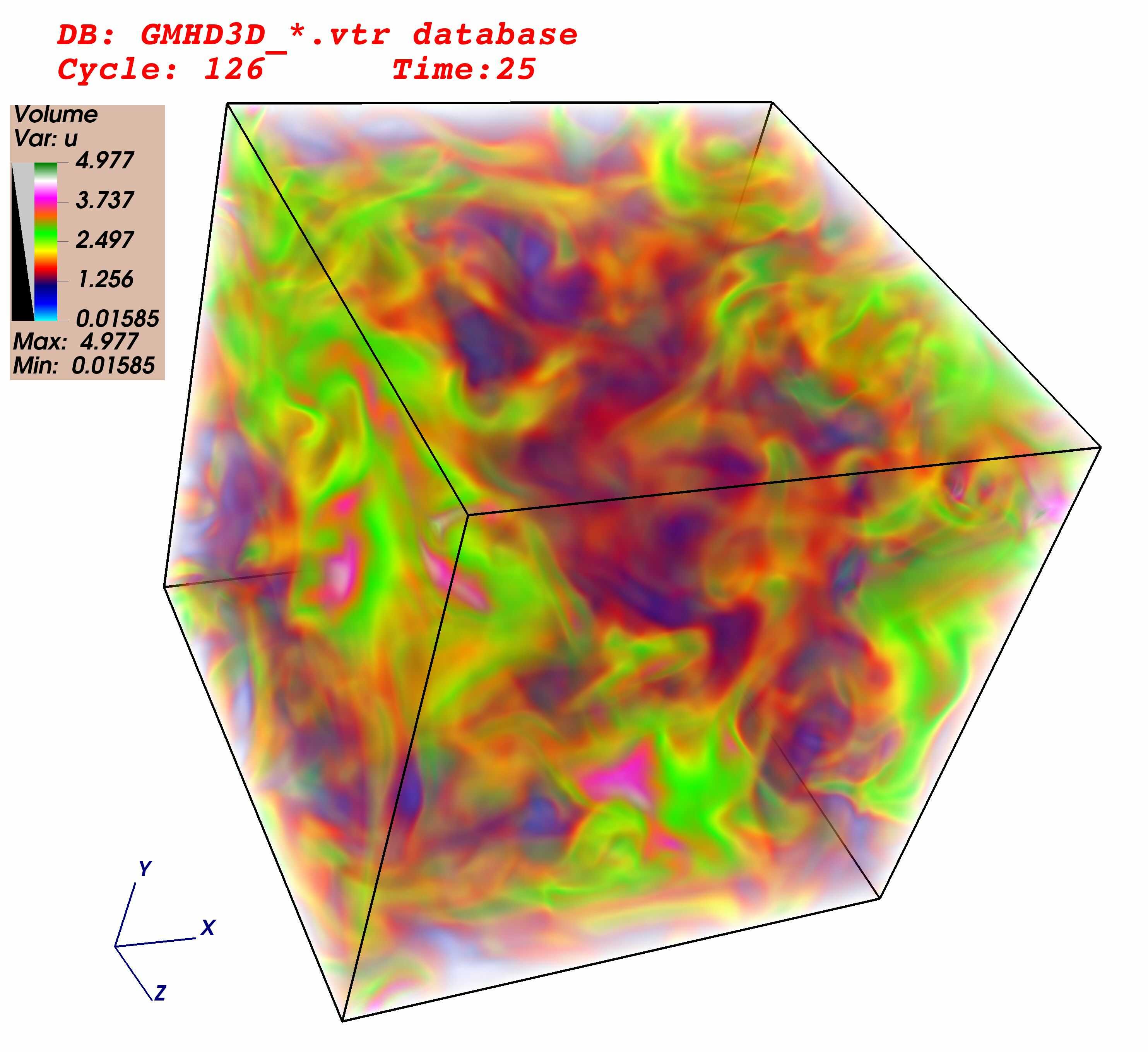}
		\caption{$P_m = 2.0$}
	\end{subfigure}
	\begin{subfigure}{0.31\textwidth}
		\centering
		\includegraphics[scale=0.05580]{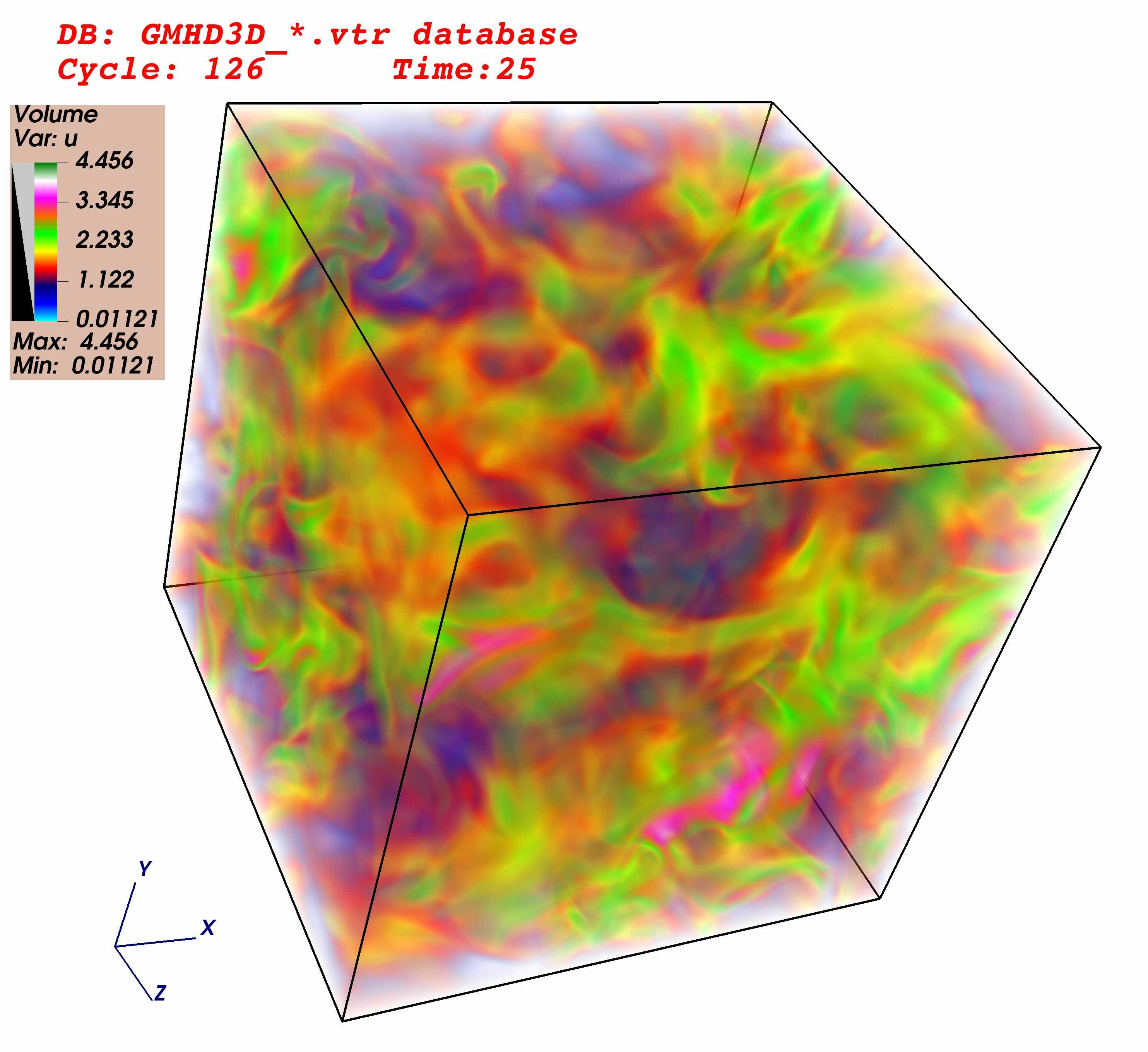}
		\caption{$P_m = 1.0$}
	\end{subfigure}
	\begin{subfigure}{0.31\textwidth}
		\centering
		\includegraphics[scale=0.05580]{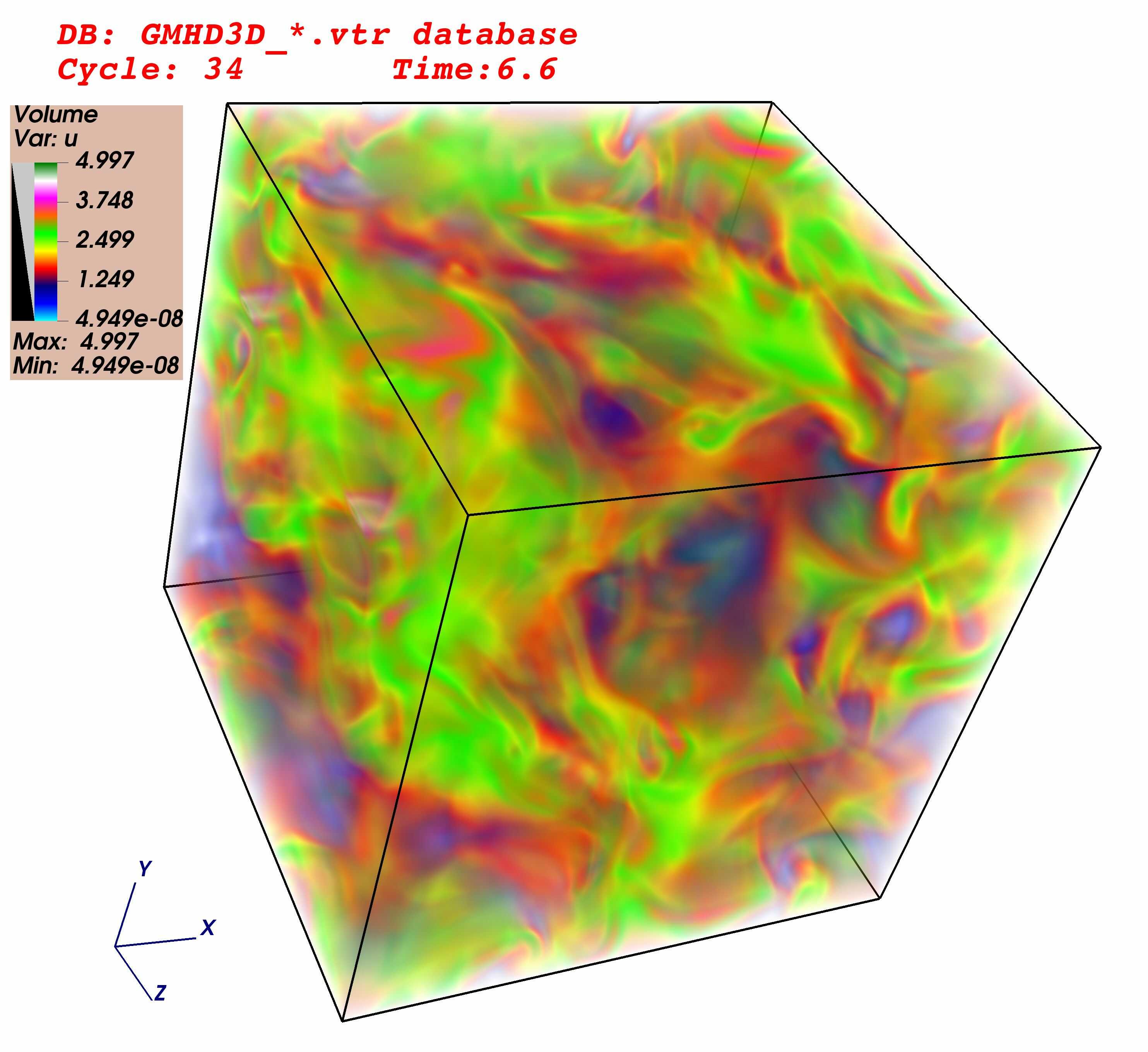}
		\caption{$P_m = 0.25$}
	\end{subfigure}
	\begin{turn}{90} 
		\large{\textbf{\textcolor{blue}{\doublebox{Saturated}}}}
	\end{turn}
	\begin{subfigure}{0.31\textwidth}
		\centering
		\includegraphics[scale=0.05580]{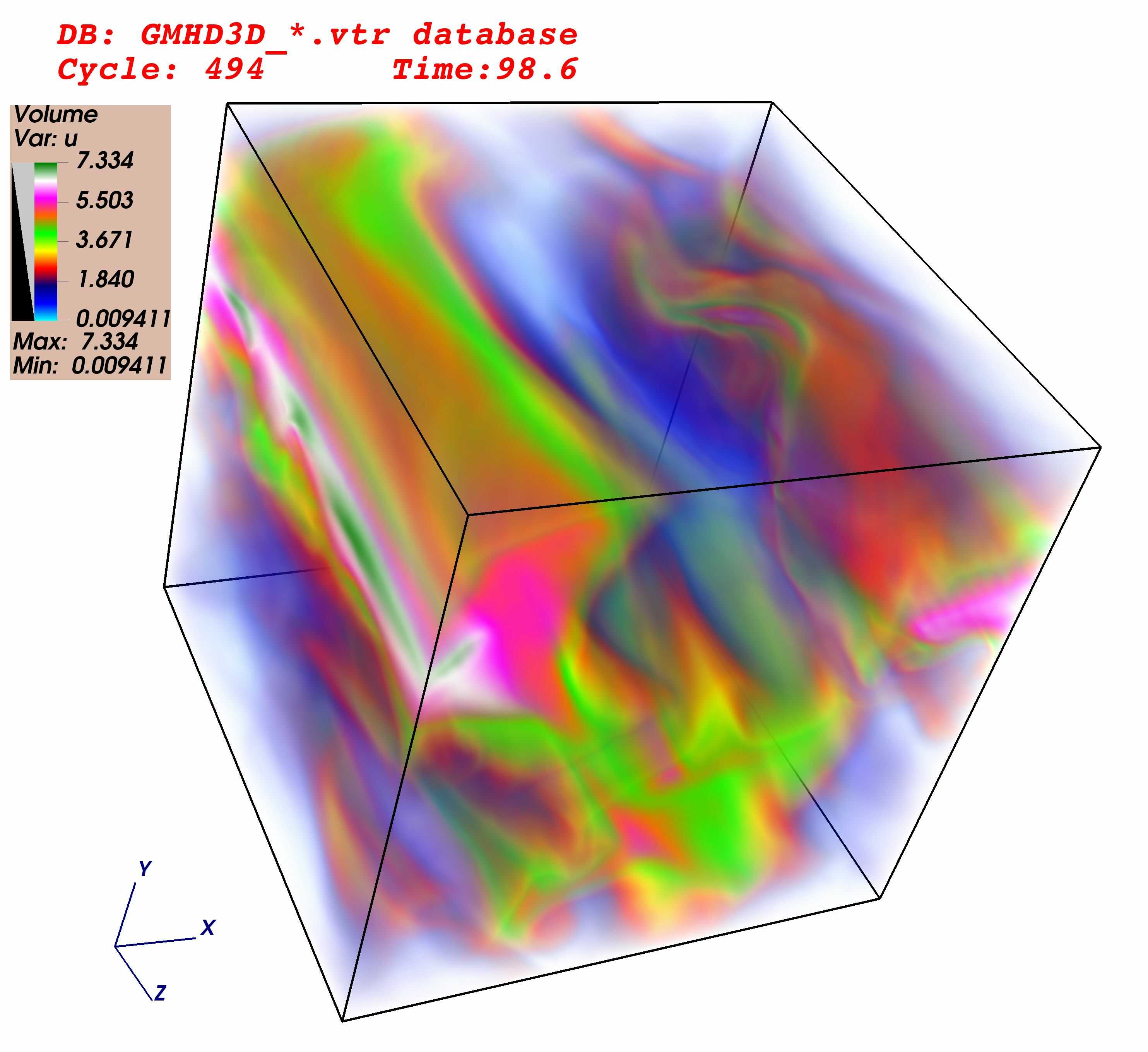}
		\caption{$P_m = 2.0$}
	\end{subfigure}
	\begin{subfigure}{0.31\textwidth}
		\centering
		\includegraphics[scale=0.05580]{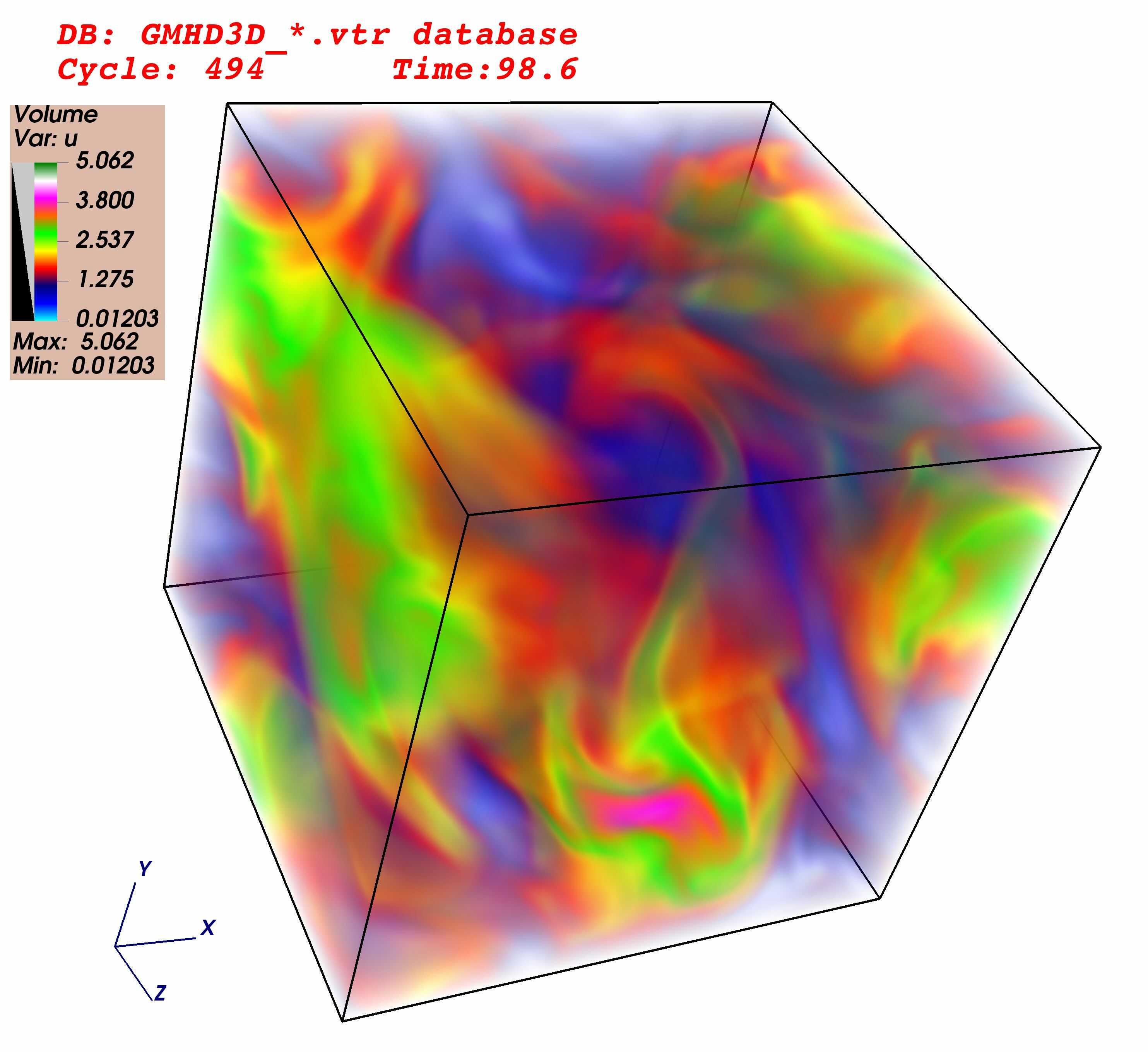}
		\caption{$P_m = 1.0$}
	\end{subfigure}
	\begin{subfigure}{0.31\textwidth}
		\centering
		\includegraphics[scale=0.05580]{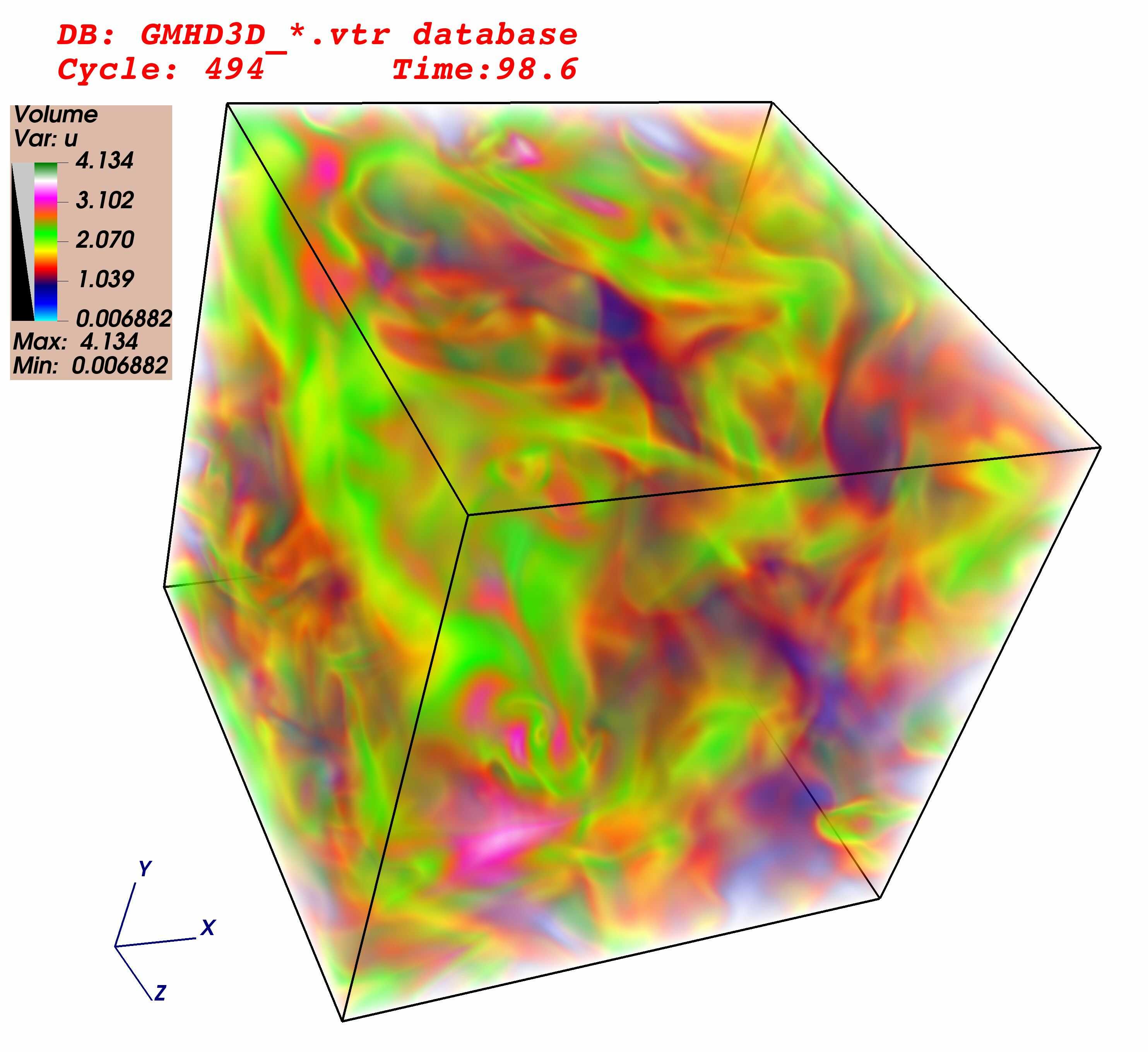}
		\caption{$P_m = 0.25$}
	\end{subfigure}
		\caption{ Volume rendering representation of the velocity field for a non-helical dynamo in the kinematic stage (upper row) and saturated stage (bottom row) with magnetic Prandtl number values of (a \& d) $P_m = 2$, (b \& e) $P_m = 1$, and (c \& f) $P_m = 0.25$. Velocity field structures increase in size as the magnetic Prandtl number ($P_m$) increases during the saturated stage. Therefore, the velocity field dynamics precisely mimic the magnetic field dynamics. Visualization is conducted using a linear scale.} 
		\label{no helical drive volume V}
\end{figure*}

Based on our previous discussion, it has been determined that in the saturated regime of non-helical dynamos, larger magnetic field structures are evident at higher $P_m$ values (refer to Figures \ref{non helical drive IsoB} \& \ref{non helical drive volume}). In order to provide clarity, we have employed a volume metric representation scheme to represent the velocity field in linear scale (refer to Figure \ref{no helical drive volume V}). It is evident from Fig. \ref{no helical drive volume V} (upper row) that the velocity field structures exhibit a high degree of similarity in kinematic phase across various magnetic $P_m$ ranges. This is due to the fact that the kinetic Reynolds numbers ($R_e$) are constant across all cases, and the magnetic fields in the kinematic phase are not strong enough to alter the velocity fields, resulting in the formation of identical structures (refer to the upper row of Figure \ref{no helical drive volume V}). The strength of magnetic fields in the saturated regime is sufficient to alter the velocity field dynamics. As a result, distinct velocity field structures are observed at various $P_m$ limits (refer to the lower row of Figure \ref{no helical drive volume V}), despite the fact that $R_e$ remains constant in all cases. It is crucial to mention that as the value of $P_m$ increases, the structures of the velocity field increase. This phenomenon precisely reflects the characteristics of the magnetic field (See the lower row of Figure \ref{no helical drive volume V}).


\begin{figure*}
	\begin{subfigure}{0.49\textwidth}
		\centering
		\includegraphics[scale=0.5]{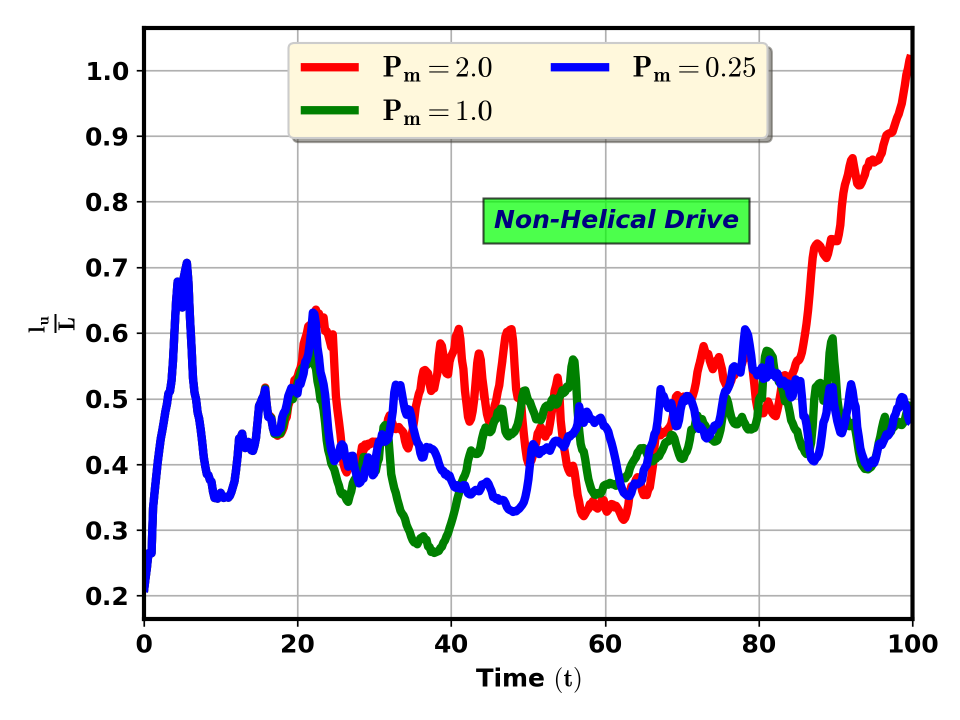}
		\caption{}
		\label{lv Non-Helical}
	\end{subfigure}
	\begin{subfigure}{0.49\textwidth}
		\centering
		\includegraphics[scale=0.5]{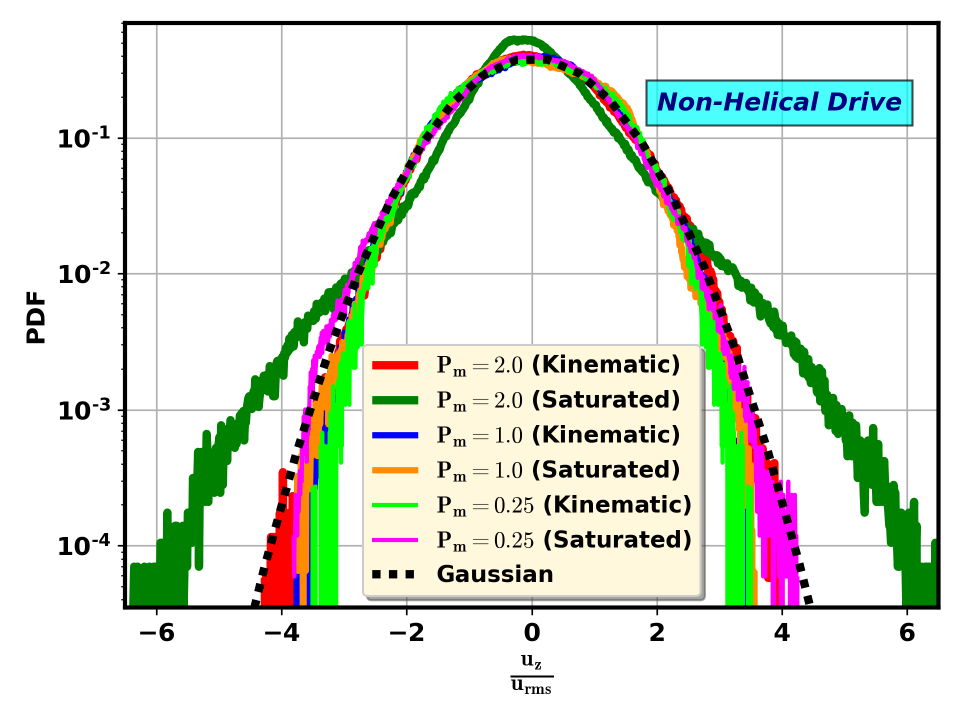}
		\caption{}
		\label{PDF Vz Non-Helical}
	\end{subfigure}
	\caption{ (a) The coherence length of the velocity field, normalized to the size of the numerical domain ($\frac{l_u}{L} = \frac{\int k^{-1} E(k) dk}{\int E(k) dk}$), has been computed for non-helically driven dynamo for various values of $P_m$. (b) The probability density function (PDFs) for the normalized $z$-component of the velocity field $\frac{u_z}{u_{rms}}$ is examined in both the kinematic and saturation stages of a non-helically driven dynamo. Here $u_{rms}$ is defined as $\sqrt{\langle u_x^2 + u_y^2 + u_z^2 \rangle}$.}
	\label{PDF and lv Non-Helical Dynamo}
\end{figure*}

After doing a more in-depth analysis of the velocity field data, we have computed the temporal evolution of the coherence length scale ($\frac{l_u}{L}$) of the velocity field and the probability density function (PDF) of one component of the velocity field. Figure \ref{lv Non-Helical} clearly demonstrates that the velocity structures have a high level of symmetry during the kinematic stage, but they are noticeably different during the saturated stage. The PDF calculation reveals that the velocity field PDF closely follows a Gaussian distribution (refer to Figure \ref{PDF Vz Non-Helical}), indicating the presence of homogeneous turbulence. Significantly, in the case of greater $P_m$ ($P_m = 2$), the intermittency is more pronounced in saturated stages, resulting in a non-Gaussian distribution (refer to Figure \ref{PDF Vz Non-Helical}). This is a noteworthy and entirely dissimilar observation compared to the previous one. The precise cause for this is not fully understood yet and necessitates additional inquiry.


Based on our previous discussion, it has been determined that while the velocity probability density functions (PDFs) exhibit a close resemblance to a Gaussian distribution, the PDFs of the magnetic field do not. In order to measure this departure from a Gaussian distribution, we have calculated the Kurtosis and Skewness parameters. The specific information can be found in Table \ref{Kurtosis Non-Helical} and Table \ref{Skewness Non Helical}.


\begin{table*}
	\centering
	\textcolor{black}{
		\begin{tabular}{ |c|c|c|c|c|c|c| }
			\hline
			\textbf{$P_m$} & \textbf{$\mathbf{K (B_z)}$ [Kin]} & \textbf{$\mathbf{K (B_z)}$ [Sat]} & \textbf{$\mathbf{K (B)}$ [Kin]} & \textbf{$\mathbf{K (B)}$ [Sat]} & \textbf{$\mathbf{K (u_z)}$ [Kin]} & \textbf{$\mathbf{K (u_z)}$ [Sat]}\\
			\hline
			2 & 5.579 &	2.557 &	8.621 & -0.207 & -0.051 &   2.761\\
			\hline
			1 & 10.581 &	2.398 &	14.209 &	1.185 & -0.243 & -0.323\\
			\hline
			0.25 & 5.395 &	9.581 &	6.030 &	13.259 & -0.370 &  0.106 \\
			\hline
		\end{tabular}
	}
	\caption{Kurtosis $K$ for the distribution of magnetic and velocity fields in a non-helically driven scenario. A high Kurtosis value for the magnetic field during the kinematic stage indicates the presence of intermittency, as seen in Figure \ref{PDF Non-Helical Dynamo}. On the other hand, Kurtosis for the velocity field is almost $0$ (except from the $P_m = 2$ scenario), which is an excellent indication of Gaussianity (See Fig. \ref{PDF Vz Non-Helical}). The Kurtosis for $P_m = 2$ has a significant value.}
	\label{Kurtosis Non-Helical}
\end{table*}

\begin{table*}
	\centering
	\textcolor{black}{
		\begin{tabular}{ |c|c|c|c|c| }
			\hline
			\textbf{$P_m$} &  \textbf{$\mathbf{S (B)}$ [Kin]} & \textbf{$\mathbf{S (B)}$ [Sat]} & \textbf{$\mathbf{S (u_z)}$ [Kin]} & \textbf{$\mathbf{S (u_z)}$ [Sat]}\\
			\hline
			2 & 2.183 &	0.666 & 0.103 &	0.310\\
			\hline
			1 & 2.485 &	1.095 & -0.009 & -0.015 \\
			\hline
			0.25 & 1.928 &	2.529 & $-3.378\times 10 ^{-10}$ & -0.094\\
			\hline
		\end{tabular}
	}
	\caption{Skewness $S$ for the velocity and magnetic field distribution in the absence of helical drive. A high Skewness value observed for the magnetic field during the kinematic stage indicates the presence of intermittency, as seen in Figure \ref{PDF Non-Helical Dynamo}. However, the Skewness of the velocity field is close to $0$ (less than $1$), indicating a strong indication of Gaussian distribution (See Figure \ref{PDF Vz Non-Helical}). A positive Skewness indicates that the distribution is skewed towards the right, whereas a negative Skewness indicates that the distribution is biased towards the left.}
	
	\label{Skewness Non Helical}
\end{table*} 

\begin{figure*}
	\centering
	\begin{subfigure}{0.32\textwidth}
		\centering
		\includegraphics[scale=0.36]{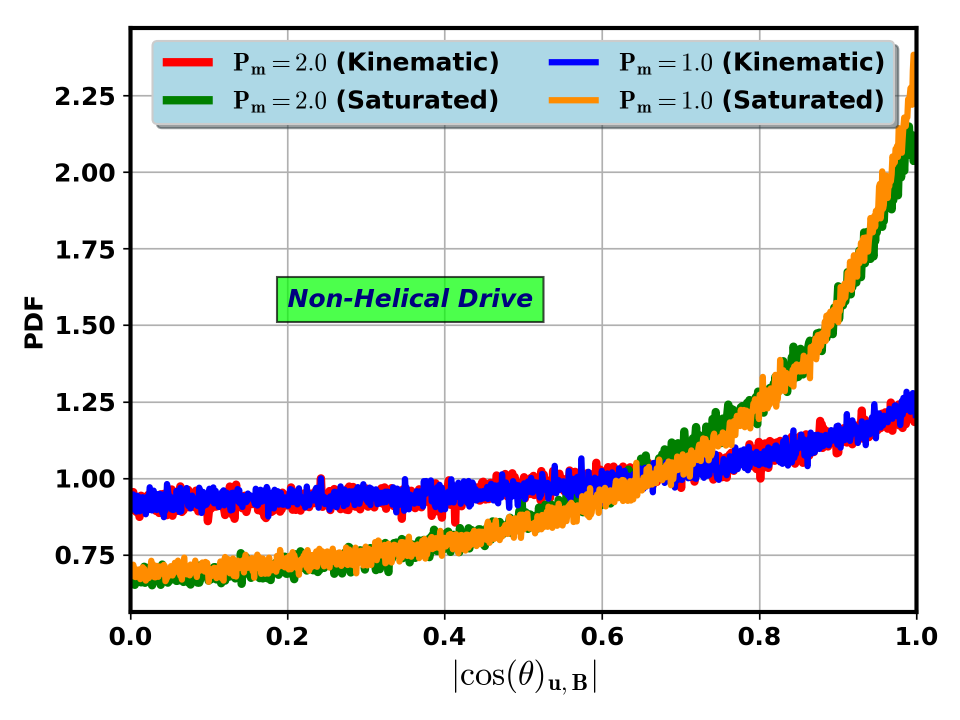}
		\caption{}
		\label{Non-Helical Cos uB}
	\end{subfigure}
	\begin{subfigure}{0.32\textwidth}
		\centering
		\includegraphics[scale=0.36]{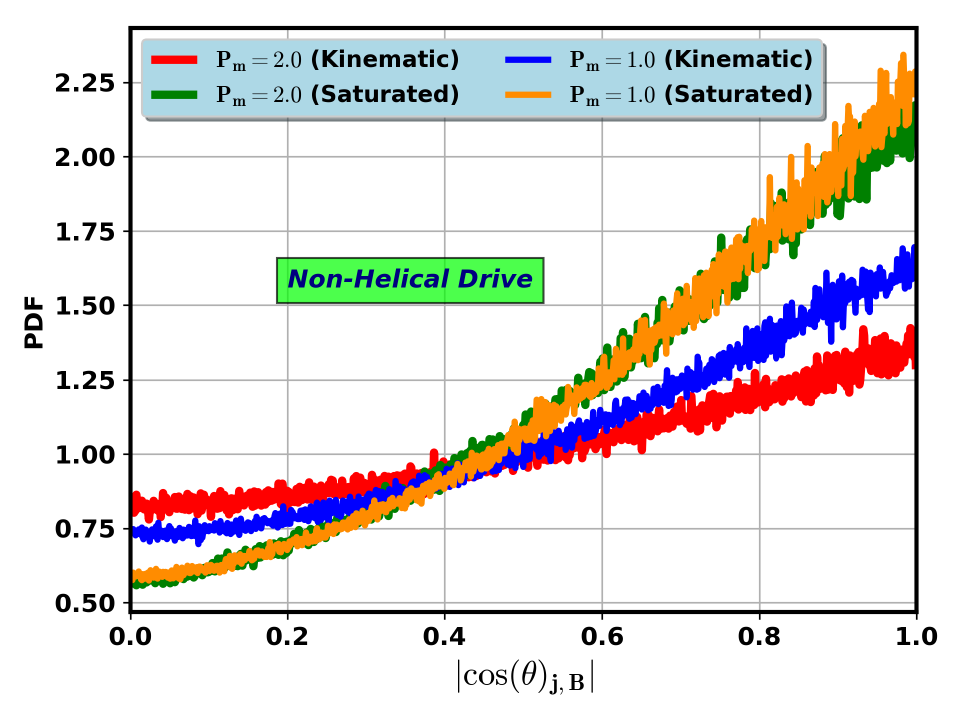}
		\caption{}
		\label{Non-Helical Cos jB}
	\end{subfigure}
	\begin{subfigure}{0.32\textwidth}
		\centering
		\includegraphics[scale=0.36]{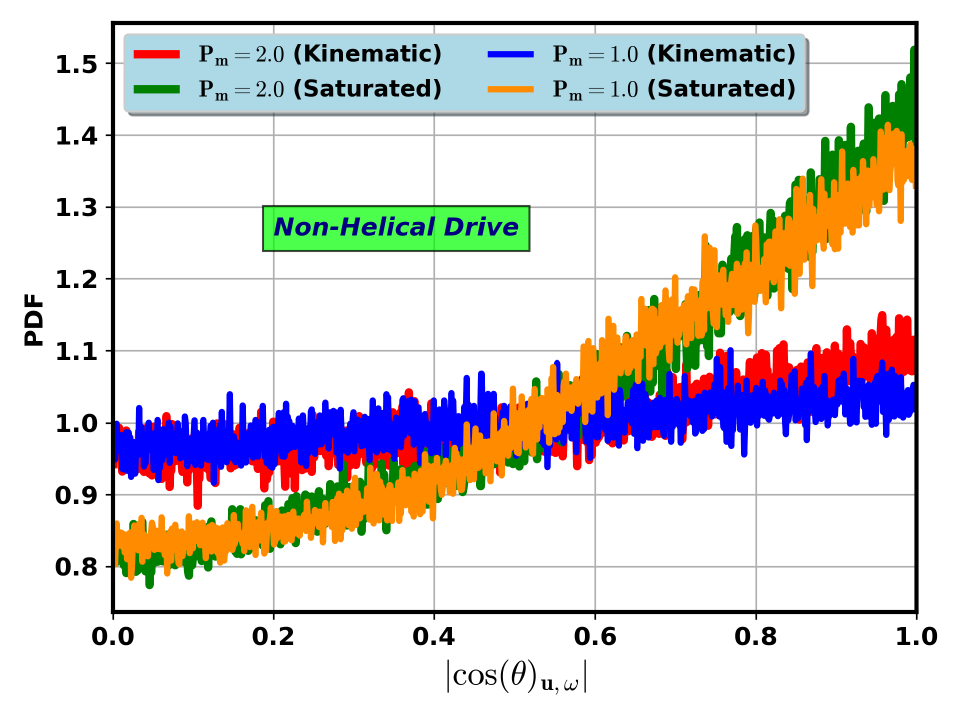}
		\caption{}
		\label{Non-Helical Cos omegaB}
	\end{subfigure}
	\caption{ PDFs are provided for the absolute value of the cosine of the angle between (a) velocity and magnetic field ($|\cos(\theta)_{u,B}|$), (b) current density and magnetic field ($|\cos(\theta)_{j,B}|$), and (c) velocity and vorticity ($|\cos(\theta)_{u,\omega}|$) for various values of $P_m$ in the presence of a non-helical drive.}
	
\end{figure*}

We have computed the probability density function (PDF) for different alignment angles, specifically the PDF for $|\cos(\theta)_{u,B}|$, $|\cos(\theta)_{j,B}|$, and $|\cos(\theta)_{u,\omega}|$, as I did before. Our numerical study reveals that in the saturated stage, the alignment between the magnetic field and the velocity field is greater, resulting in a reduction of the induction effects (i.e., $\vec{\nabla} \times (\vec{u} \times \vec{B}) \to 0$), compared to the kinematic stage (refer to Fig. \ref{Non-Helical Cos uB}). We have observed that the current density ($\vec{j}$) and the magnetic field ($\vec{B}$) are aligned to a high degree in the saturated stage. This alignment results in a minimum Lorentz force, as shown in Figure \ref{Non-Helical Cos jB}.  In other words, when the alignment between vector $\vec{u}$ and vector $\vec{B}$ is improved, it results in a reduction in the transfer of energy from the flow to the magnetic field. On the other hand, when vector $\vec{j}$ and vector $\vec{B}$ are aligned, it leads to a drop in the magnetic feedback. All of these factors ultimately lead to the dynamo becoming saturated. In the saturated stage, it is noticed that the $\vec{u}$ and $\omega$ fields are parallel to each other (refer to Figure \ref{Non-Helical Cos omegaB}).


\begin{figure*}
	\centering
	\begin{subfigure}{0.32\textwidth}
		\centering
		\includegraphics[scale=0.36]{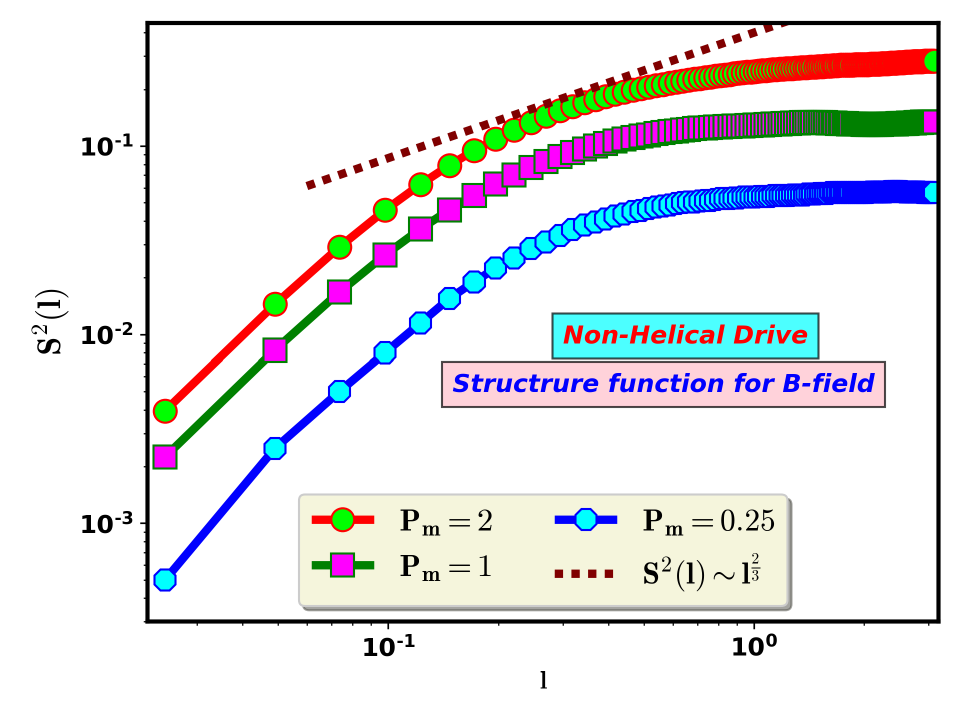}
		\caption{}
	\end{subfigure}
	\begin{subfigure}{0.32\textwidth}
		\centering
		\includegraphics[scale=0.36]{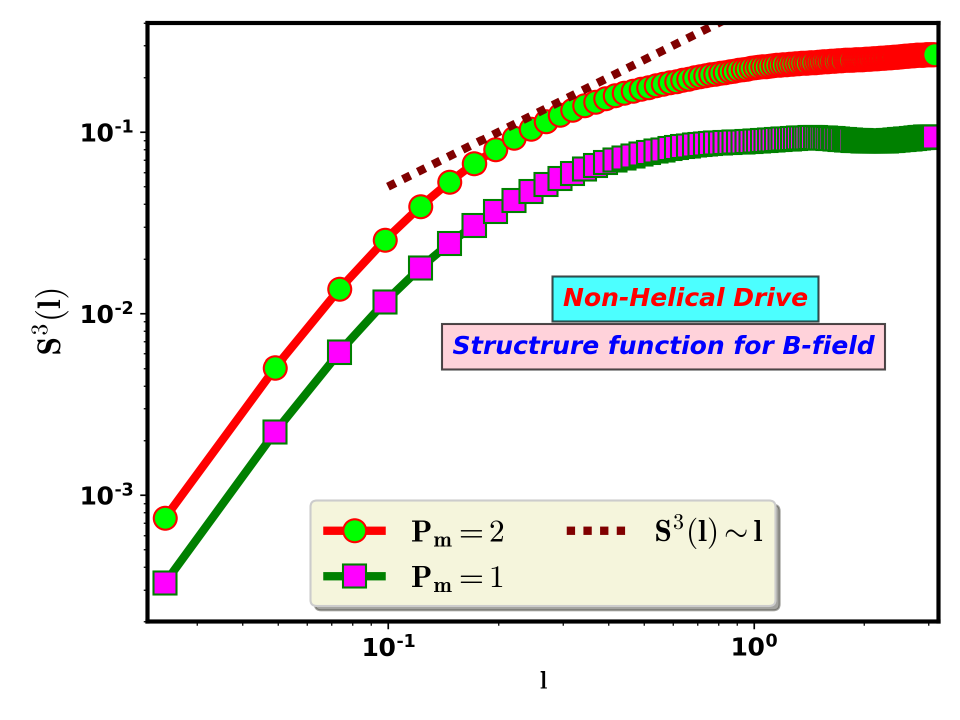}
		\caption{}
	\end{subfigure}
	\begin{subfigure}{0.32\textwidth}
		\centering
		\includegraphics[scale=0.36]{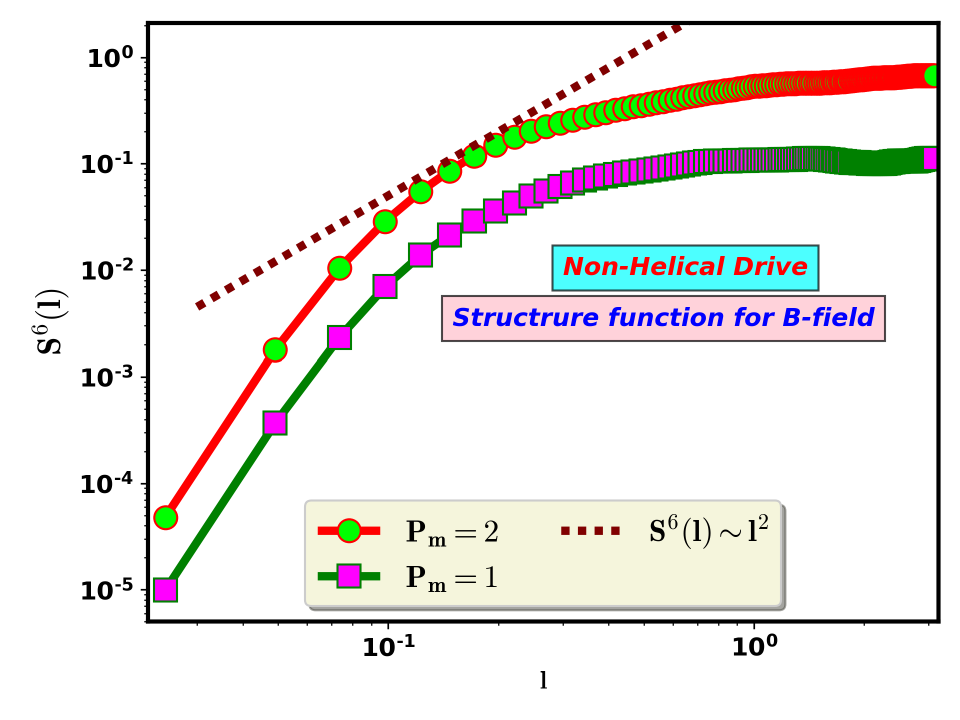}
		\caption{}
	\end{subfigure}
	\begin{subfigure}{0.32\textwidth}
		\centering
		\includegraphics[scale=0.36]{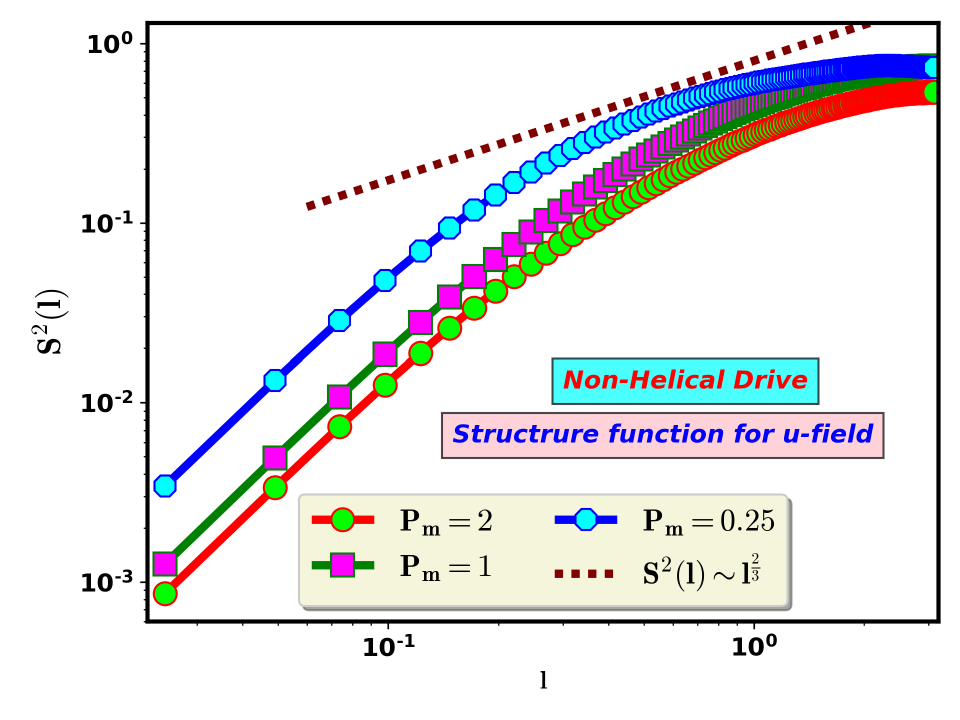}
		\caption{}
	\end{subfigure}
	\begin{subfigure}{0.32\textwidth}
		\centering
		\includegraphics[scale=0.36]{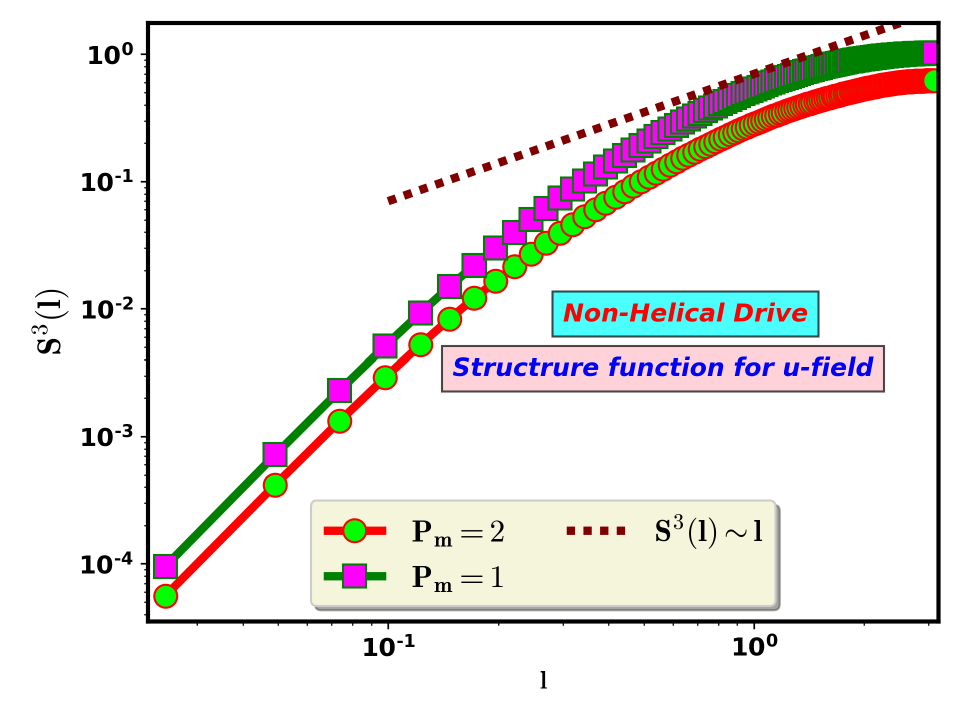}
		\caption{}
	\end{subfigure}
	\begin{subfigure}{0.32\textwidth}
		\centering
		\includegraphics[scale=0.36]{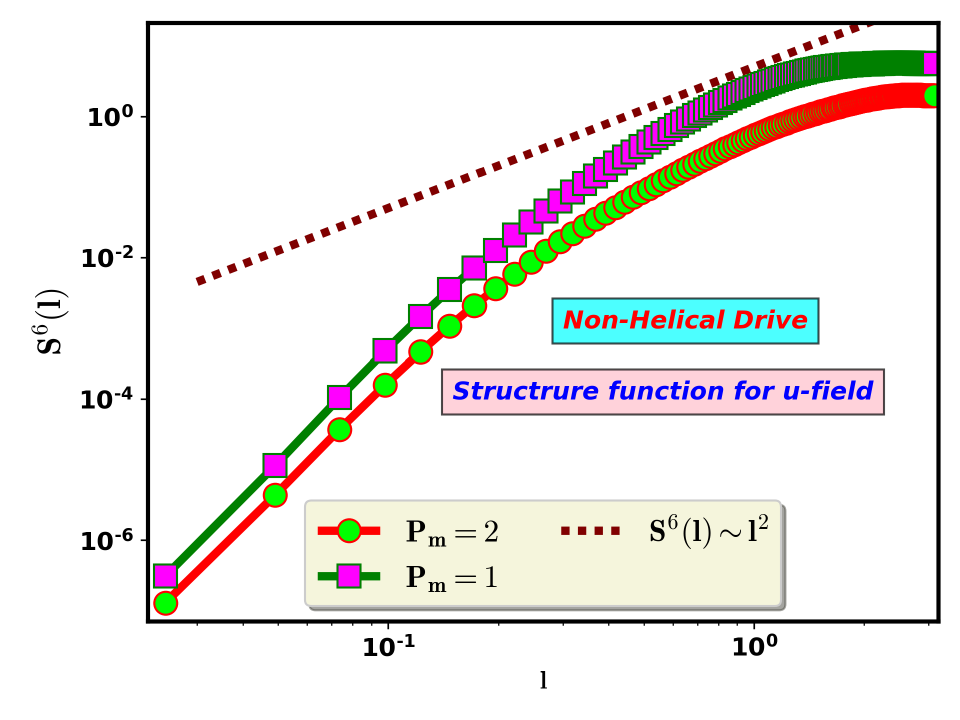}
		\caption{}
	\end{subfigure}
	\caption{ The plots display the structure functions of three distinct orders, namely $p = 2, 3$, and $6$, as a function of $l$ for velocity and magnetic field at various $P_m$ values in the context of a non-helical dynamo. The numerical analysis confirms that Kolmogrov's scalings \cite{kolmogorov:1941}, specifically $S^2 (l) \sim l^\frac{2}{3}$, $S^3 (l) \sim l$, and $S^6 (l) \sim l^2$, hold true for both the velocity and magnetic field.}
	
	\label{Structure Function Non-Helical}
\end{figure*}

Similar to the helical example, we have computed the second ($S^2 (l)$), third ($S^3 (l)$), and sixth ($S^6 (l)$) order structure function of the magnetic field (refer to Fig. \ref{Structure Function Non-Helical} [upper row]) and velocity (refer to Fig. \ref{Structure Function Non-Helical} [bottom row]) for the non-helical dynamo as well (refer to Fig. \ref{Structure Function Non-Helical}).  Our numerical analysis reveals that the predictions made by \cite{kolmogorov:1941} are evident in both lower and higher order computations (i.e, $p = 2, 3$, and $6$), similar to a helical dynamo. Interestingly, $S^2 (l)$, $S^3 (l)$, and $S^6 (l)$ are found to be independent of the magnetic Prandtl number ($P_m$).


\begin{figure*}
	\begin{subfigure}{0.49\textwidth}
		\centering
		\includegraphics[scale=0.49]{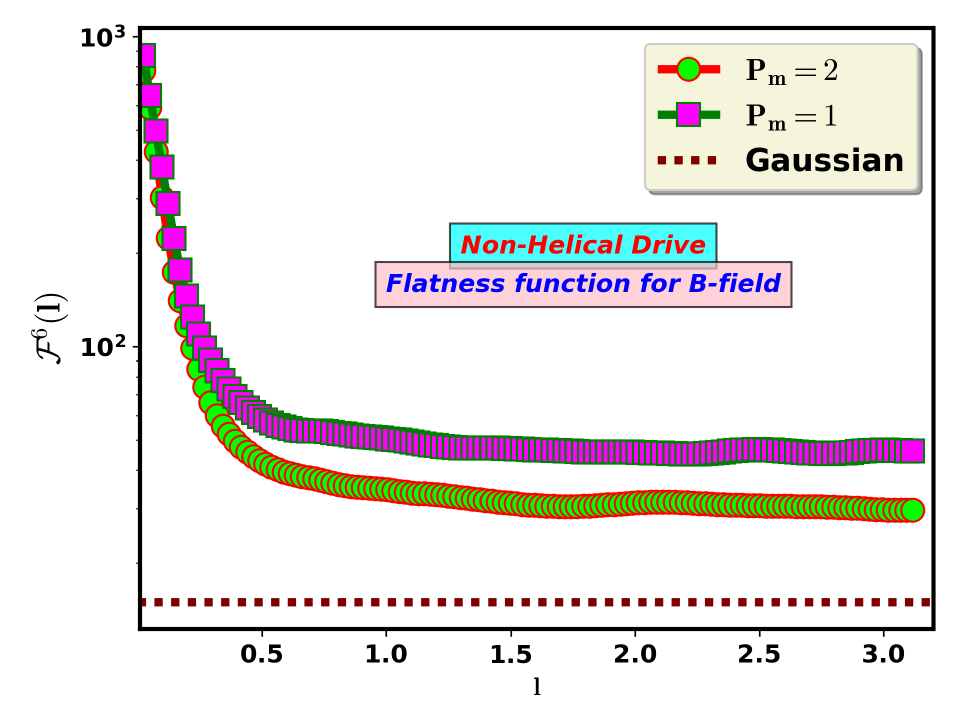}
		\caption{}
	\end{subfigure}
	\begin{subfigure}{0.49\textwidth}
		\centering
		\includegraphics[scale=0.49]{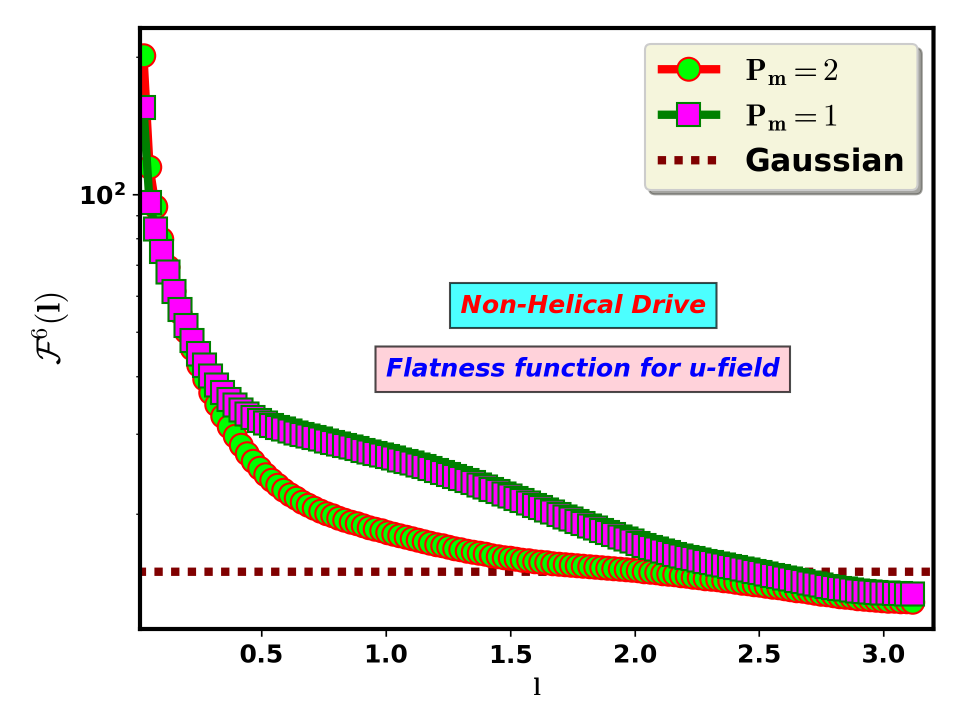}
		\caption{}
	\end{subfigure}
	\caption{ The hyper-flatness $\mathcal{F}^6(l)$ is calculated as a function of $l$ for both (a) magnetic field and (b) velocity field  at various values of $P_m$ for a non-helical dynamo.}
	\label{Flatness Non-Helical}
\end{figure*}

For a non-helical dynamo, we illustrate the hyper-flatness $\mathcal{F}^6(l)$ versus $l$ for both the magnetic field and velocity field (refer to Fig. \ref{Flatness Non-Helical}). This plot demonstrates that the hyper-flatness function, denoted as $\mathcal{F}^6(l)$, increases as the length scale $l$ approaches zero. This suggests the presence of intermittent behavior at small scales within the dissipation range.


\begin{figure*}
	\begin{subfigure}{0.32\textwidth}
		\centering
		\includegraphics[scale=0.39]{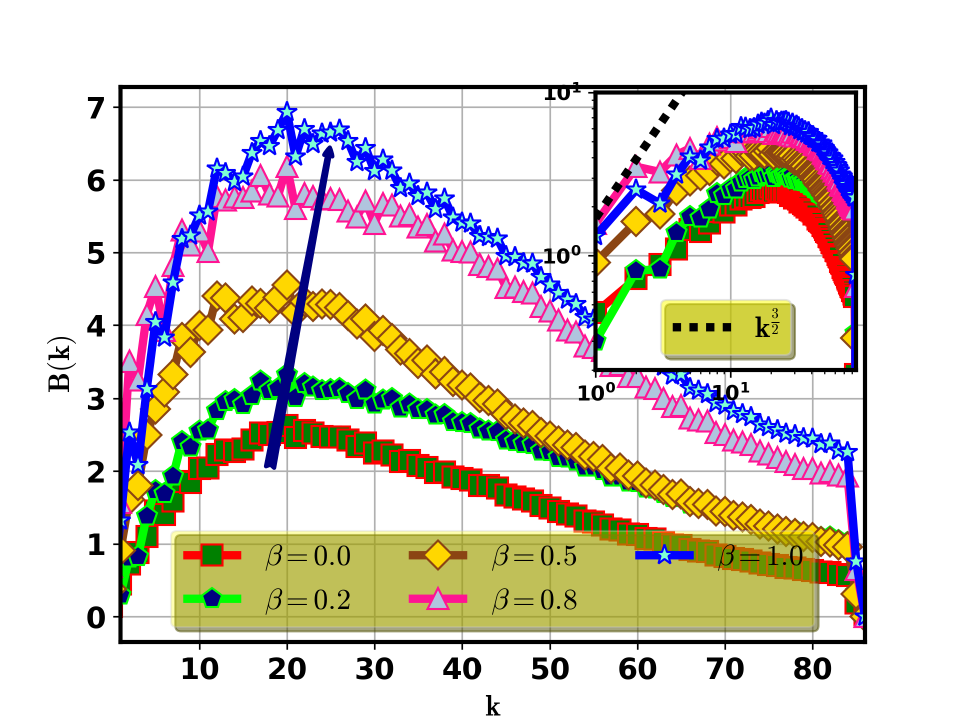}
		\caption{}
	\end{subfigure}
	\begin{subfigure}{0.32\textwidth}
		\centering
		\includegraphics[scale=0.39]{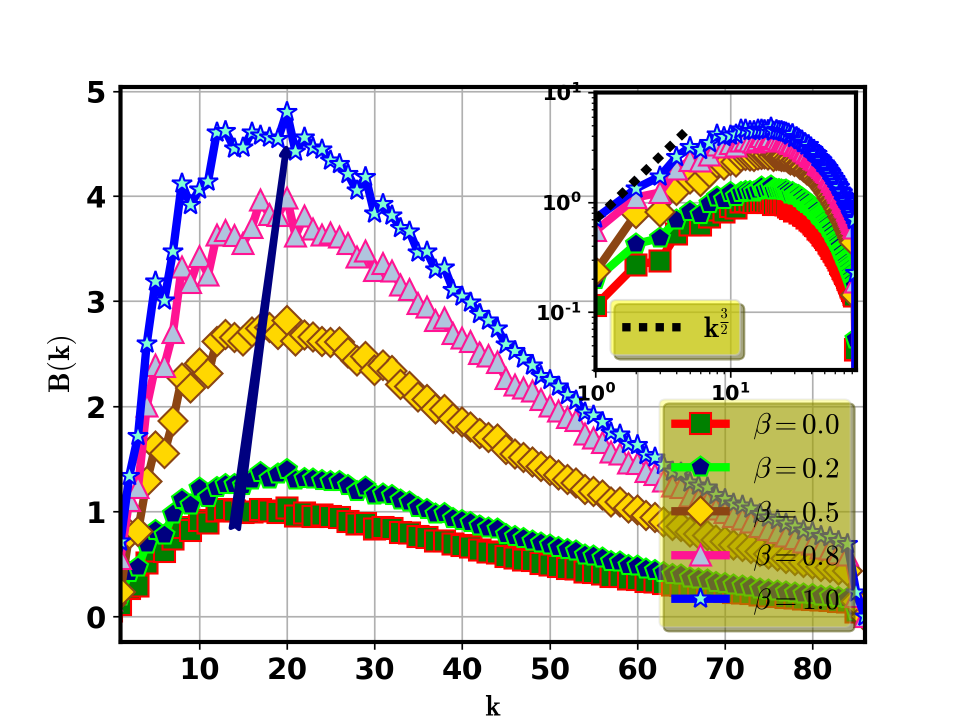}
		\caption{}
	\end{subfigure}
	\begin{subfigure}{0.32\textwidth}
		\centering
		\includegraphics[scale=0.39]{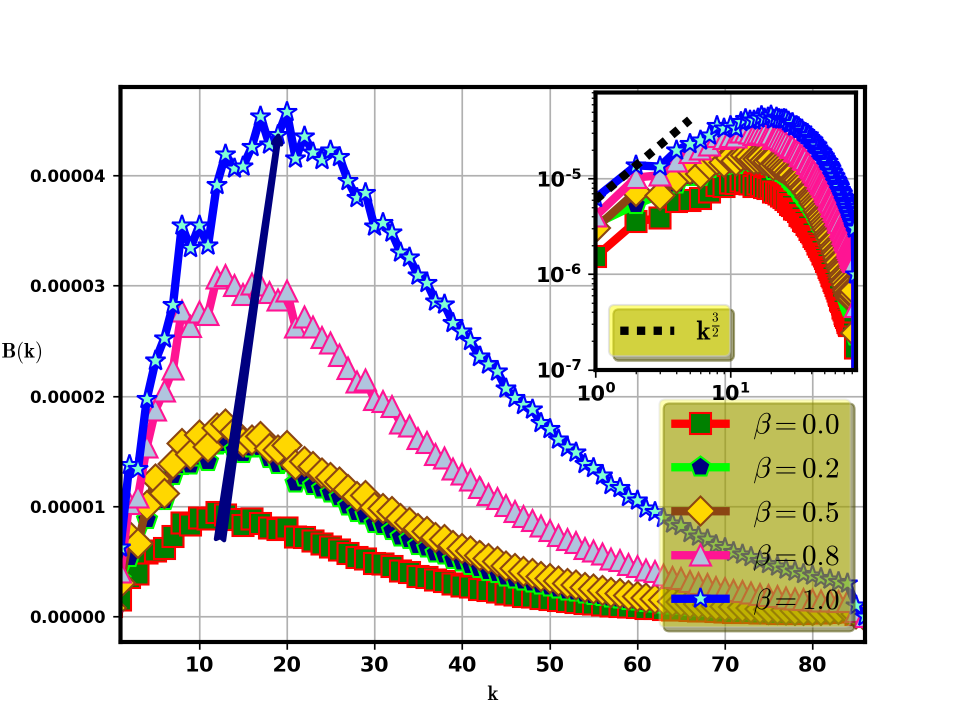}
		\caption{}
	\end{subfigure}
	\caption{ Magnetic energy spectral density estimation for various classes of Yoshida-Morrison (YM) flow drives with $P_m$ values of (a) $2.0$, (b) $1.0$, and (c) $0.25$, such that $\int |B(k, t)|^2 dk$ represents the total energy at time t and $k = \sqrt{k_x^2 + k_y^2 + k_z^2}$. It is evident that the dynamos are small scale dynamos (SSD) suitable for all kinds of drives. Undoubtedly, the SSD spectral structure is ultimately impacted by the helical configuration of the drive. Plots are represented on both a linear and log-log scale (inset view).
} 
		\label{controled helicity spectra}
\end{figure*}

Additionally, it is mentioned earlier that the Yoshida-Morrison (YM) flow family is utilized as the driving force in our simulation. An essential feature of this YM flow is that finite fluid helicity injection into the system is possible by regulating a specific physical parameter ($\beta$) prior to simulation initiation. The influence of driven controlled fluid helicity injection on self-consistent dynamo activity is now being examined. We compute the magnetic field spectral density ($B(k)$) for various $\beta$ values and plot it against $k$ over a variety of $P_m$ values. It is evident from Fig. \ref{controled helicity spectra} that the dynamos observed for each form of drive are, in essence, small-scale dynamos (SSD). Additionally, the magnetic field is observed to adhere to the widely recognized Kazantsev $k^\frac{3}{2}$ scaling \cite{Kazantsev:1968} . Furthermore, it remains unaffected by the helicity strength of the drive (refer to Figure \ref{controled helicity spectra}).
The helical structure of the flow seems to impact the SSD spectral structure throughout a variety of $P_m$ values, as seen in Fig. \ref{controled helicity spectra} (Also see Appendix \ref{Appen C}). This observation is found to coincide with a recent study on kinematic dynamos \cite{Biswas_Scripta:2023}.


\begin{figure*}
	\begin{subfigure}{0.32\textwidth}
		\centering
		\includegraphics[scale=0.39]{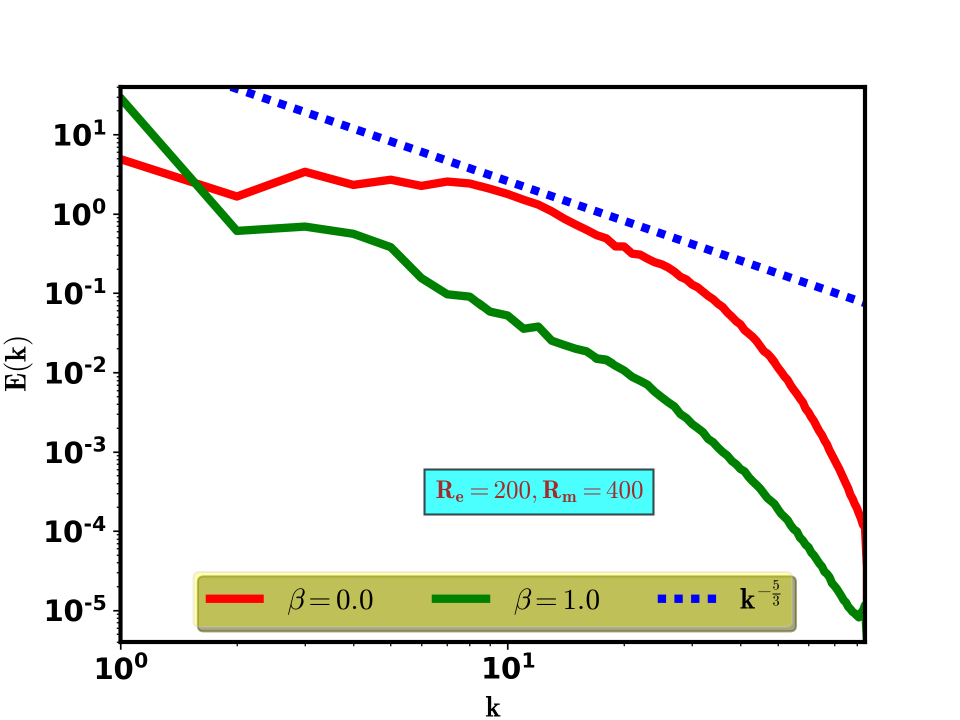}
		\caption{}
	\end{subfigure}
	\begin{subfigure}{0.32\textwidth}
		\centering
		\includegraphics[scale=0.39]{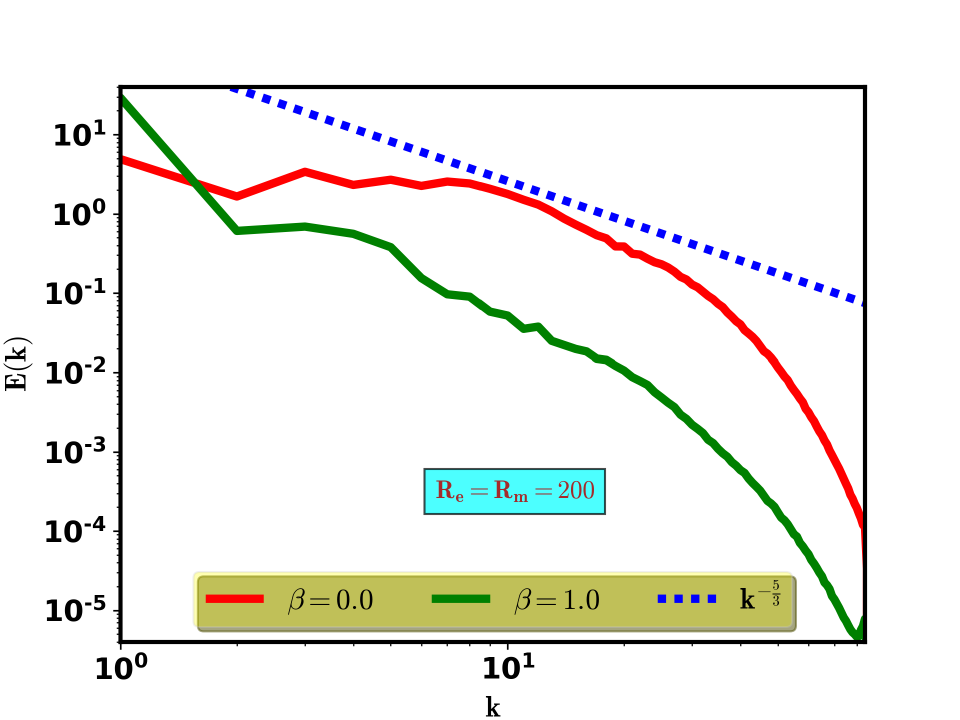}
		\caption{}
	\end{subfigure}
	\begin{subfigure}{0.32\textwidth}
		\centering
		\includegraphics[scale=0.39]{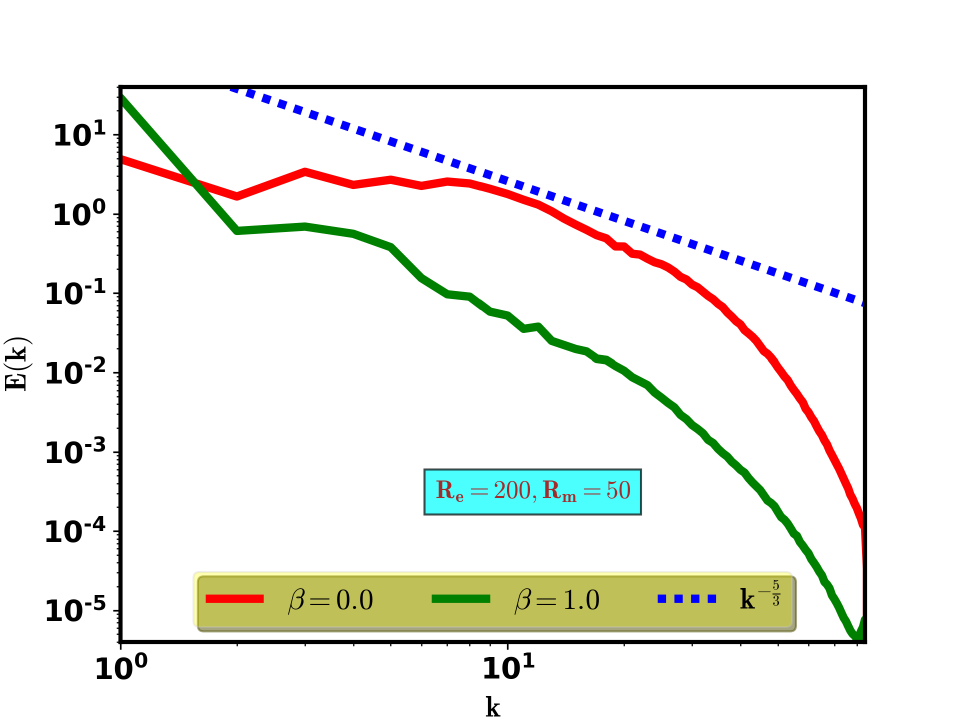}
		\caption{}
	\end{subfigure}
	\caption{ Calculation of the kinetic energy spectral density ($E(k)$) for helical ($\beta = 1.0$) and non-helical ($\beta = 0.0$) dynamos, where $k = \sqrt{k_x^2 + k_y^2 + k_z^2}$ and $\int |E(k, t)|^2 dk$ represents the total energy at time t. We have computed it for $P_m$ values (a) $2.0$, (b) $1.0$, and (c) $0.25$. For all the examples addressed here, we have found that the kinetic energy spectrum is close to the Kolmogorov spectrum, that is, $E(k) \propto k^{-\frac{5}{3}}$.} 
		\label{controled helicity injection KE}
\end{figure*}

Additionally, we have computed the spectral density of kinetic energy denoted as $E(k)$, where $\int |E(k, t)|^2 dk$ represents the total energy at time t and $k = \sqrt{k_x^2 + k_y^2 + k_z^2}$. Figure \ref{controled helicity injection KE} illustrates the relationship between $E(k)$ and $k$ for various $\beta$ values and $P_m$ values of $0.25$, $1.0$, and $2.0$. The kinetic energy spectrum closely resembles the Kolmogorov spectrum, with $E(k) \propto k^{-\frac{5}{3}}$ scaling (See Fig. \ref{controled helicity injection KE}). The observed scaling also remains consistent across a broad range of $P_m$ values and is considered to be independent of the helical structure of the flow (See Fig. \ref{controled helicity injection KE}).

\begin{figure*}
	\begin{subfigure}{0.49\textwidth}
		\centering
		\includegraphics[scale=0.55]{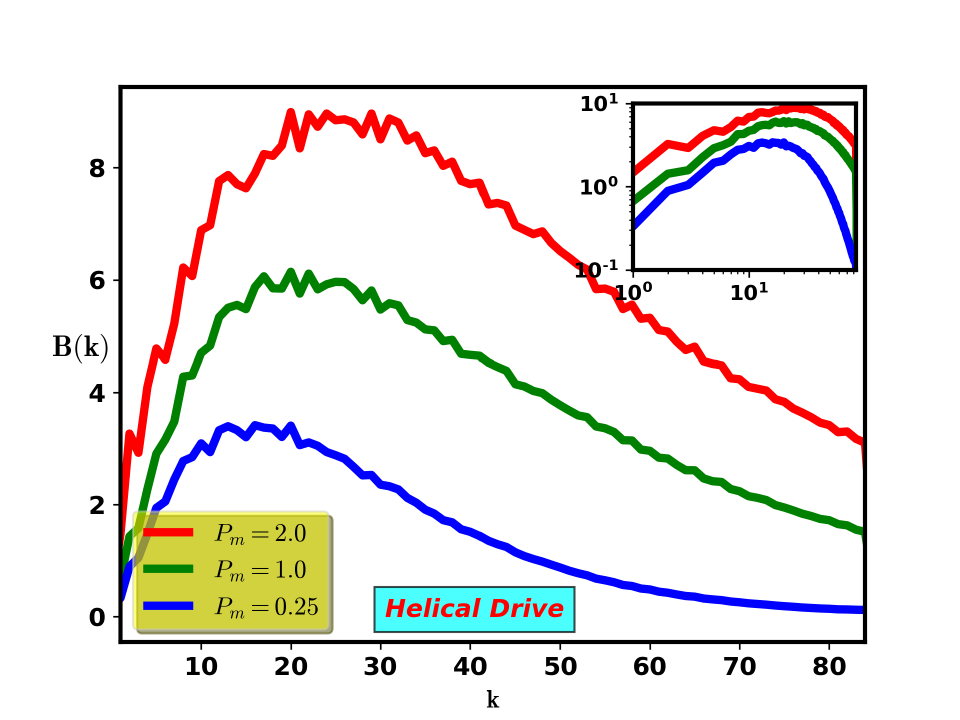}
		\caption{}
	\end{subfigure}
	\begin{subfigure}{0.49\textwidth}
		\centering
		\includegraphics[scale=0.55]{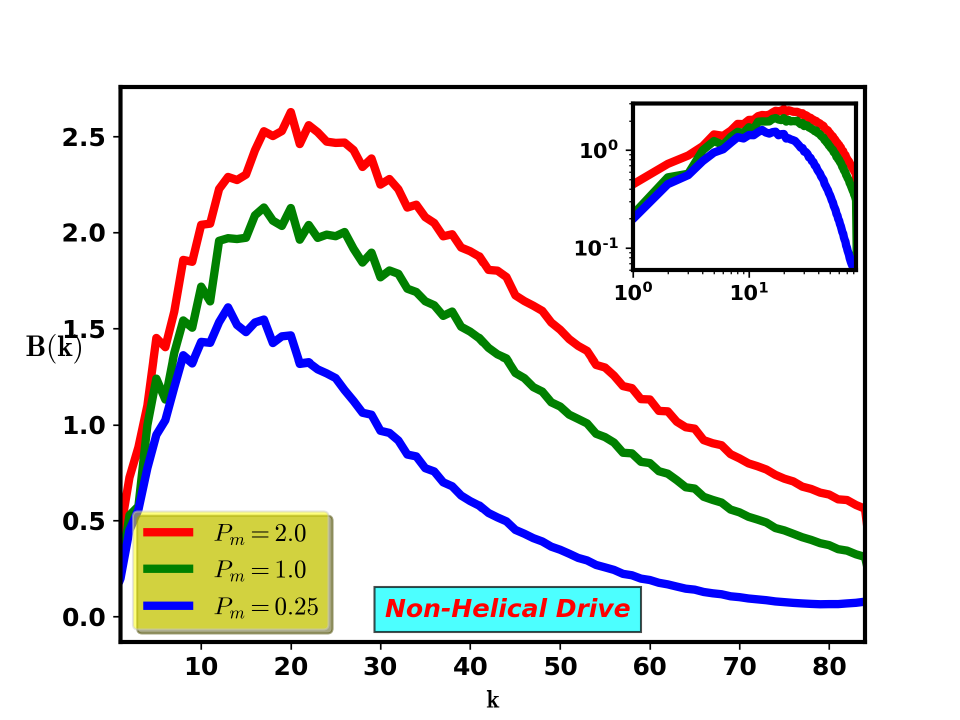}
		\caption{}
	\end{subfigure}
	\caption{ Estimation of the magnetic energy spectral density $B(k)$ at $P_m$ values of $2.0$, $1.0$, and $0.25$ for (a) helical-drive and (b) non-helical-drive, where $\int |B(k, t)|^2 dk$ is the total energy at time t and $k = \sqrt{k_x^2 + k_y^2 + k_z^2}$. It is evident that as the value of $P_m$ is reduced, the peak of the spectral energy density ($B(k)$) shifts towards a lower mode ($k$). Plots are represented on both a linear and log-log scale (inset view).} 
		\label{Spectra at different Pm for helical and nonhelical drive}
\end{figure*}

Figure \ref{Spectra at different Pm for helical and nonhelical drive} provides a comparison of the magnetic energy spectral density ($B(k)$) for various values of magnetic Prandtl numbers ($P_m$). It is evident that as the value of $P_m$ decreases, the peak of $B(k)$ shifts towards decreasing wave numbers. The rationale for this is that a decrease in $P_m$ while holding the kinetic Reynolds number ($R_e$) constant signifies an increase in the resistivity of the plasma, which causes the resistive scale to kicks of earlier in modes.


\section{Summary and Conclusion}
Direct numerical simulations (DNS) of self-consistent dynamos have been performed in this study utilizing a three-dimensional magnetohydrodynamic model. We have explored several aspects of self-consistent dynamos based on the type of driving force present in the system. The newly proposed Yoshida-Morrison (YM) flow drive has been taken into consideration as a potential driver in our simulation. One distinguishing feature of this drive is that its helicity can be altered by changing the magnitude of a physically significant parameter. When driven in a helical manner, we classify the dynamo as a helical dynamo, and when driven non-helically, we refer to it as non-helical dynamos.\\


Our major findings are: \\

$\bullet$ The differences between helical and non-helical dynamos has been investigated. Our numerical simulation has shown that helical dynamos exhibit a significantly greater growth rate of magnetic energy compared to non-helical dynamos. The strong ``back-reaction'' of the magnetic field on the velocity field resulting from this high growth rate causes non-linearity to manifest much earlier in the helical case and causes the exponential growth rate to decelerate earlier in comparison to the non-helical case. As a result, the helical turbulent dynamos has been observed to reach saturation state prior to the non-helical turbulent dynamos.


$\bullet$ It is observed that the initial seed magnetic field has no influence on the time evolution of volume-averaged magnetic energy. It is observed that the saturation value and the growth rate of magnetic energy in the kinematic stage are not influenced by the seed field. This observation holds true in the case of helical and non-helical dynamos alike. It appears that all memory of the original seed field has been lost.


$\bullet$ It is observed that the Alfven speed has a substantial impact on both helical and non-helical dynamos. It has been noted that the prominence of the dynamo activity strengthens in the super Alfvenic regime. This observation is consistent with helical and non-helical dynamos.


$\bullet$ For both helical and non-helical driven dynamos, the impact of the magnetic Prandtl number ($P_m$) has been studied. Our numerical experiments have revealed that dynamo action is suppressed at lower $P_m$ limits for both helical and non-helical dynamos similarly.


$\bullet$ This investigation also encompasses the dynamics of the magnetic field iso-surface (IsoB) at various $P_m$ limits. The findings indicate that during the kinematic stage, small scale structures predominate, whereas during the saturated stage, relatively large scale structures are observed. It appears that this finding holds true for both helical and non-helical drives.


$\bullet$ It has been determined through magnetic field iso-surface (IsoB) visualization that the magnetic field structure for helical dynamos becomes more pronounced at lower $P_m$ values. The emergence of such enormous structures is a consequence of the system's increasing magnetic resistivity, which subsequently eliminates fluctuations.


$\bullet$ It is observed that the variation of magnetic field structures for non-helical dynamos with respect to $P_m$ follows the same pattern that has been observed for helical dynamos, but only during the kinematic stage. During the saturation stage, the observation is diametrically opposite; as $P_m$ increases, the magnetic field structures are observed to reveal large-scale structures. The velocity field has also been visualized, and it has been observed that as $P_m$ increases, the structures of the velocity field become more pronounced; this is consistent with the dynamics of the magnetic field. This observation is noteworthy in comparison to previous works.


$\bullet$ For both helical and non-helical dynamos, the spectral distribution of the magnetic field has also been presented. It has been observed that the majority of power is concentrated in higher mode numbers for each scenario examined in this study; therefore, the dynamos are small-scale dynamos (SSD). The magnetic field exhibits the established Kazantsev scaling ($B(k) \propto k^\frac{3}{2}$), which remains consistent for both helical and non-helical dynamos. It is also observed that the variation of $P_m$ has no effect on the Kazantsev scaling. This is another significant finding of the investigation.


$\bullet$ Controlled helicity injection has been observed to impact the spectral structure of small scale dynamo (SSD) in a variety of $P_m$ values in our numerical experiments. Without magnetic back-reaction, the same observation has been reported. The presence of magnetic back-reaction does not impact the nature. Likewise, this is a noteworthy finding. 


$\bullet$ According to our numerical experiments, as $P_m$ decreases, the peak of the magnetic energy spectral density ($B(k)$ migrates toward smaller modes.


$\bullet$ The kinetic energy spectral density ($E(k)$) has been calculated for both helical and non-helical dynamos. Our spectral analysis shows that the kinetic energy spectrum closely adheres to Kolmogorov scaling ($E(k) \propto k^{-\frac{5}{3}}$). This Kolmogorov scaling justifies the turbulent nature of the velocity field. 

$\bullet$ We have also developed various diagnostic tools such as those utilized to determine the coherence length scale [$l_u(t)$ \& $l_B(t)$], examine the probability density function (PDF) of velocity and magnetic field, evaluate the PDF of alignment angle between different fields, calculate the Skewness and Kurtosis [$S(f)$ \& $K(f)$], determine the structure functions of different orders [$S^p(l)$], and evaluate the hyper-flatness [$\mathcal{F}^6 (l)$]. The PDF calculation reveals that the magnetic field structure displays higher levels of intermittency during the kinematic stage in comparison to the self-consistent stage. This increased intermittency is also indicated by the higher values of Kurtosis and Skewness. Through structure function calculation, it is observed that Kolmogorov's hypothesis has been satisfied for both helical and non-helical dynamos


In conclusion, we have studied the impact of helical and non-helical drive on the development of fully self-consistent dynamos through numerical simulation. The findings of our numerical analysis indicate that small scale dynamos (SSD) are generated by both helical and non-helical drives. Kazantsev's $k^\frac{3}{2}$ scaling is observed to persist for both helical and non-helical dynamos, and this scaling remains consistent regardless of the magnetic Prandtl number ($P_m$). In the nonlinear limit, our investigation also demonstrates that the helical nature of the drive does influence the dynamics of small-scale dynamos which are significant in various astrophysical objects, particularly in the interstellar and intergalactic medium, stars, planets, and liquid metal experiments.


We have carried out the analysis for $P_m<1$ regime, which is significant for stars, planets, and liquid metal experiments, as well as $P_m>1$ regime, which is relevant to fluctuation dynamo in the interstellar and intergalactic medium. The study can be further extended through various means. An immediate extension would involve conducting the same analysis for dynamos utilizing a Hall MHD model \cite{Gomez:2004}. In the nonlinear limit, the effect of flow shear on dynamo activity has also not been investigated. Therefore, an investigation of the impact of flow shear \cite{Shishir_POF:2022} on the activity of the self-consistent dynamo would be of considerable interest. We have implemented the MHD approximation, although plasma effects could also be significant. Comparing our results with those of the plasma dynamo would be intriguing to observe how the relationship between velocity and magnetic fields, as well as the magnetic field structure, change when plasma effects are taken into account. Plasma effects are significant for gases with weak collisions and partially ionized mediums such as the interstellar medium (ISM). The MHD model is inadequate for describing these plasmas due to its inability to differentiate the relative motion among various species. Therefore, the incorporation of two fluid effects \cite{Tilley_Balsara_MNRAS:2011, Xu_Balsara_APJ:2019} is necessary. We intend to address these issues in our upcoming work.


\section{SUPPLEMENTARY MATERIAL}
The movie descriptions can be found in the supplementary material.

\section{ACKNOWLEDGMENTS}
The simulations and visualizations presented here are performed on GPU nodes and visualization nodes of Antya cluster at the Institute for Plasma Research (IPR), INDIA. One of the authors, S.B, acknowledges Prof. A. Alexakis of ENS in Paris, France, for a valuable discussion during his visit to the ENS lab in Paris. S.B is thankful to HPC support team of IPR for extending their help related to ANTYA cluster.


\section{\textbf{DATA AVAILABILITY}}
The data underlying this article will be shared on reasonable request to the corresponding author.\\

\section{\textbf{Conflict of Interest}}
The authors have no conflicts to disclose.


\bibliography{references}

\onecolumngrid
\appendix
\section{\textbf{Initial condition invariance test for non-helical case}}\label{Appen A}
\begin{figure*}[h!]
		\centering
		\includegraphics[scale=0.55]{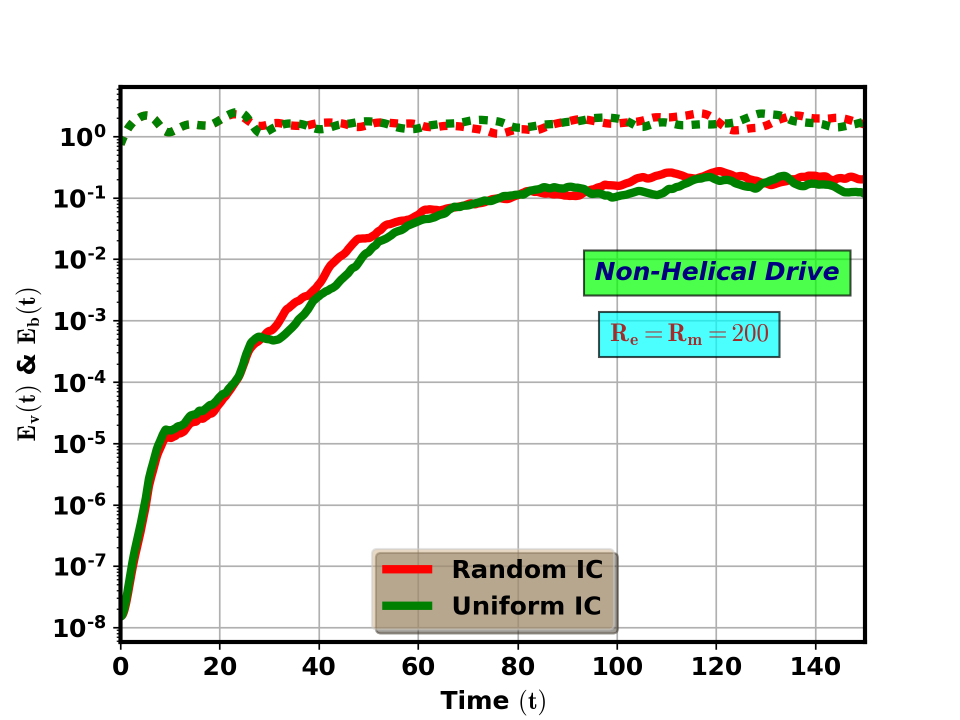}
	\caption{Time evolution of kinetic and magnetic energy for two distinct initial conditions: random seed (red) \& uniform (green) in a non-helical dynamo. The evolution of magnetic and kinetic energy is observed to be identical in both cases.}
	\label{Initial condition independency nonhelical}
\end{figure*}

We have verified the independency of the initial condition for the non-helically diven case. It is noted that the growth rate and saturated value of the magnetic field are similar for both the uniform and random seed fields in the non-helical driven case, as shown in Figure \ref{Initial condition independency nonhelical}. Therefore, it can be inferred from Figure \ref{Initial condition independency nonhelical} that the memory of the initial seed field is lost, similar to the helical case discussed previously.

\section{\textbf{Effect of $P_m$ on growth rate \& saturation state of magnetic energy}}\label{Appen B}
\begin{figure*}[h!]
	\begin{subfigure}{0.49\textwidth}
	\centering
	\includegraphics[scale=0.55]{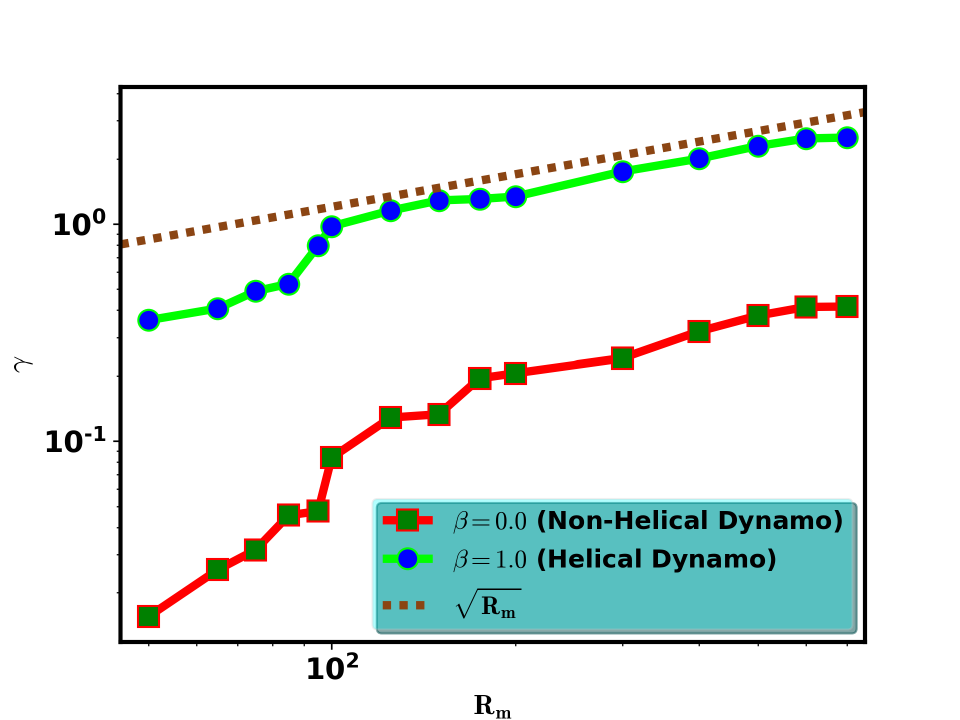}
		\caption{}
	\end{subfigure}
	\begin{subfigure}{0.49\textwidth}
		\centering
		\includegraphics[scale=0.55]{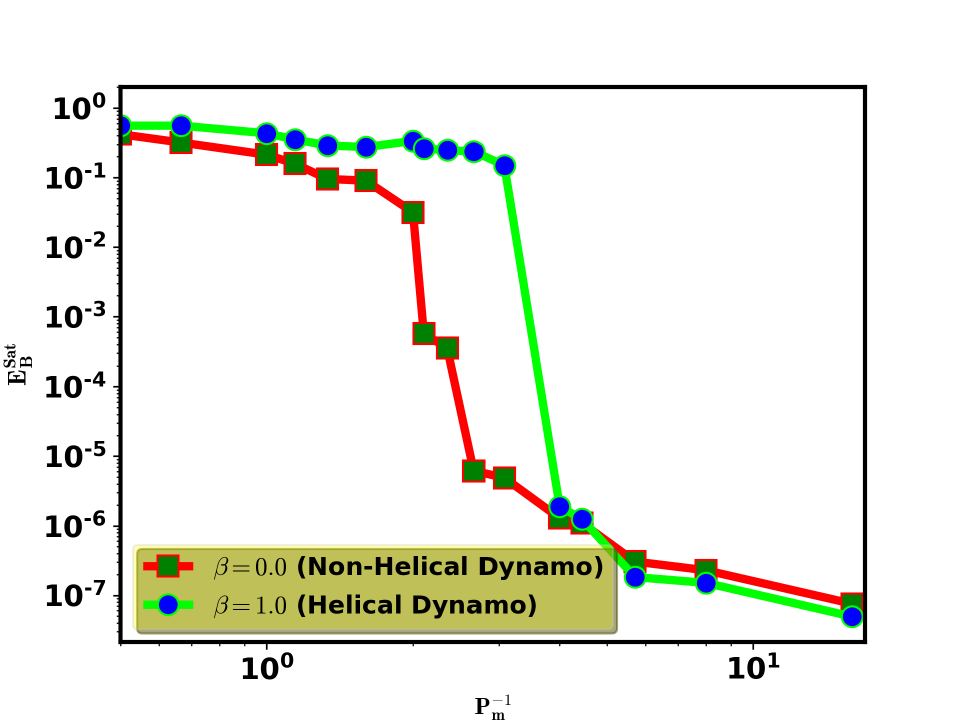}
		\caption{}
	\end{subfigure}
	\caption{ (a) The growth rate of magnetic energy ($\gamma = \frac{d}{dt}(\ln E_B(t))$) has been computed for various values of $R_m$. In the case of both helical and non-helical drives, the growth rate ($\gamma$) is determined to be proportional to the square root of $R_m$ (i.e, $\gamma \propto \sqrt{R_m}$).  (b) The relationship between the saturation value of the magnetic energy ($E_B^{Sat}$) and $P_m^{-1}$ for helical and non-helical dynamos. It is evident from both figures that dynamo action is more difficult to achieve at lower values of the magnetic Prandtl number ($P_m$). We have tweaked the variable $R_m$ by adjusting $P_m$ while maintaining $R_e$ constant. Thus, we can also view these two graphs as illustrating the magnetic Prandtl number's influence on $\gamma$ and $E_B^{Sat}$.}
		\label{growth rate and saturation}
\end{figure*}

The magnetic energy growth rate ($\gamma = \frac{d}{dt}(\ln E_B(t))$) and saturation value of magnetic energy ($E_B^{Sat}$) for both helical and non-helical dynamos have been computed and plotted as functions of $P_m$ [See Fig. \ref{growth rate and saturation}]. It is evident from Fig. \ref{growth rate and saturation} that dynamos are more difficult to excite at low $P_m$ limits.  We have calculated the values of $\gamma$ and $E_B^{Sat}$ based on the data presented in figures \ref{helical Rm} \& \ref{nonhelical Rm}.

\section{\textbf{Effect of helicity injection on magnetic field dynamics}}\label{Appen C}
\begin{figure*}[h!]
	\begin{subfigure}{0.32\textwidth}
		\centering
		\includegraphics[scale=0.087]{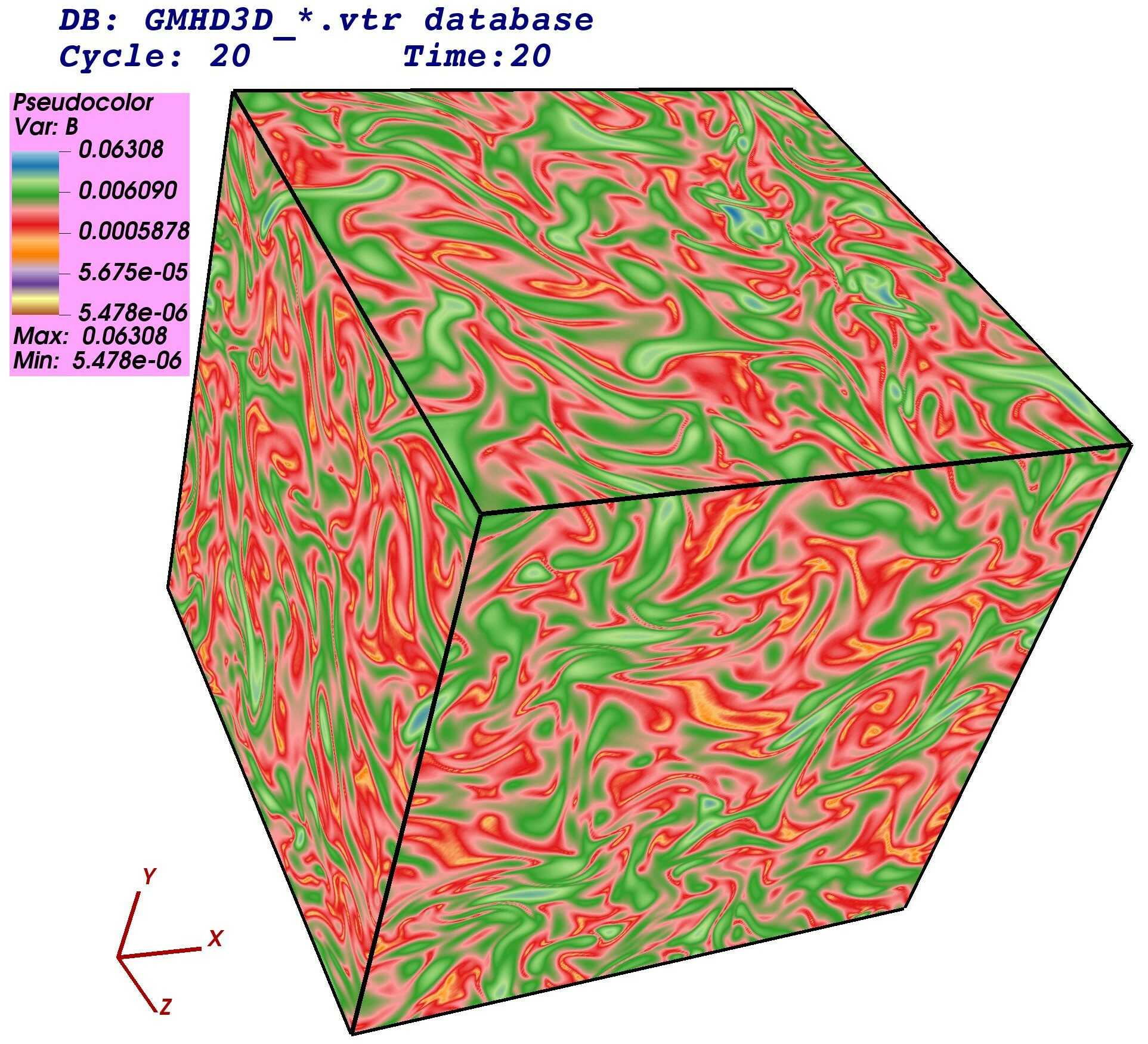}
		\caption{}
	\end{subfigure}
	\begin{subfigure}{0.32\textwidth}
		\centering
		\includegraphics[scale=0.087]{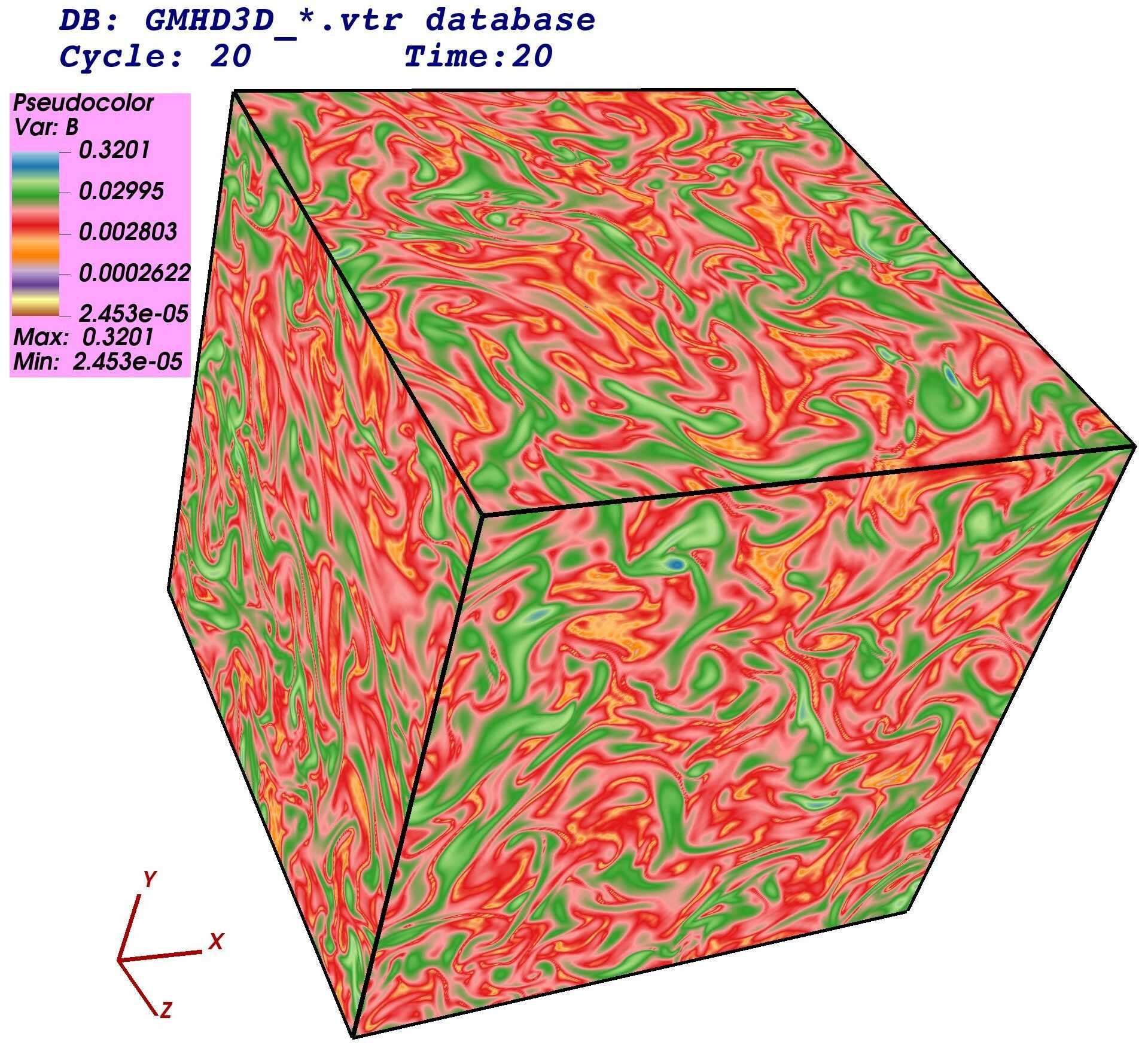}
		\caption{}
	\end{subfigure}
	\begin{subfigure}{0.32\textwidth}
		\centering
		\includegraphics[scale=0.087]{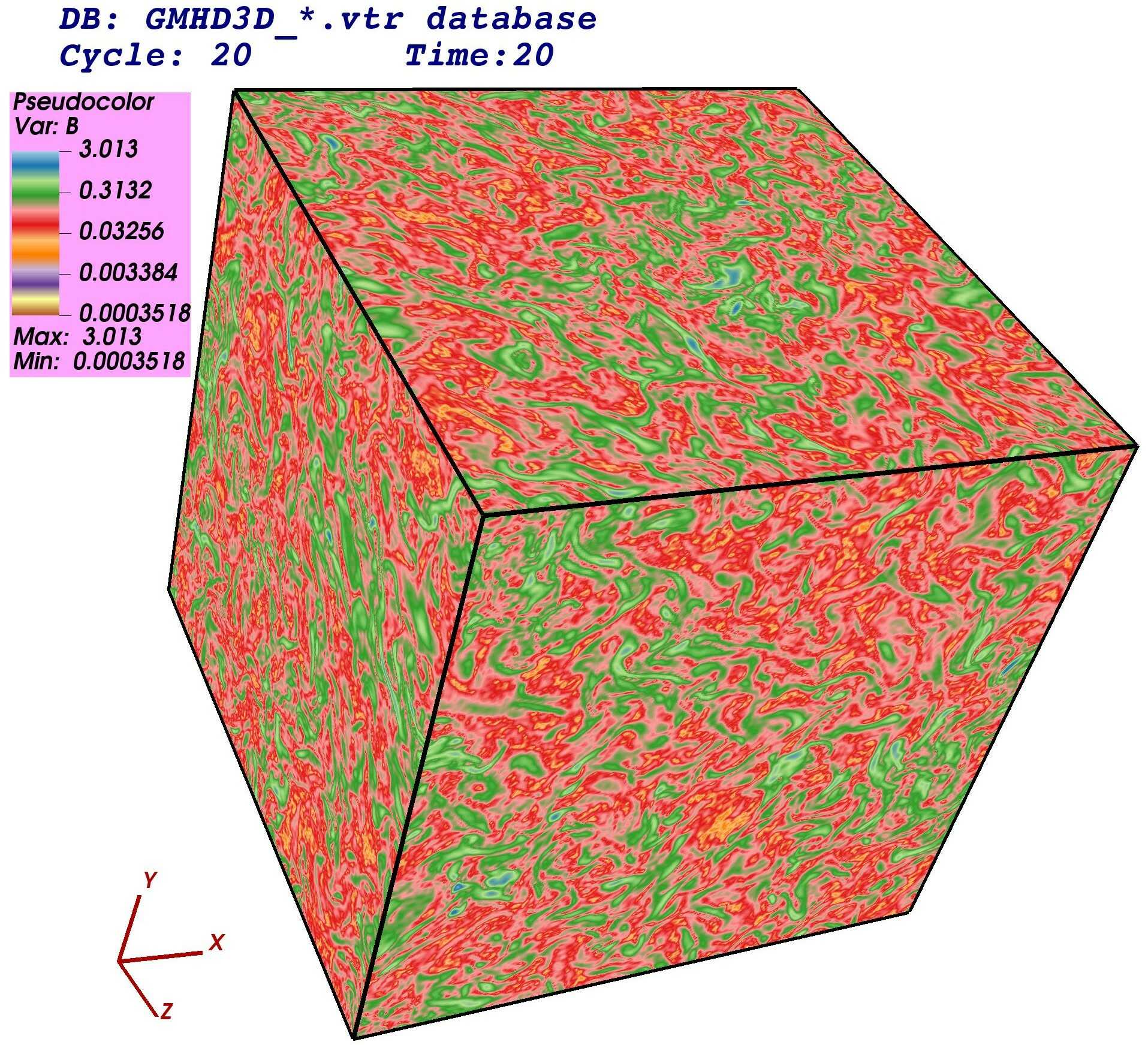}
		\caption{}
	\end{subfigure}
	\caption{ Visualization of the magnetic field in the kinematic stage at (a) $\beta = 0.0$, (b) $\beta = 0.5$, and (c) $\beta = 1.0$. The helical configuration of the YM flow drive is regulated by the parameter $\beta$. With increasing $\beta$, the magnetic field structures become more smaller in size. Visualization is conducted using a logarithmic scale.} 
	\label{controled helicity visulization}
\end{figure*}

The magnetic field structures for the YM flow drive classes have been visualized. Through the implementation of controlled helicity injection, more finer magnetic field structures are generated, as demonstrated by our numerical experiments (See Fig. \ref{controled helicity visulization}). This observation suggests that the helical structure of the flow drive influences the SSD structure. 

\end{document}